\documentclass[%
preprint,
 amsmath,amssymb,
 aps,prab,
 onecolumn,
]{revtex4-2}

\usepackage{graphicx}
\usepackage{graphics}
\usepackage{dcolumn}
\usepackage{bm}
\usepackage{amsmath}
\usepackage{amssymb}
\usepackage{tikz}
\usepackage[english]{babel}

\begin{document}

\title{Circular-Mode Optics for High-Intensity Beams and High-Luminosity Colliders}

\author{Onur~Gilanliogullari}
\email{onur.gilanliogullari@physics.uu.se}
\affiliation{Department of Physics, Illinois Institute of Technology,
Chicago, Illinois 60616, USA}
\affiliation{Physics Division, Argonne National Laboratory,
Lemont, Illinois 60439, USA}
\altaffiliation{Present address: Department of Physics and Astronomy,
FREIA Division, Uppsala University, Uppsala, Sweden}

\author{Brahim~Mustapha}
\affiliation{Physics Division, Argonne National Laboratory,
Lemont, Illinois 60439, USA}

\author{Pavel~Snopok}
\affiliation{Department of Physics, Illinois Institute of Technology,
Chicago, Illinois 60616, USA}
\date{\today}

\begin{abstract}
    We study circular-mode optics as a framework for transporting and accelerating intrinsically flat beams while maintaining a round transverse projection in physical space. Motivated by the strong degradation of flat beams under low-energy space-charge forces, we develop a coupled-optics description of circular modes, analyze their formation and preservation, and derive lattice design conditions for conserving canonical angular momentum and coupling phase. These conditions motivate rotationally invariant transport and ring cells based on solenoids, symmetric alternating-gradient structures, and indexed dipoles. We further derive space-charge tune-shift and tune-spread expressions for circular-mode beams in uncoupled and coupled lattices, showing that they preserve the leading-order space-charge behavior of round beams while retaining intrinsic eigenemittance flatness. Particle-in-cell simulations with TRACK and ImpactX confirm the generation, transport, and acceleration of circular-mode beams in representative lattices, including regimes with strong space charge. Circular-mode optics therefore provides a promising route toward high-intensity hadron beams and high-luminosity collider applications.
\end{abstract}

\maketitle

\section{Introduction}
Particle accelerators have been essential for scientific breakthroughs and for a wide range of medical and industrial uses. Advances in accelerator technology have paved the way for exploring higher-energy scales in high-energy physics (the energy frontier). 
However, discoveries involving rare events, as well as physics insights and future applications, require a significant increase in beam brightness and collision luminosity in future colliders (intensity frontier).

The brightness of a beam, $\mathcal{B}$, and the collision luminosity of two beams, $\mathcal{L}$, are defined respectively as
\begin{equation}
   \mathcal{B}=\frac{I}{8\pi^{2}\epsilon_{x}\epsilon_{y}}, \quad
   \mathcal{L}= \frac{f_{0}N_{1}N_{2}}{4\pi\sigma_{x}^{*}\sigma_{y}^{*}}, 
   \label{Eq:qualityparameters}
\end{equation}
where $I$ is the beam intensity, $\epsilon_{x,y}$ are the beam emittances, $\sigma_{x,y}^{*}=\sqrt{\beta_{x,y}^{*}\epsilon_{x,y}}$ are the beam sizes at the interaction point, $f_{0}$ is the repetition rate, and $N_{1,2}$ are the numbers of particles in each of the colliding beams. Unless otherwise specified, $\epsilon_x,\epsilon_y,\epsilon_{1,2}$ denote geometric rms emittances. In accelerating simulations, normalized emittances are used and are stated explicitly. Clearly, both beam brightness and luminosity are related to beam emittances. Luminosity and beam brightness can be defined as beam quality parameters~\cite{nagaitsev2021accelerator}. One of the research and development goals in accelerator physics is to increase the beam quality towards the quantum degeneracy limit~\cite{nagaitsev2021accelerator}. This objective can be realized by designing high-brightness beams that support high intensity while keeping the emittance small. This process begins at low energies, where collective effects, such as space charge, are dominant and often detrimental to beam quality. To enhance the quality parameters of Eq.~\eqref{Eq:qualityparameters}, the intensity of the beam should increase, or the beam emittance should decrease, or both. These are often limited by space-charge effects, especially in proton and heavy-ion beams. 

Flat beams, where the emittance in one of the transverse planes is much smaller than in the other plane (typically $\epsilon_{y}\ll\epsilon_{x}$), are good candidates for enhancing the beam quality, leading to higher beam brightness and collision luminosity~\cite{burov2013circular}. Compared to a round beam with equal emittances, $\epsilon_{x}=\epsilon_{y}=\epsilon_{0}$, a flat beam with the same $\epsilon_x$ emittance and $\epsilon_y$ smaller by a factor $\mathcal{R}$, $\epsilon_{y}=\epsilon_{x}/\mathcal{R}$, has an overall 4D emittance reduced by a factor of $\mathcal{R}$.
The comparison of brightness and luminosity between round and flat beams with a flatness ratio $\mathcal{R}$ is given by
\begin{equation}
    \frac{\mathcal{B}_{\text{f}}}{\mathcal{B}_{\text{r}}}=\mathcal{R}, \qquad 
    \frac{\mathcal{L}_{\text{f}}}{\mathcal{L}_{\text{r}}} = \sqrt{\mathcal{R}}.
    \label{Eq:enhancement}
\end{equation}
Therefore, the enhancement in beam brightness is equal to $\mathcal{R}$, while for collision luminosity it is only $\sqrt{\mathcal{R}}$. This is because $\mathcal{B}$ is proportional to beam emittances, while $\mathcal{L}$ depends on the beam sizes (square root of emittances) at the collision point. Hence, an order of magnitude improvement in beam brightness can be achieved for a beam flatness ratio of $R = 10$, while an $R=100$ is required for an order of magnitude enhancement in collision luminosity. This comparison assumes equal beam intensity and equal optics at the interaction point.

However, at low energies, direct space-charge forces dominate the transverse dynamics of intense proton and heavy-ion beams. The collective defocusing grows rapidly as energy decreases---roughly scaling with the line current as $\lambda_{z}/(\beta^{2}\gamma^{3})$, where $\lambda_z$ is line current density, $\beta,\gamma$ the relativistic parameters---so the space-charge regime is most severe during injection and early acceleration. For flat beams ($\sigma_{y}\ll\sigma_{x}$), charge density and self-field gradients are concentrated in the thin plane, making envelope control, emittance preservation, and loss mitigation particularly challenging. As a result, direct space-charge forces tend to reduce the transverse aspect ratio, driving initially flat beams toward a more round transverse profile.

\begin{figure}[tbp]
    \centering
        \raisebox{-.42\height}{\includegraphics[width=0.45\linewidth]{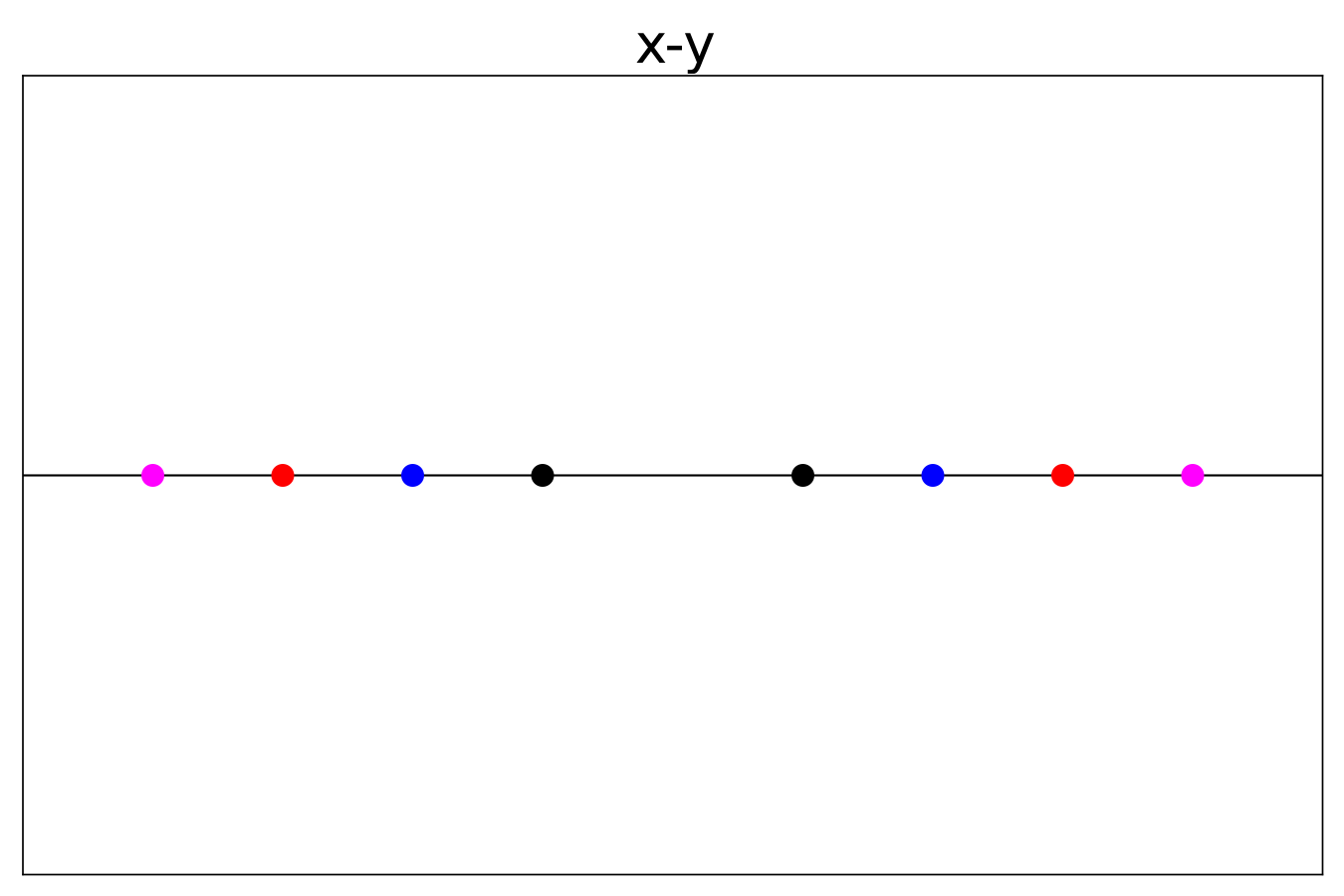}}
        {\huge $\Rightarrow$}
        \raisebox{-.42\height}{\includegraphics[width=0.45\linewidth]{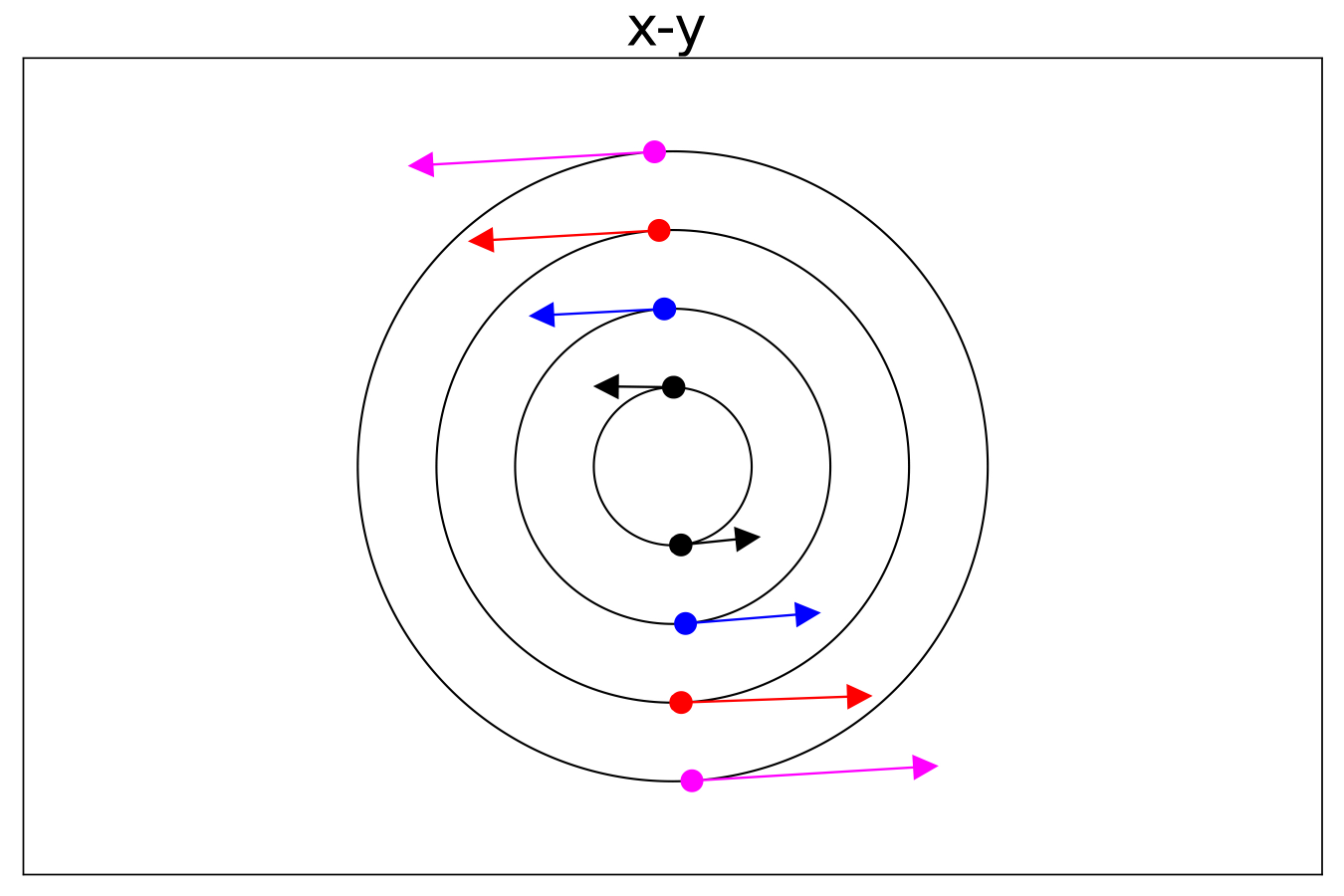}}
        \raisebox{-.42\height}{\includegraphics[width=0.45\linewidth]{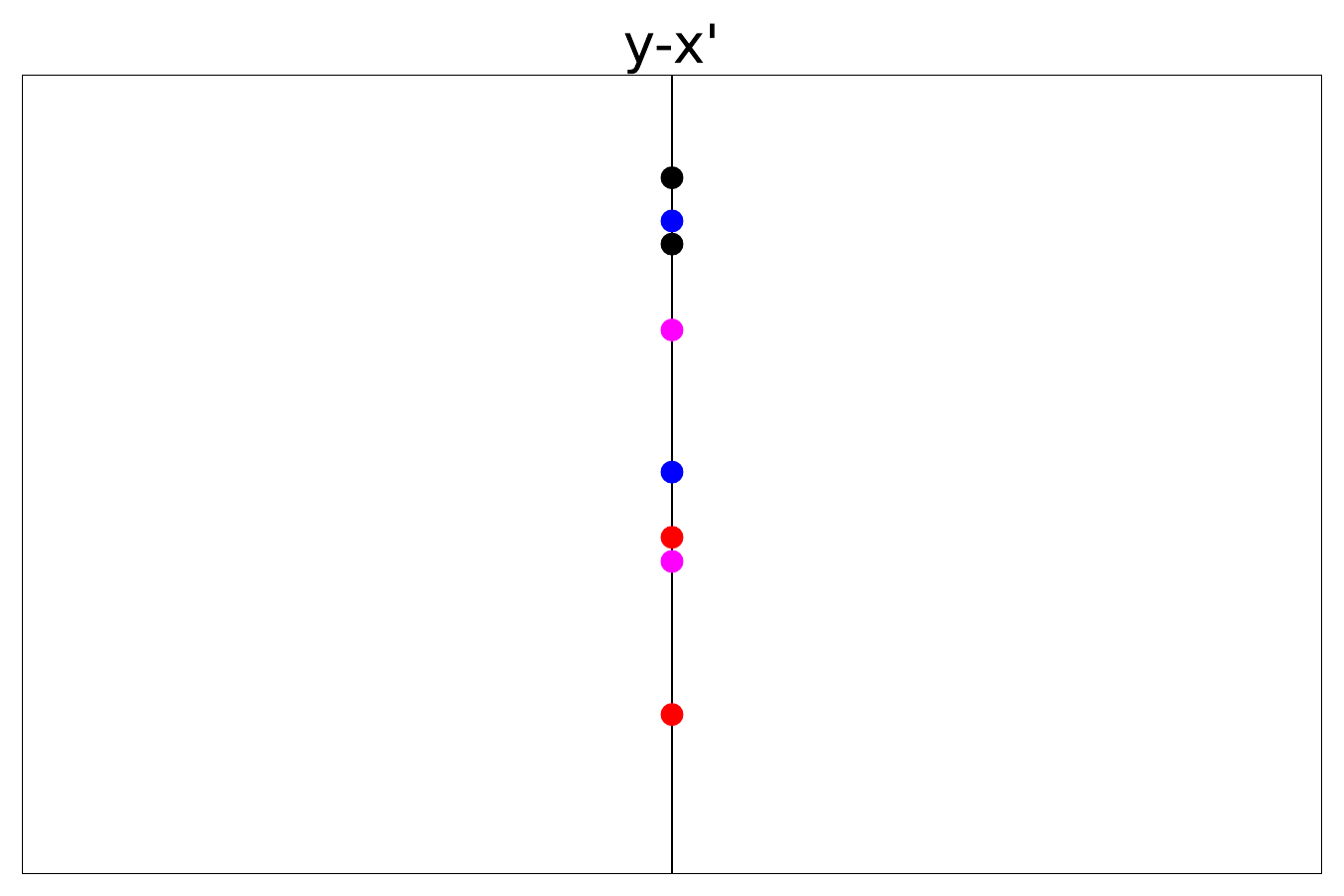}}
        {\huge $\Rightarrow$}
        \raisebox{-.42\height}{\includegraphics[width=0.45\linewidth]{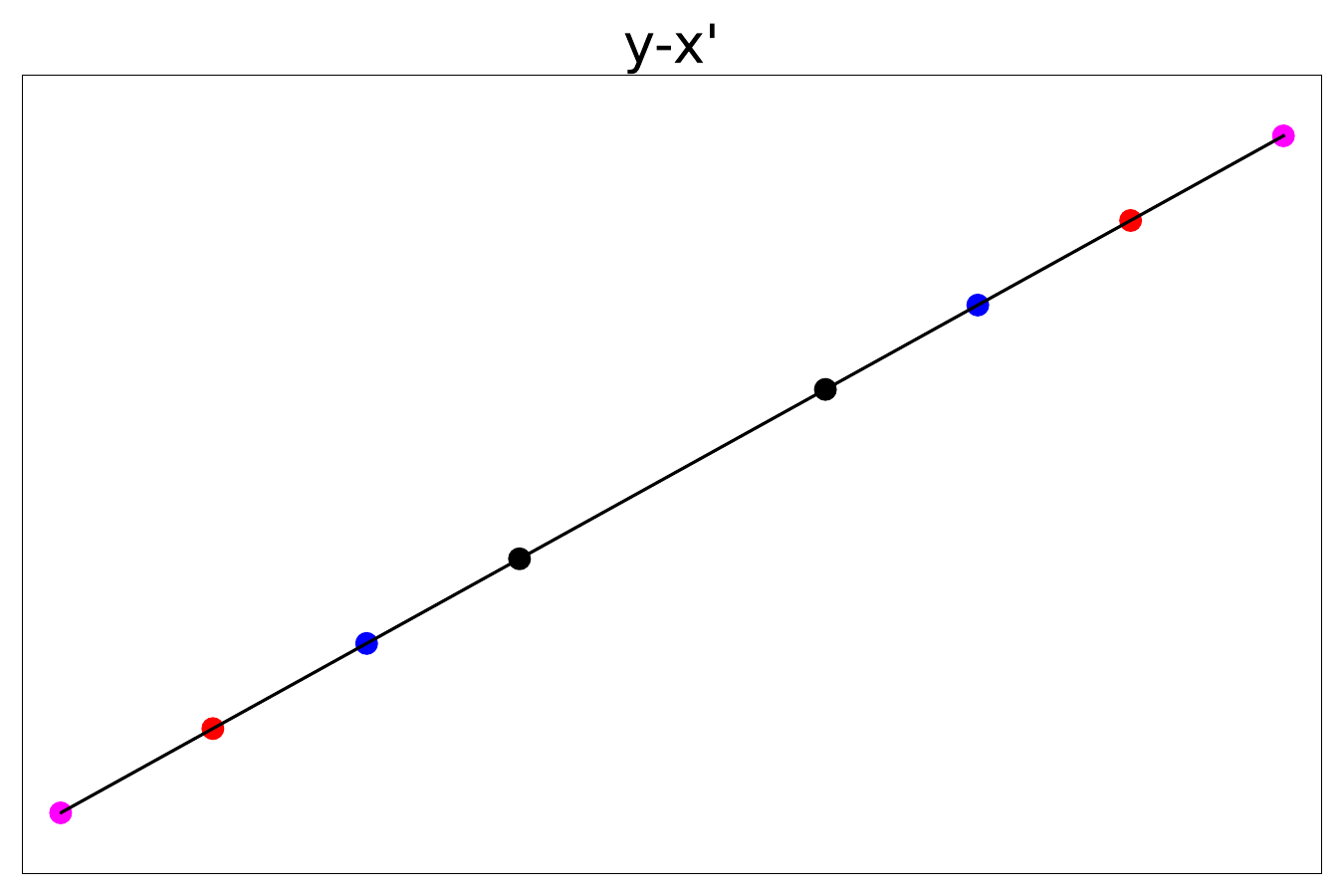}}
    \caption{Circular-mode beam creation using Derbenev’s adapter. A physically flat beam is transformed into a beam with a round (x,y) projection but strong intrinsic flatness in the eigenmode/mixed phase-space planes with angular momentum dominance, $(y,x')$ plane.}
    \label{fig:exampletransformation}
\end{figure}

Circular-mode beams provide a possible way to address this limitation. Circular-mode beams are strongly-coupled and angular-momentum-dominated beams with a round cross-section and intrinsic flatness. They require a relatively simple transformation~\cite{derbenev1998adapting} to become flat in physical space when needed. Circular modes were initially proposed by Derbenev~\cite{derbenev1998adapting} for a high-energy cooling experiment, where a flat beam is converted to a round beam with non-zero angular momentum using three skew quadrupoles. Angular momentum is generated via the coupled section comprising three skew quadrupoles, known as Derbenev's Adapter. The Derbenev Adapter establishes strong coupling between the $x$ and $y$ planes, mapping the beam's physical flatness to the eigenmode planes while producing a round beam in the transverse plane. The experimental demonstration of flat-to-round converters is shown in~\cite{kim2025experimental}. Figure~\ref{fig:exampletransformation} illustrates this transformation from a physically flat beam to a rotating round beam. The physical flatness of the initial flat beam is converted into intrinsic flatness. The flatness is moved to the $(y,x')$ and $(x,y')$ phase planes which produces the angular-momentum-dominated structure. 

Magnetized beams satisfying the vortex condition~\cite{groening2021particle} provide another important example of circular modes, alongside certain self-consistent beam distributions~\cite{danilov2003self,holmes2018injection}. The vortex condition, introduced in~\cite{derbenev1998adapting,groening2021particle}, leads to intrinsically coupled transverse dynamics. Because such beams carry nonzero canonical angular momentum, their motion is coupled in the full four-dimensional transverse phase space and must therefore be described using coupled beam optics, as discussed in the theory section. In most conventional lattice designs, transverse coupling is usually minimized because it complicates both the beam dynamics and the phase-space parametrization. In contrast, circular modes arise precisely in systems with strong transverse coupling. For uncoupled lattices, the standard Courant–Snyder parametrization~\cite{courant1958theory} remains applicable; however, circular-mode optics in rings generally require solenoidal focusing and therefore a coupled-optics description.


The theory of circular modes has already been established in~\cite{burov2002circular,derbenev1998adapting,kim2003round}.
The new contribution of this work is not the introduction of circular modes themselves, but rather a unified accelerator-design framework for preserving circular modes through space-charge-dominated transport and acceleration. In particular, we identify practical lattice conditions for maintaining the circular-mode state, compare solenoid-based and symmetric alternating-gradient implementations, derive tune-shift and tune-spread expressions in the circular-mode limit, and verify the preservation and eventual decoupling of high-flatness eigenemittance beams with self-consistent PIC tracking.

This manuscript is organized as follows. Section~\ref{sec:theory} introduces the theory of circular-mode beams and summarizes their main properties. Section~\ref{sec:creationofcircmode} presents methods for generating circular-mode beams. Section~\ref{sec:preservationofcircularmodes} discusses the conditions required to preserve circular modes and canonical angular momentum. Section~\ref{sec:spacechargeeffects} examines space-charge effects, including tune shift and space-charge-induced tune spread. Finally, Section~\ref{sec:Results} presents applications to transport lines and ring designs that preserve circular modes, and evaluates their performance under space charge using the PIC codes TRACK and ImpactX.

\section{Theory}\label{sec:theory}


This section reviews the established theory of circular modes ~\cite{derbenev1998adapting,burov2002circular,kim2003round} and specializes the Lebedev–Bogacz coupled-optics parametrization~\cite{lebedev2010betatron} to the circular-mode limit. The derivations presented here and in Appendix A are included for pedagogical completeness and to establish the notation and relationships among optical parameters, eigenemittances, and canonical angular momentum used throughout this work. This section thus provides the theoretical background for the subsequent analyses of circular-mode preservation, space-charge effects, and lattice design.

In a transversely coupled system with rotational symmetry, the linear particle motion can be decomposed into two eigenmodes with opposite senses of rotation. These are referred to as circular modes~\cite{burov2002circular,kim2003round}. The two modes have opposite rotational handedness and therefore carry angular momenta of opposite signs. In the high-flatness circular-mode state considered in this work, one mode has a substantially larger eigenemittance than the other. The beam then has a preferred rotational handedness, nonzero net canonical angular momentum, and strong antisymmetric correlations between the horizontal coordinate and vertical momentum and between the vertical coordinate and horizontal momentum.

A circular-mode beam is therefore fundamentally different from an ordinary uncorrelated round beam. Both can have a round projected density in the physical $(x,y)$ plane, but an uncorrelated round beam generally has no net transverse angular momentum and has comparable transverse eigenemittances. A circular-mode beam, by contrast, can have a round physical projection while retaining a large separation between its two eigenemittances.

The ideal single-mode state can be summarized by the vortex relation
\begin{equation}
    \begin{pmatrix}
        y\\
        y'
    \end{pmatrix}
    =
    \begin{pmatrix}
        0 & -\beta_r\\
        1/\beta_r & 0
    \end{pmatrix}
    \begin{pmatrix}
        x\\
        x'
    \end{pmatrix}.
    \label{eq:vortexreltheory}
\end{equation}
Here, \(\beta_r\) is the circular-mode beta function, and \((x,x',y,y')^T\) denotes the transverse phase-space vector. Equation~\eqref{eq:vortexreltheory} implies \(y=-\beta_r x'\) and \(y'=x/\beta_r\). Thus, the horizontal and vertical oscillations are separated by a phase of \(\pi/2\), and the particle trajectory has a circular projection in the \((x,y)\) plane at an ideal circular-mode reference location. Reversing the signs of the off-diagonal elements produces the oppositely rotating mode. A geometric derivation of the vortex relation and its connection to the Courant--Snyder parametrization are given in Appendix~\ref{subsec:appendixAvortexrel}.

The vortex relation provides an intuitive description of an ideal beam occupying a single circular mode. A general beam, however, can contain both counter-rotating eigenmodes, and their projections onto the laboratory coordinates evolve as the beam propagates through the lattice. A general coupled-optics description is therefore required to determine the projected beam shape, the canonical angular momentum, and the relative contribution of each mode.

Several coupled-optics formalisms are available, including the Edwards--Teng~\cite{edwards1973parametrization}, Mais--Ripken~\cite{willeke1989methods}, and Lebedev--Bogacz~\cite{lebedev2010betatron} parametrizations. We adopt the Lebedev--Bogacz parametrization because it expresses the physical coordinates directly in terms of the eigenvectors of a general linear symplectic \(4\times4\) transfer matrix and makes the coupling strength and coupling phases explicit.

\subsection{Coupled eigenmode description}\label{subsec:theoryeigenmodedesc}

To construct the general coupled-optics description, consider the transverse phase-space vector$\vec{z}=(x,x',y,y')^T$. For stable linear motion, the one-period \(4\times4\) symplectic transfer matrix has two pairs of complex-conjugate eigenvectors. Each pair defines an independent transverse normal mode. Although the two modes evolve independently in normal-mode coordinates, each mode generally has nonzero projections onto both the horizontal and vertical laboratory coordinates. At a given longitudinal position, the transverse phase-space vector can be parametrized as
\begin{equation}
    \vec{z}
    =
    \frac{1}{2}\sqrt{2J_1}\,\vec{v}_1 e^{-i\psi_1}
    +
    \frac{1}{2}\sqrt{2J_2}\,\vec{v}_2 e^{-i\psi_2}
    +
    \frac{1}{2}\sqrt{2J_1}\,\vec{v}_1^{*} e^{i\psi_1}
    +
    \frac{1}{2}\sqrt{2J_2}\,\vec{v}_2^{*} e^{i\psi_2}.
    \label{eq:phasespaceparamLB}
\end{equation}
The complex-conjugate terms ensure that \(\vec{z}\) is real. Here, \(J_1\) and \(J_2\) are the single-particle actions, \(\psi_1\) and \(\psi_2\) are the corresponding betatron phases, and \(\vec{v}_1\) and \(\vec{v}_2\) are the symplectically normalized eigenvectors of the transverse transfer matrix. In an ideal linear lattice, the actions are invariants, whereas the betatron phases advance through the lattice. For a beam ensemble, the phase-space averages of the actions define the rms eigenemittances, \(\epsilon_j=\langle J_j\rangle\), for \(j\in\{1,2\}\).

Using the Lebedev--Bogacz parametrization~\cite{lebedev2010betatron}, the eigenvectors are written as
\begin{equation}
    \vec{v}_1=
    \begin{pmatrix}
        \sqrt{\beta_{1x}} \\[1mm]
        -\dfrac{i(1-u)+\alpha_{1x}}{\sqrt{\beta_{1x}}} \\[2mm]
        \sqrt{\beta_{1y}}e^{i\nu_1} \\[1mm]
        -\dfrac{iu+\alpha_{1y}}{\sqrt{\beta_{1y}}}e^{i\nu_1}
    \end{pmatrix},
    \qquad
    \vec{v}_2=
    \begin{pmatrix}
        \sqrt{\beta_{2x}}e^{i\nu_2} \\[1mm]
        -\dfrac{iu+\alpha_{2x}}{\sqrt{\beta_{2x}}}e^{i\nu_2} \\[2mm]
        \sqrt{\beta_{2y}} \\[1mm]
        -\dfrac{i(1-u)+\alpha_{2y}}{\sqrt{\beta_{2y}}}
    \end{pmatrix}.
    \label{eq:eigenvectorsgen}
\end{equation}
For eigenmode \(j\), the functions \(\beta_{jx}\) and \(\beta_{jy}\) describe its projections onto the horizontal and vertical coordinates, respectively. The corresponding functions \(\alpha_{jx}\) and \(\alpha_{jy}\) describe the slopes of the projected phase-space ellipses. The coupling phases \(\nu_1\) and \(\nu_2\) specify the relative phases between the horizontal and vertical components of the two eigenmodes. Together with the coupled beta functions and eigenemittances, these phases determine whether the intrinsic eigenmode flatness appears as a round, tilted, or flat projection in the physical \((x,y)\) plane.

The scalar parameter \(u\) is an optics parameter that characterizes the distribution of the normal-mode motion between the two physical planes. It should not be confused with the relative population of the two modes, which is determined by the eigenemittances. The limiting values of \(u\) have the following interpretations:
\begin{itemize}
    \item \(u=0\): the decoupled limit, in which eigenmode~1 reduces to purely horizontal motion in \((x,x')\), while eigenmode~2 reduces to purely vertical motion in \((y,y')\);

    \item \(u=1\): the decoupled limit with the mode labels exchanged, so that eigenmode~1 reduces to vertical motion and eigenmode~2 to horizontal motion;

    \item \(u=\tfrac{1}{2}\): the maximally coupled value in the Lebedev--Bogacz convention. This is one of the conditions required for the ideal circular-mode limit.
\end{itemize}

The condition \(u=\tfrac{1}{2}\) is necessary but not sufficient to define an ideal circular mode. The coupled beta functions, alpha functions, and coupling phases must also satisfy the circular-mode conditions introduced in the following subsection. Furthermore, whether one circular mode dominates the beam is determined by the eigenemittance ratio \(\epsilon_1/\epsilon_2\), rather than by \(u\).

Equation~\eqref{eq:phasespaceparamLB} and the eigenvectors in Eq.~\eqref{eq:eigenvectorsgen} provide a general description of stable linear coupled motion. Circular modes correspond to a particular set of optical conditions within this general parametrization. We now specialize the coupled eigenvectors to this limit and show how the resulting rotational correlations, canonical angular momentum, and eigenemittance separation are related.

\subsection{Circular-mode limit and intrinsic flatness}
\label{subsec:circmodelimittheorysection}

The parametrization introduced in Sec.~\ref{subsec:theoryeigenmodedesc} describes general stable linear coupled motion. Circular modes correspond to a particular limit in which the horizontal and vertical components of each eigenmode have equal amplitudes and remain in phase quadrature. At a circular-mode reference location, the required optical conditions are
\begin{equation}
    \begin{split}
        &\beta_{1x}=\beta_{2x}=\beta_{1y}=\beta_{2y}
        \equiv\beta_0,
        \qquad
        \alpha_{1x}=\alpha_{2x}=\alpha_{1y}=\alpha_{2y}=0,\\
        &\nu_1=\nu_2=\frac{\pi}{2}\;(\mathrm{mod}\,\pi),
        \qquad
        u=\frac{1}{2}.
    \end{split}
    \label{eq:circmodeconditionsonoptics}
\end{equation}
The condition \(u=1/2\) gives maximal transverse coupling in the Lebedev--Bogacz convention. Equality of the coupled beta functions gives equal horizontal and vertical mode amplitudes, while \(\nu_{1,2}=\pi/2\) places the two components of each eigenmode in phase quadrature. The vanishing alpha functions define the reference location at which the projected phase-space ellipses are upright. Together, these conditions define the ideal circular-mode state. Their derivation from the symplectic normalization conditions is given in Appendix~\ref{subsec:appendixAcircularmodecondLB}.

The connection between Eq.~\eqref{eq:circmodeconditionsonoptics} and the vortex relation in Eq.~\eqref{eq:vortexreltheory} can be seen by first considering a particle occupying only eigenmode~1. Substitution of the circular-mode conditions into Eqs.~\eqref{eq:phasespaceparamLB} and \eqref{eq:eigenvectorsgen} gives
\begin{equation}
    \begin{aligned}
        x  &= A_1\cos\psi_1,
        &\qquad
        x' &= -\frac{A_1}{\beta_r}\sin\psi_1,\\
        y  &= A_1\sin\psi_1,
        &
        y' &= \frac{A_1}{\beta_r}\cos\psi_1,
    \end{aligned}
    \qquad
    A_1=\sqrt{2J_1\beta_0},
    \qquad
    \beta_r=2\beta_0.
    \label{eq:mode1circularcoordinates}
\end{equation}
The coordinates in Eq.~\eqref{eq:mode1circularcoordinates} satisfy
\[
    y=-\beta_r x',
    \qquad
    y'=\frac{x}{\beta_r},
\]
and therefore recover the vortex relation in Eq.~\eqref{eq:vortexreltheory}. They also give \(x^2+y^2=A_1^2\), showing that a particle occupying eigenmode~1 has a circular projection in the physical \((x,y)\) plane at the reference location. Eigenmode~2 satisfies the corresponding relations with the opposite phase ordering and therefore rotates with the opposite handedness. Thus, the general coupled-eigenmode parametrization reproduces the ideal single-mode picture introduced at the beginning of Sec.~II.

A beam generally contains particles in both eigenmodes. Under the conditions of Eq.~\eqref{eq:circmodeconditionsonoptics}, the relevant transverse second moments are
\begin{equation}
    \begin{aligned}
        \langle x^2\rangle
        &=
        \langle y^2\rangle
        =
        (\epsilon_1+\epsilon_2)\beta_0,\\
        \langle x'^2\rangle
        &=
        \langle y'^2\rangle
        =
        \frac{\epsilon_1+\epsilon_2}{4\beta_0},\\
        \langle xy\rangle
        &=
        \langle x'y'\rangle
        =
        0,\\
        \langle xy'\rangle
        &=
        \frac{\epsilon_1-\epsilon_2}{2},
        \qquad
        \langle yx'\rangle
        =
        -\frac{\epsilon_1-\epsilon_2}{2}.
    \end{aligned}
    \label{eq:circularmodesecondmoments}
\end{equation}
The derivation of Eq.~\eqref{eq:circularmodesecondmoments} from the general Lebedev--Bogacz second moments is given in Appendix~\ref{subsec:appendixAsecondmoments}. The relations
\[
    \langle x^2\rangle=\langle y^2\rangle,
    \qquad
    \langle xy\rangle=0
\]
show that the rms beam projection is round and untilted in the physical \((x,y)\) plane. Importantly, this projected roundness depends on the sum \(\epsilon_1+\epsilon_2\) and does not require the eigenemittances to be equal. Their difference, \(\epsilon_1-\epsilon_2\), instead appears in the antisymmetric mixed-plane correlations.

\begin{figure}[t]
    \centering
    \includegraphics[width=\columnwidth]
    {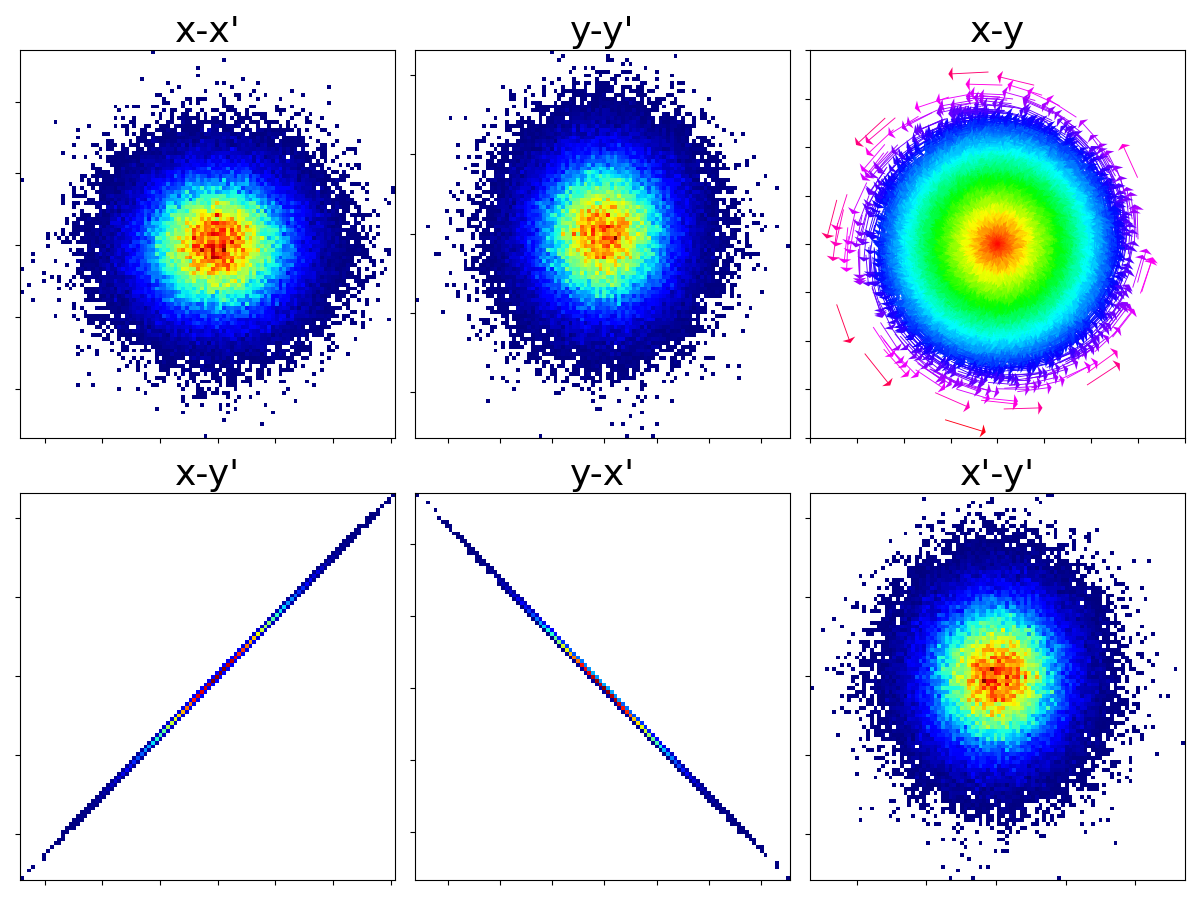}
    \caption{Six transverse phase-space projections of a mode-1-dominated circular-mode beam at an ideal circular-mode reference location. The \((x,y)\) projection is round, and the arrows indicate the local direction of transverse motion. The narrow distributions in the mixed planes \((x,y')\) and \((y,x')\) display the vortex correlations \(y'=x/\beta_r\) and \(x'=-y/\beta_r\). In the pure single-mode limit, these projections become straight lines; a finite secondary eigenemittance broadens them into narrow ellipses.}
    \label{fig:circularmodephasespaces}
\end{figure}

The second moments in Eq.~\eqref{eq:circularmodesecondmoments} therefore do not imply that the beam is isotropic in its full four-dimensional phase space. As illustrated in Fig.~\ref{fig:circularmodephasespaces}, the same-plane projections \((x,x')\) and \((y,y')\) are upright, while the mixed projections \((x,y')\) and \((y,x')\) exhibit strong positive and negative correlations, respectively. In the pure mode-1 limit, the vortex relations \(y'=x/\beta_r\) and \(x'=-y/\beta_r\) cause the mixed-plane projections to collapse to straight lines. A finite contribution from eigenmode~2 broadens these lines into narrow ellipses. Thus, the beam can be round in configuration space while retaining strong intrinsic flatness in its mixed-plane correlations.

In field-free regions, where \(x'\) and \(y'\) represent normalized transverse canonical momenta, the beam-averaged canonical angular momentum is
\begin{equation}
    L_{z,\mathrm{rms}}
    \equiv
    \left\langle xy'-yx'\right\rangle
    =
    \epsilon_1-\epsilon_2.
    \label{eq:circularmodeangularmomentum}
\end{equation}
Inside a solenoid, the canonical momenta additionally contain the magnetic-vector-potential contribution. Equation~\eqref{eq:circularmodeangularmomentum} is the corresponding field-free representation used to describe the beam correlations outside the solenoid body.

Equation~\eqref{eq:circularmodeangularmomentum} shows that the two counter-rotating eigenmodes contribute angular momenta of opposite signs. If \(\epsilon_1=\epsilon_2\), their contributions cancel, and the beam has zero net angular momentum even though the optics remain maximally coupled. If one eigenemittance is substantially larger than the other, this cancellation is incomplete, and the beam acquires a preferred rotational handedness.

\begin{figure}[t]
    \centering
    \includegraphics[width=0.95\columnwidth]
    {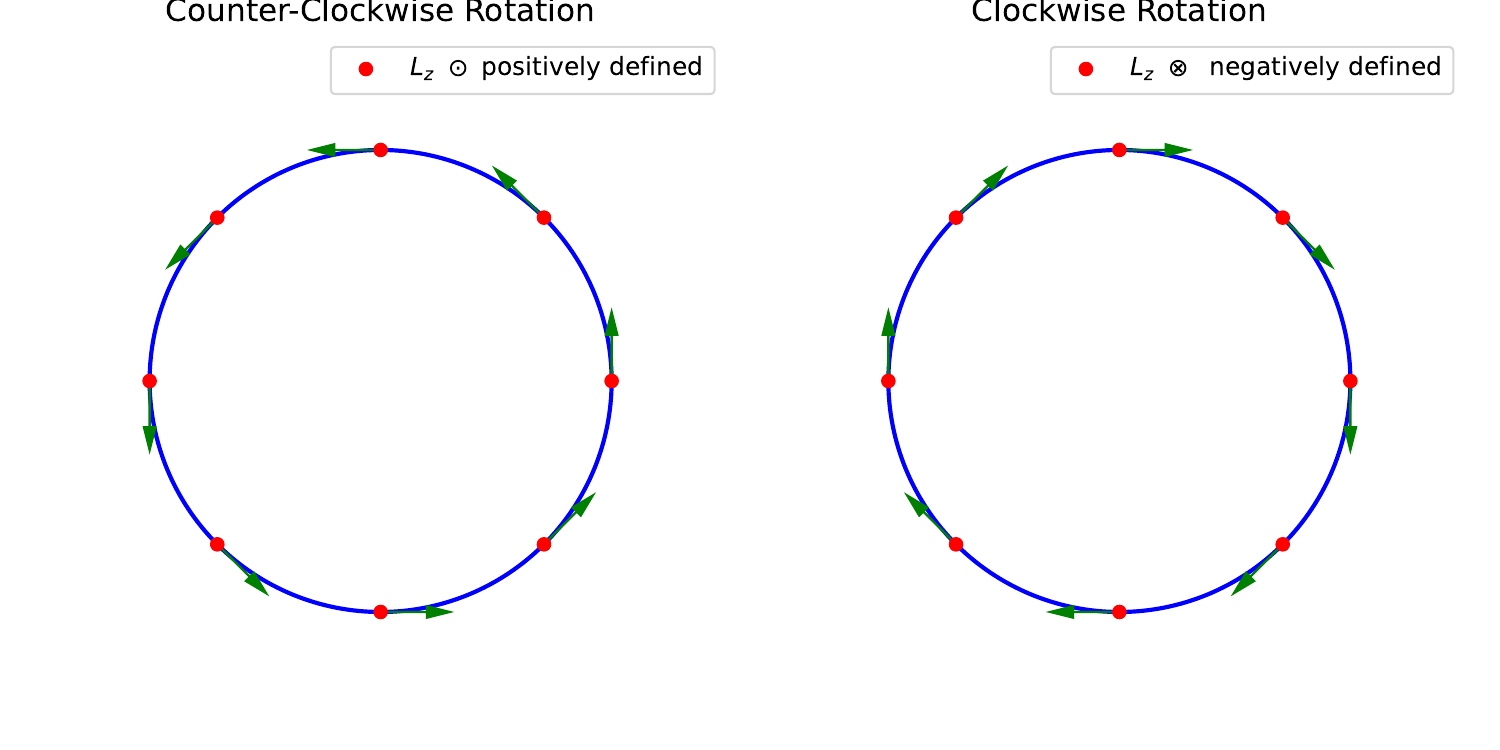}
    \caption{Rotational handedness of the two ideal circular eigenmodes. With the positive longitudinal axis directed out of the page and \(L_z=xy'-yx'\), counter-clockwise rotation corresponds to \(L_z>0\), while clockwise rotation corresponds to \(L_z<0\). The red markers indicate representative particle positions, and the green arrows indicate the direction of transverse motion.}
    \label{fig:circularmodehandedness}
\end{figure}

The sign convention adopted in this work is illustrated in Fig.~\ref{fig:circularmodehandedness}. Counter-clockwise rotation in the \((x,y)\) plane, viewed along the positive longitudinal axis, is defined to have positive angular momentum. With this convention, eigenmode~1 carries positive angular momentum, whereas eigenmode~2 carries negative angular momentum.

The separation between the eigenemittances is characterized by the intrinsic flatness ratio
\begin{equation}
    \mathcal{R}
    =
    \frac{\epsilon_1}{\epsilon_2}.
    \label{eq:intrinsicflatnessratio}
\end{equation}
For the mode-1-dominated case considered throughout this work, \(\mathcal{R}\gg1\). Combining Eqs.~\eqref{eq:circularmodeangularmomentum} and \eqref{eq:intrinsicflatnessratio} gives
\begin{equation}
    L_{z,\mathrm{rms}}
    =
    \epsilon_1
    \left(
        1-\frac{1}{\mathcal{R}}
    \right).
    \label{eq:angularmomentumflatness}
\end{equation}
As \(\mathcal{R}\) increases, the contribution of eigenmode~2 becomes progressively smaller, and the beam approaches the pure mode-1 circular limit. Conversely, a mode-2-dominated beam has the opposite rotational handedness and negative angular momentum under the convention adopted here.

A circular-mode beam is therefore round in its physical transverse projection but intrinsically flat in its eigenmode structure. The intrinsic flatness can subsequently be converted into unequal projected emittances by a decoupling transformation, such as the inverse Derbenev adapter, thereby producing a physically flat beam.

\section{Creation of Circular Modes}\label{sec:creationofcircmode}
In this section, we discuss three approaches for generating circular modes: Derbenev adapter, magnetized beams, and phase-space painting.

\subsection{Derbenev's Adapter}\label{subsec:derbenevadapter}
The Derbenev Adapter is a system of three skew quadrupoles; it facilitates the transformation of flat beams into circular modes~\cite{derbenev1998adapting}. The detailed formalism for the flat-to-round adapter is presented in~\cite{burov2002circular,kim2003round,gilanliogullari2025formalism,kim2025experimental}. To form a circular-mode beam, the Derbenev Adapter uses a skew quadrupole triplet satisfying the following conditions:
\begin{equation}
		\mathcal{M}=R(\pi/4)\begin{pmatrix}
			M & 0 \\
			0 & N
		\end{pmatrix}R^{-1}(\pi/4), \quad
		N=\begin{pmatrix}
			0 & -\beta_{r} \\ 
			\frac{1}{\beta_{r}} & 0
		\end{pmatrix}\cdot M.
\end{equation}
Here, $M$ and $N$ are the $(x,x')$ and $(y,y')$ phase space transfer matrices, respectively. $R$ is $4\times 4$ rotation matrix and $\mathcal{M}$ is the $4\times 4$ transverse transfer matrix. This relation between the transfer matrices $N$ and $M$ ensures that the transformed beam satisfies the vortex condition. Derbenev's adapter maps the uncoupled $(x, y)$ plane emittances onto the eigenmode emittances, $\epsilon_{x}\rightarrow\epsilon_{1}$ and $\epsilon_{y}\rightarrow\epsilon_{2}$, by creating a strong coupling between $x$ and $y$ phase spaces. To create a circular-mode beam, the initial uncoupled beam must be a flat beam, with a large emittance ratio $\epsilon_{y}\ll\epsilon_{x}$. 

\begin{figure}[tbp]
    \centering
    \includegraphics[width=0.49\linewidth]{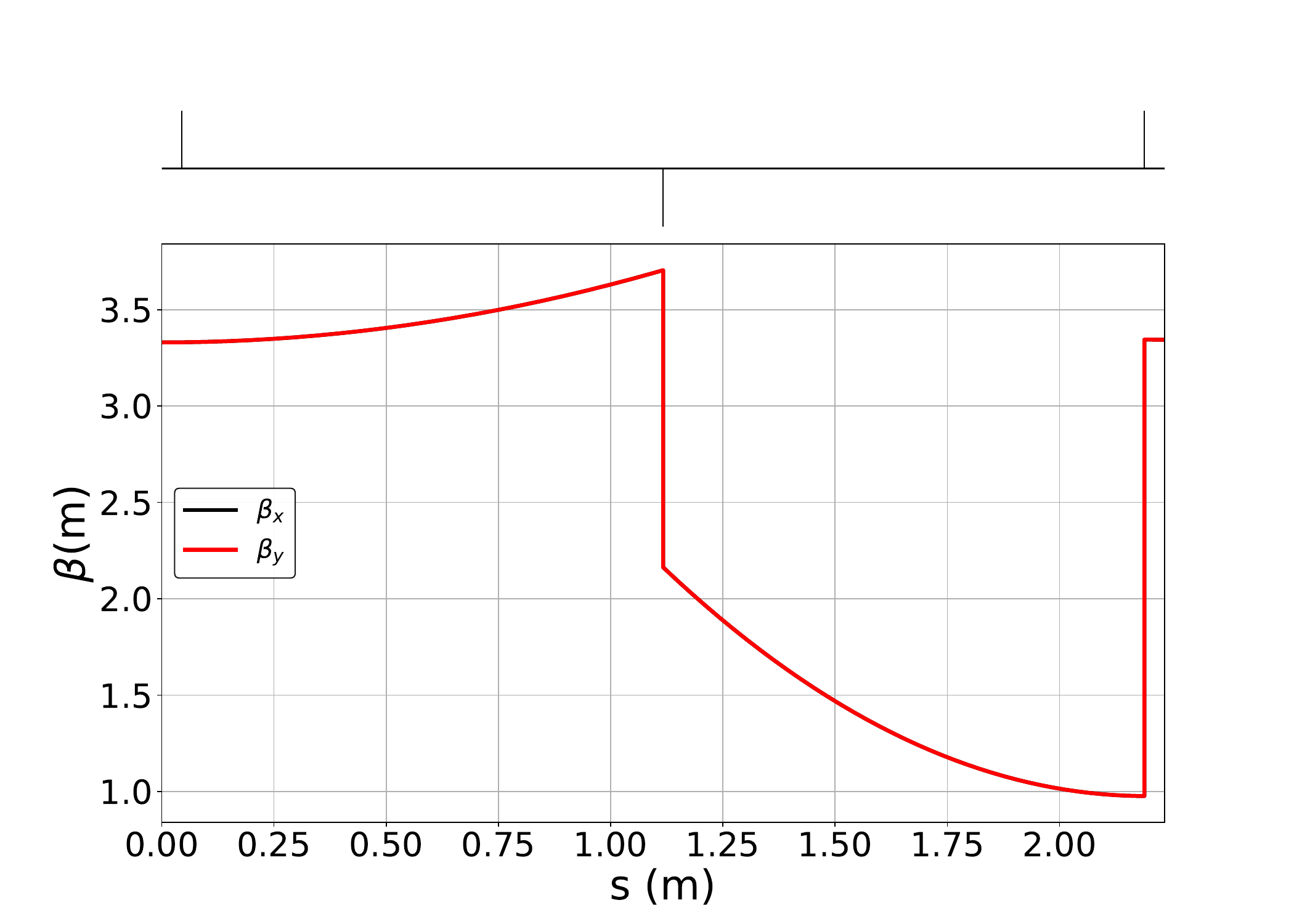}
    \includegraphics[width=0.49\linewidth]{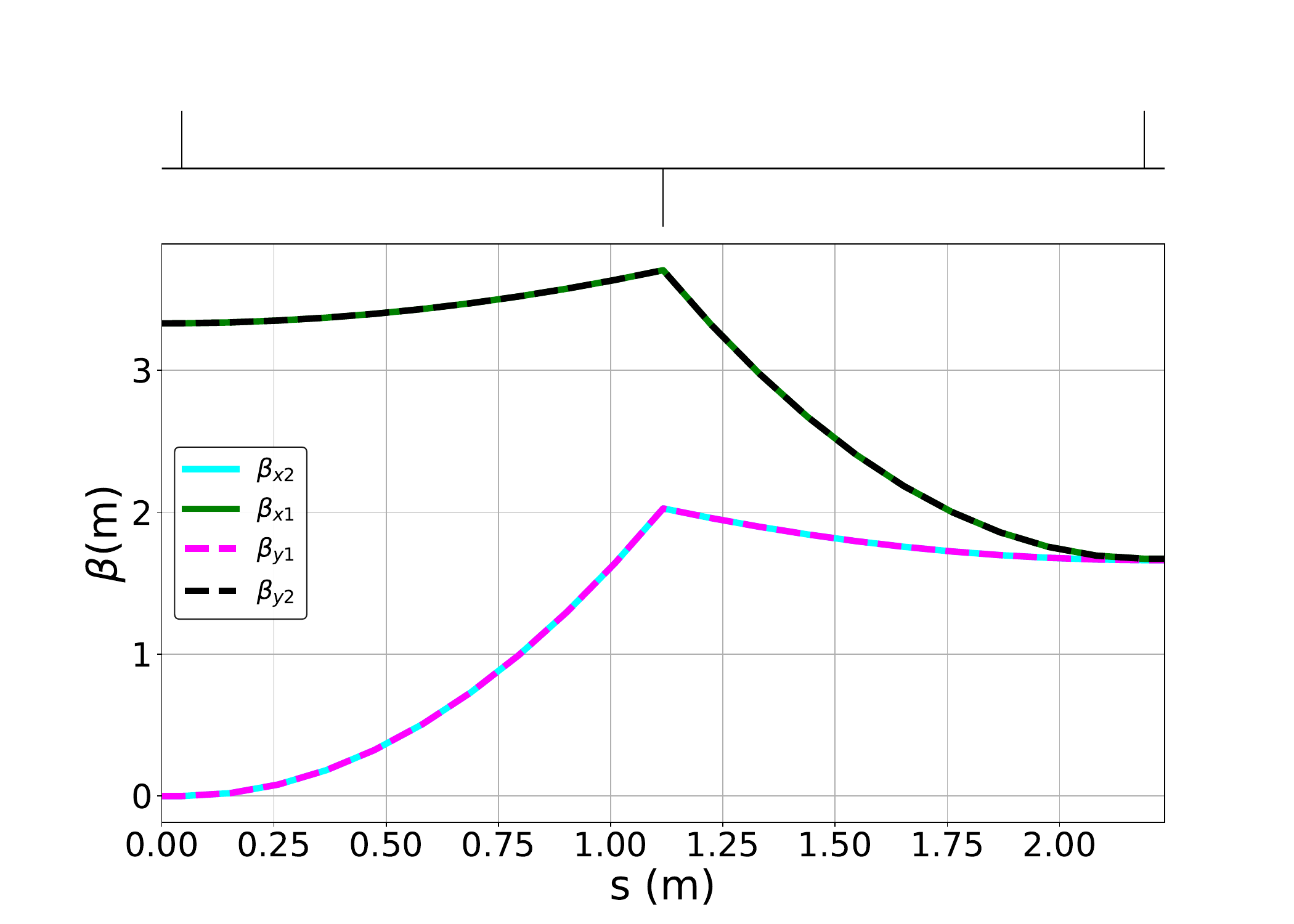}
    \caption{Derbenev's adapter optics functions. Left: decoupled $\beta_x$ and $\beta_y$. Right: coupled (eigenmode) $\beta$ functions in the Lebedev--Bogacz parametrization.}
    \label{fig:derbenevadapteroptics}
\end{figure}

\begin{figure}[tbp]
    \centering
    \includegraphics[width=0.49\linewidth]{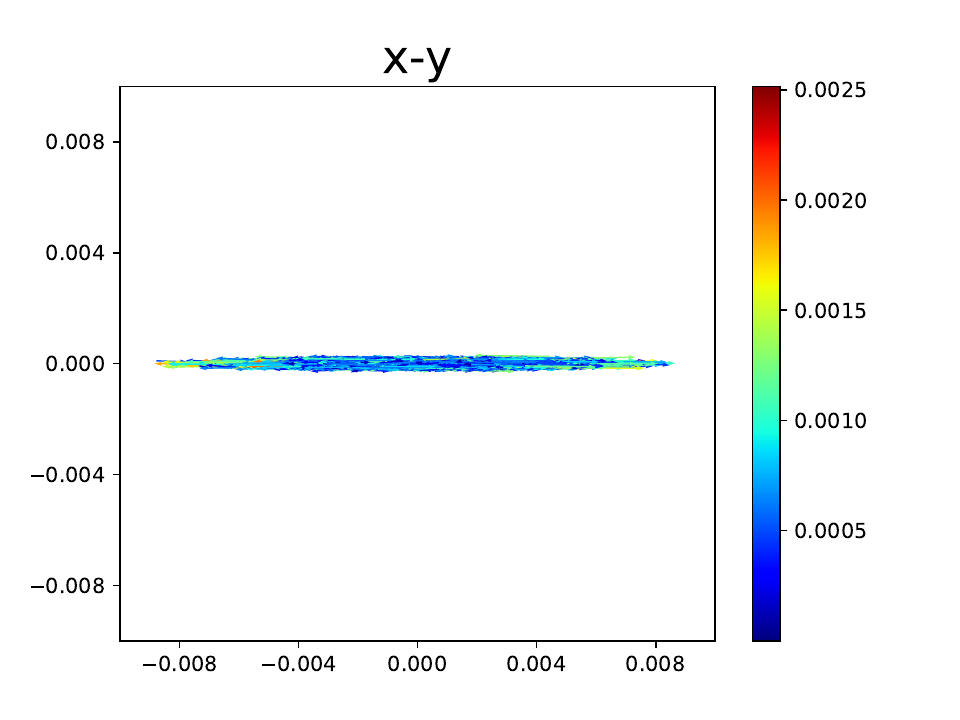}
    \includegraphics[width=0.49\linewidth]{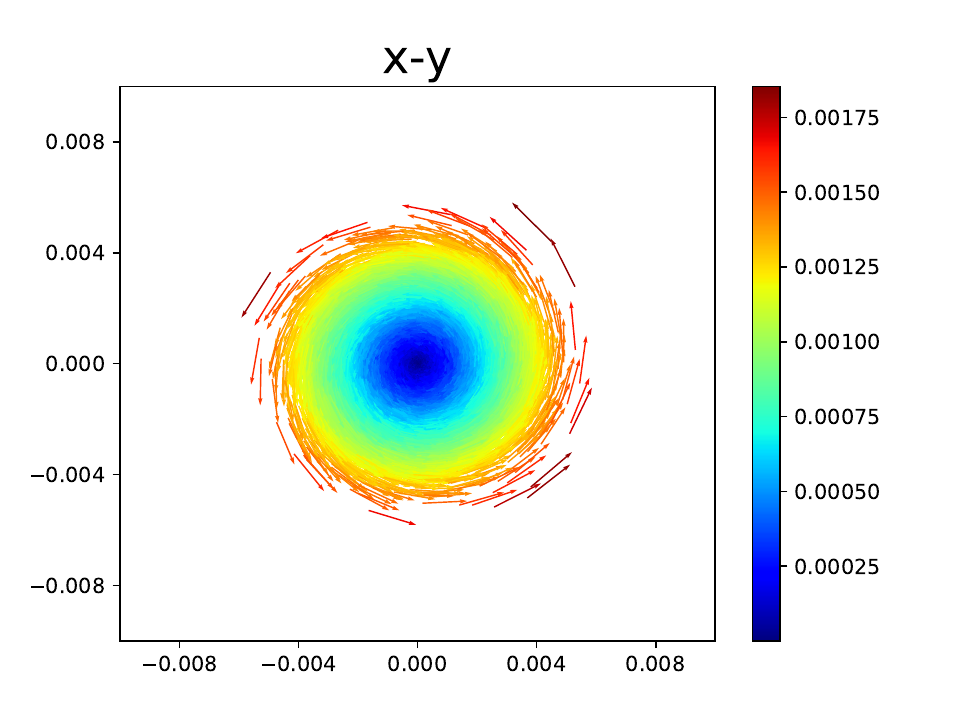}
    \caption{Derbenev's adapter beam transformation. Left: initial uncoupled flat beam. Right: coupled circular-mode beam.}
    \label{fig:derbenevtransformbeam}
\end{figure}

The optics of the flat-to-round transformer are shown in Fig.~\ref{fig:derbenevadapteroptics}. The left plot shows the decoupled $\beta$ functions, the right plot shows coupled $\beta$ functions, where the decoupled $\beta$ functions are matched to the adapter. The coupled $\beta$ functions start with the dominant-mode functions equal to the uncoupled $\beta$ functions with $\beta_{1x}=\beta_{x}$ and $\beta_{2y}=\beta_{y}$. As the beam passes through skew quadrupoles, coupling is created and, at the end, all coupled $\beta$ functions are equal. The transformation from an uncoupled flat beam to a circular-mode beam is illustrated in Fig.~\ref{fig:derbenevtransformbeam}. The inverse adapter corresponds to flipped signs of the skew quadrupoles, which will decouple the circular-mode beam in the right plot of Fig.~\ref{fig:derbenevtransformbeam} to a flat beam in the left plot within the same figure.

\subsection{Magnetized Beams}\label{subsec:magnetizedbeams}
\begin{figure}[tbp]
	\centering
	\includegraphics[width=0.49\linewidth,clip,trim={20 20 30 30}]{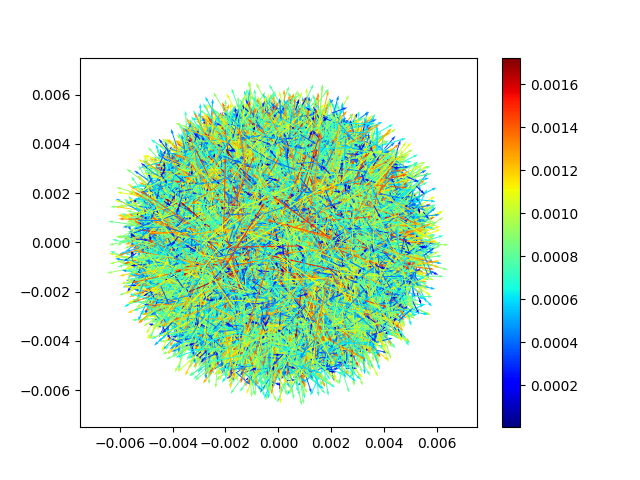}
	\includegraphics[width=0.49\linewidth,clip,trim={20 20 30 30}]{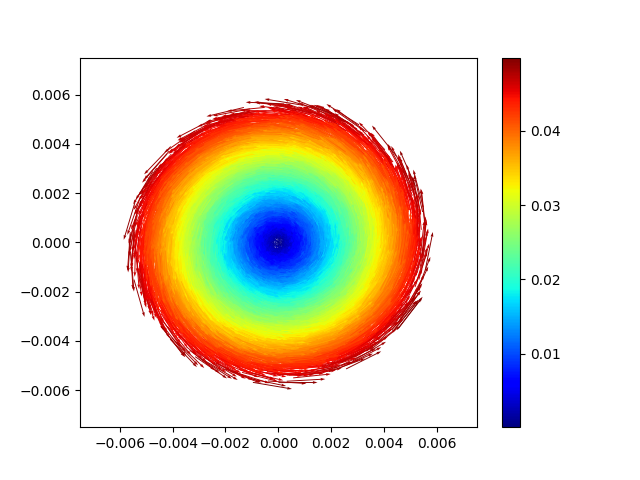}
	\caption{$(x,y)$ plane: beam generated inside a solenoid (left), magnetized beam after exiting the solenoid (right). The color bar shows the magnitude of transverse momentum. The initial beam on the left is a round, uncorrelated beam, and the arrows indicate random directions; the transformed magnetized beam on the right shows a correlated circular-mode beam. The angular speed increases as you go away from the center on the right plot.}
	\label{Fig:magnetizedformation}
\end{figure}

Beams generated inside a longitudinal magnetic field, such as a solenoid, are referred to as ``magnetized beams''. When a beam originates inside a solenoid, it is predominantly subject to the central magnetic field. Upon exiting the solenoid, the beam experiences only the exit fringe field, which generates the angular momentum. The propagation of such a beam can be described using the following $4 \times 4$ transfer matrix product:
\begin{equation}
    \begin{split}
        &\mathcal{M} = M_{exit}\cdot M_{body} = \\
        &\!\begin{pmatrix}
        1 & 0 & 0 & 0 \\
        0 & 1 & k & 0 \\
        0 & 0 & 1 & 0 \\
        -k & 0 & 0 & 1
        \end{pmatrix}\!\begin{pmatrix}
        \frac{1+\cos(lk)}{2} & \frac{\sin(lk)}{k} & \frac{\sin(lk)}{2} & \frac{1-\cos(lk)}{k} \\
        \frac{-k\sin(lk)}{4} & \frac{1+\cos(lk)}{2} & \frac{-k}{4}(1-\cos(lk)) & \frac{\sin(kl)}{2} \\
        -\frac{\sin(kl)}{2} & -\frac{(1-\cos(kl))}{k} & \frac{1+\cos(kl)}{2} & \frac{\sin(kl)}{k} \\
        \frac{k(1-\cos(kl))}{4} & -\frac{\sin(kl)}{2} & -\frac{k(\sin(kl))}{4} & \frac{(1+\cos(kl))}{2}
        \end{pmatrix}.
    \end{split}
    \label{Eq:solenoidmagnetized}
\end{equation}
Here, $k=\frac{B_{0}}{2B\rho}$ is the solenoid focusing index and $l$ is the solenoid length. If the initial transverse divergence is small, i.e., $x_0'\approx y_0'\approx0$, then after the transport through the solenoid, the beam satisfies the vortex condition:
\begin{equation}
	\begin{pmatrix}
		y \\
		y'
	\end{pmatrix} = \begin{pmatrix}
		0 & 1/k \\
		-k & 0
	\end{pmatrix}\begin{pmatrix}
		x \\
		x'
	\end{pmatrix}.
\end{equation}
As discussed in Sec.~\ref{sec:theory}, the rms canonical angular momentum is $L_{z,\mathrm{rms}}=\epsilon_{1}-\epsilon_{2}$. Squaring this relation gives
$L_{z,\mathrm{rms}}^{2}=\epsilon_{1}^{2}+\epsilon_{2}^{2}-2\epsilon_{1}\epsilon_{2}$, showing an explicit dependence on the 4D emittance $\epsilon_{4D}\equiv\epsilon_{1}\epsilon_{2}$. For linear transport, $\epsilon_{4D}$ is invariant; thus $\epsilon_{4D}=\epsilon_{1f}\epsilon_{2f}=\epsilon_{xi}\epsilon_{yi}$, where $i$ denotes the initial uncoupled beam emittances and $f$ denotes the final (coupled) eigenemittances after the solenoid. Using $L_{z,\mathrm{rms}}=\epsilon_1-\epsilon_2$, we can express the eigenmode emittances in terms of $L_z$ as follows:
\begin{equation}
    \begin{split}
        \epsilon_{1}&=\frac{1}{2}L_{z,\mathrm{rms}} + \frac{1}{2}\sqrt{L_{z,\mathrm{rms}}^{2}+4\epsilon_{4D}}, \\
        \epsilon_{2}&=-\frac{1}{2}L_{z,\mathrm{rms}} + \frac{1}{2}\sqrt{L_{z,\mathrm{rms}}^{2}+4\epsilon_{4D}}.
    \end{split}
    \label{eq:magnetizedbeameigenmodeemitcomputation}
\end{equation}
Since the beam is generated inside a solenoid, the canonical angular momentum is conserved~\cite{groening2018extension} (Busch's theorem). In contrast with Derbenev's adapter, the initial beam does not need to be flat for this transformation. The emergence of the circular mode in the context of a magnetized beam is depicted in Fig.~\ref{Fig:magnetizedformation}, consistent with the description in~\cite{hannon2019transverse}.

\begin{figure}[tbp]
	\centering
	\includegraphics[width=0.328\linewidth]{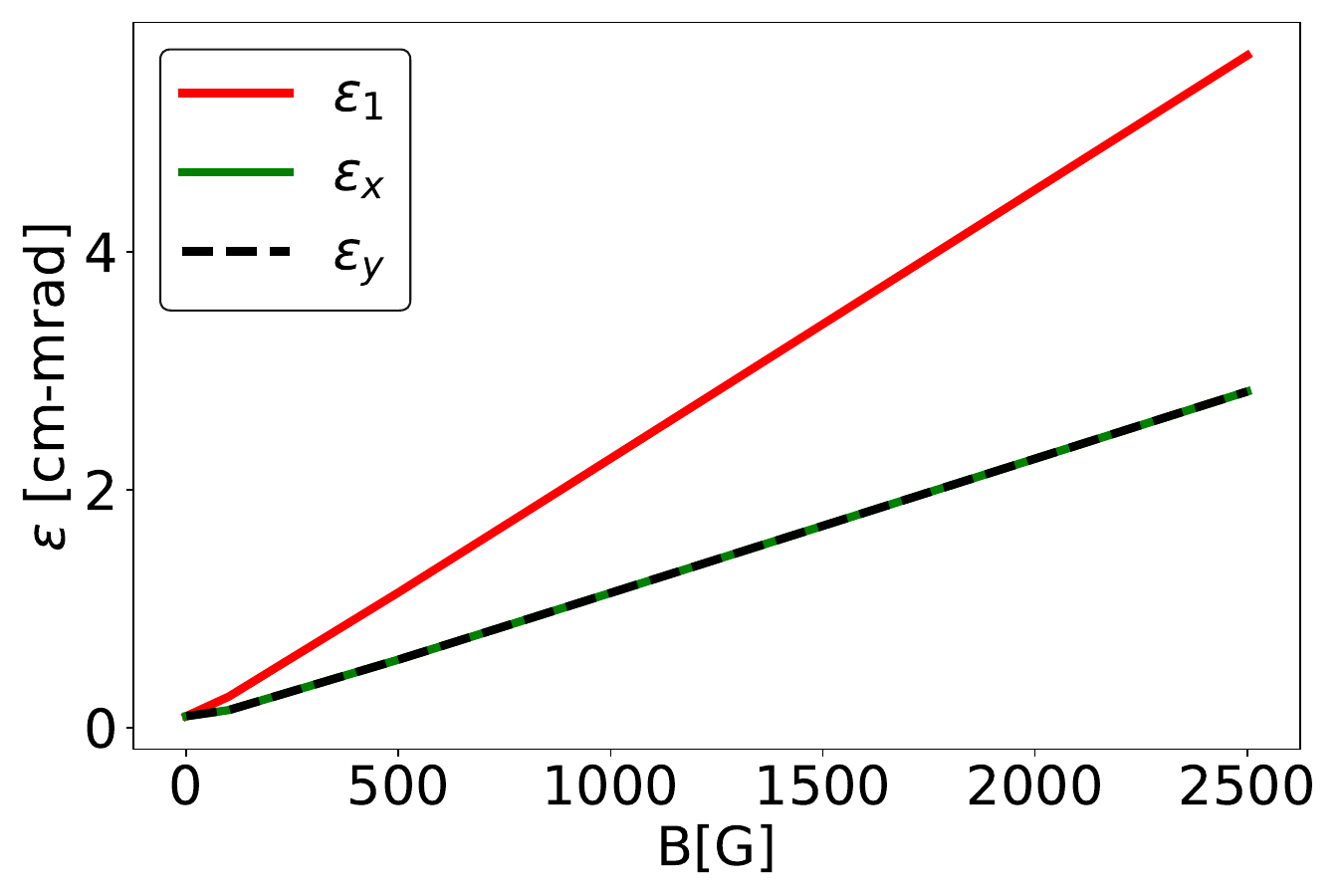}
	\includegraphics[width=0.328\linewidth,trim={0 0 0 0}]{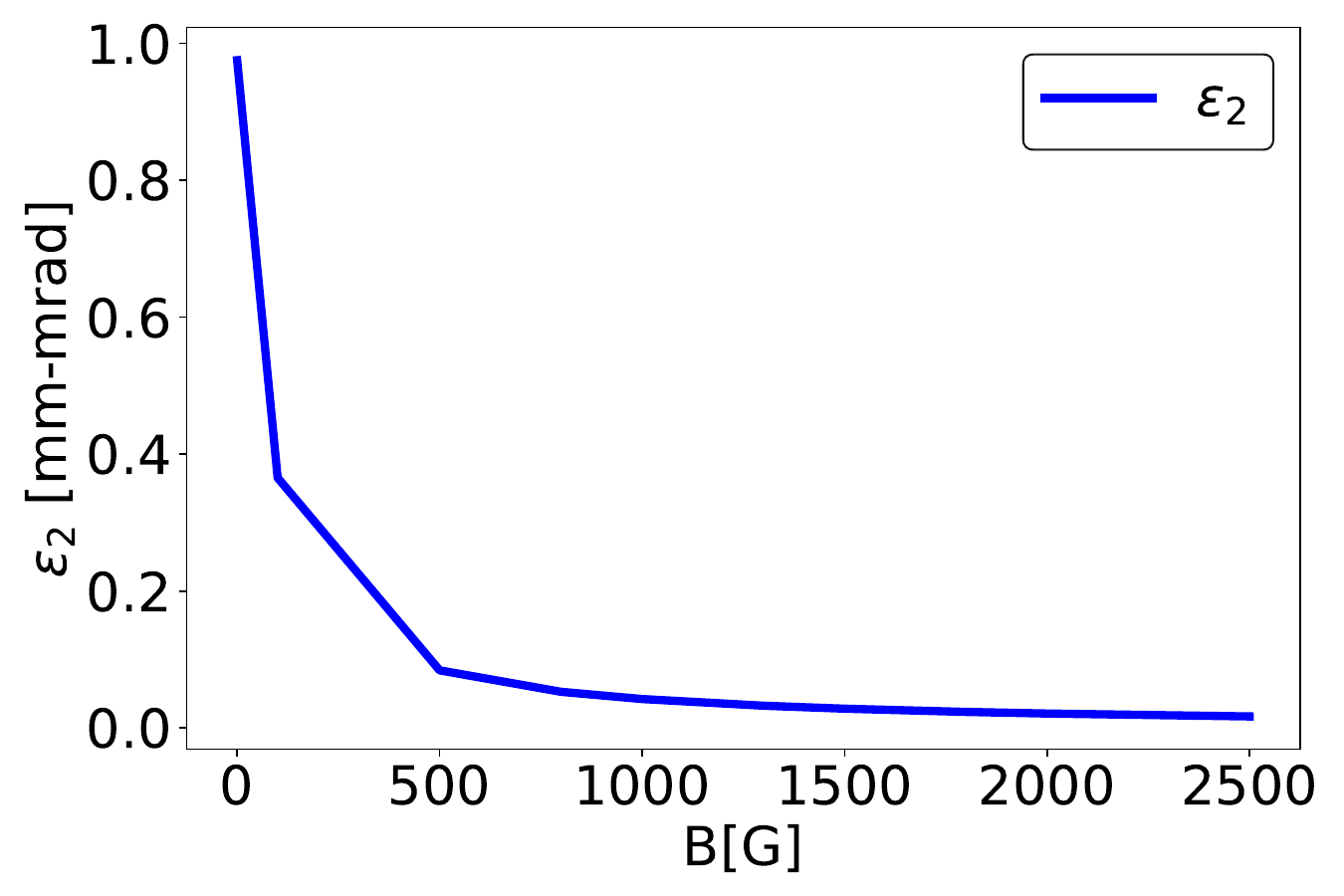}
	\includegraphics[width=0.328\linewidth,trim={0 0 0 0}]{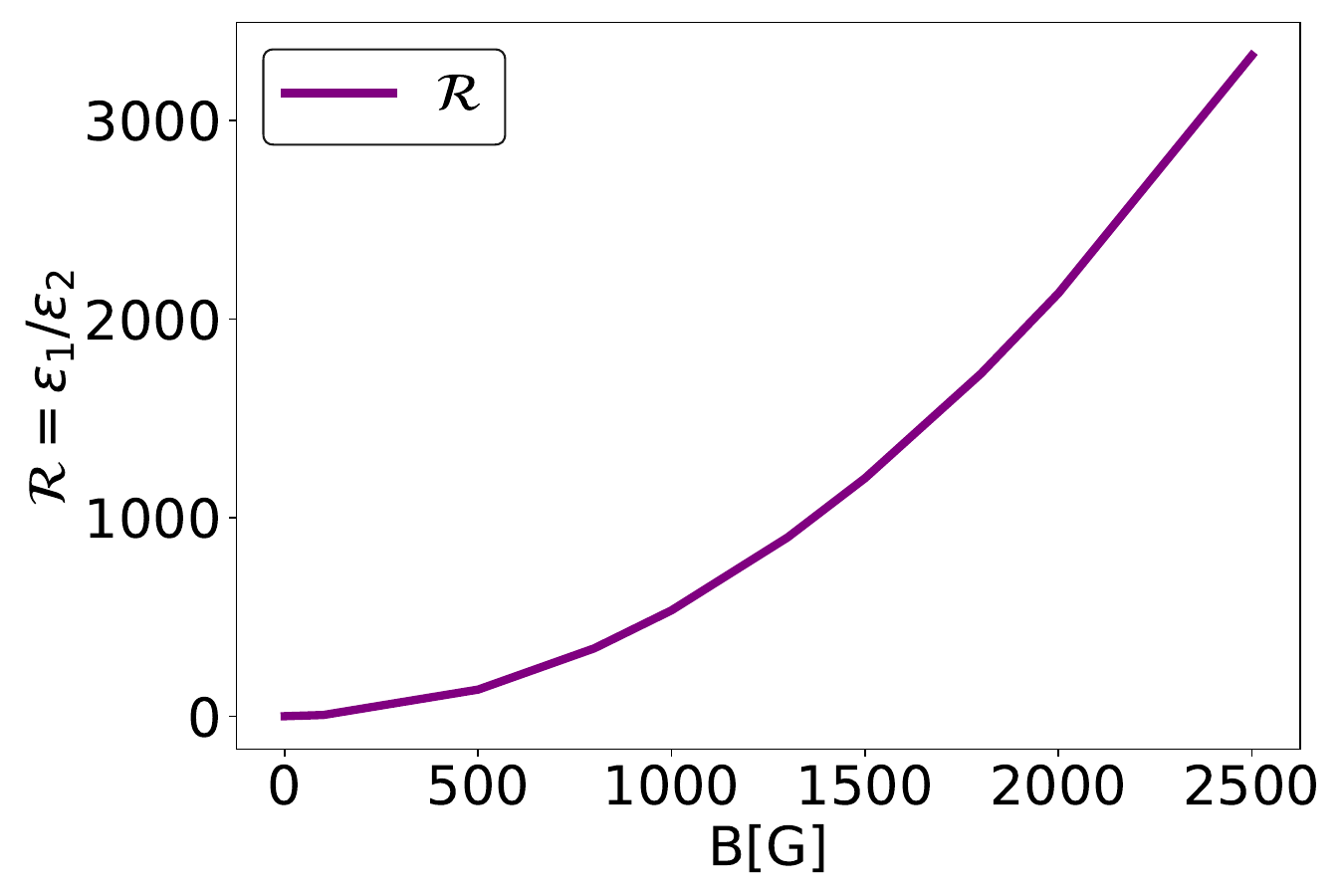}
	\caption{Mode-1 eigenemittance, $\epsilon_{1}$, and apparent emittances $\epsilon_{x,y}$ with solenoid magnetic field, $B$ (left), eigenmode emittance $\epsilon_{2}$ (middle), ratio of eigenmode emittances $\mathcal{R}=\epsilon_{1}/\epsilon_{2}$ (right). The magnetization process is simulated using TRACK~\cite{TRACKweb}.}
	\label{Fig:magnetizedparameters}
\end{figure}

The eigenmode emittances can be deduced from the initial 4D distribution by using Eq.~\eqref{eq:magnetizedbeameigenmodeemitcomputation}. A graph showing the beam parameters as functions of the solenoidal magnetic field is provided in Fig.~\ref{Fig:magnetizedparameters}. Our postulate regarding the coupling parameter, $u=1/2$, is corroborated by the correlation between the apparent emittances, $\epsilon_{x,y}$, and the eigenmode emittance of mode~1. 
As the solenoidal field increases, the canonical angular momentum increases and the eigenemittances become increasingly separated. The four-dimensional emittance remains invariant under linear Hamiltonian transport, while $\epsilon_1$ grows and $\epsilon_2$ decreases such that $\epsilon_1\epsilon_2=\epsilon_{4D}$. This increases the intrinsic flatness ratio $\mathcal{R}=\epsilon_1/\epsilon_2$.
We note that electron beams are typically generated from cathodes within solenoids, as noted in \cite{brinkmann2001low, halavanau2017magnetized}. Successful trials have also been conducted to strip $H^{-}$ ions inside a solenoid \cite{groening2011concept, groening2014emtex,cheon2021mitigation}. After the magnetized beam exits the solenoid, it is typically transformed into a flat beam downstream, thereby removing the coupling. Other methods of coupling removal have also been studied in the literature; however, we focus on maintaining strong coupling until the low-energy region is passed. 

\subsection{Phase Space Painting}\label{subsec:phasespacepainting}
Phase space painting is an advanced technique used to ``draw" or form a particular beam distribution within the acceptance of an accelerator ring. It is particularly effective when using a pencil beam from a linac, injecting and carefully placing it within the phase-space beamlets, which are significantly smaller than the ring's acceptance. For instance, this method was used to create self-consistent beams following the so-called Danilov distribution \cite{danilov2003self}, which has attracted substantial attention recently \cite{holmes2018injection,hoover2021computation,hoovereigenpainting}. In the idealized model, the Danilov distribution has delta-function structure in the four-dimensional phase space and carries nonzero angular momentum. The method used to craft such a distribution, known as elliptical painting, is detailed in the cited literature. A hallmark of this technique is aligning the beamlets along the $(y',x)$ plane, as inferred from the vortex condition (see Eq.~\eqref{eq:phasespacecoordvortex}). 

\begin{figure}[tbp]
	\centering
	\includegraphics[width=0.49\linewidth,clip,trim={70 10 80 40}]{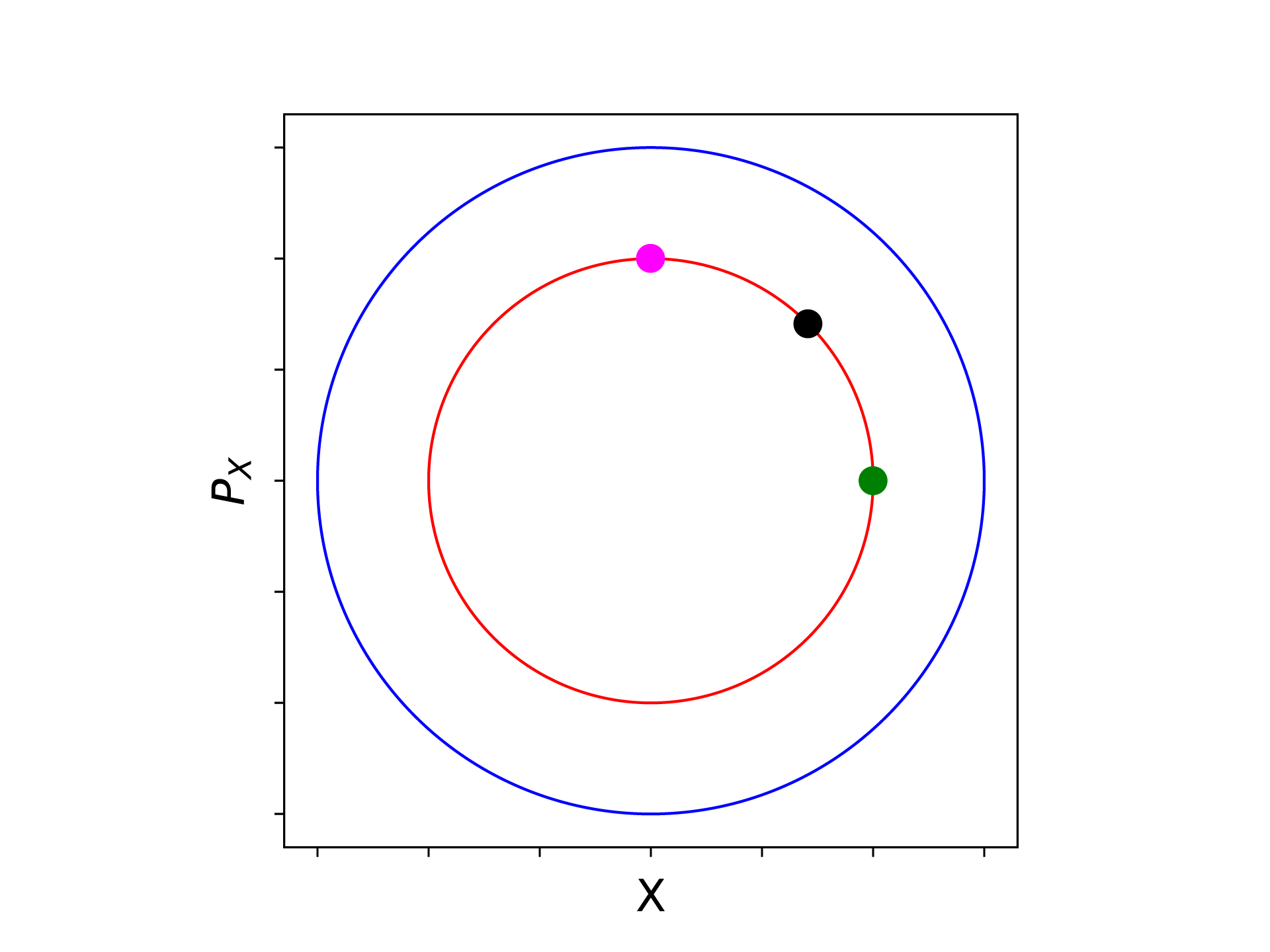}
    \includegraphics[width=0.49\linewidth,clip,trim={70 10 80 40}]{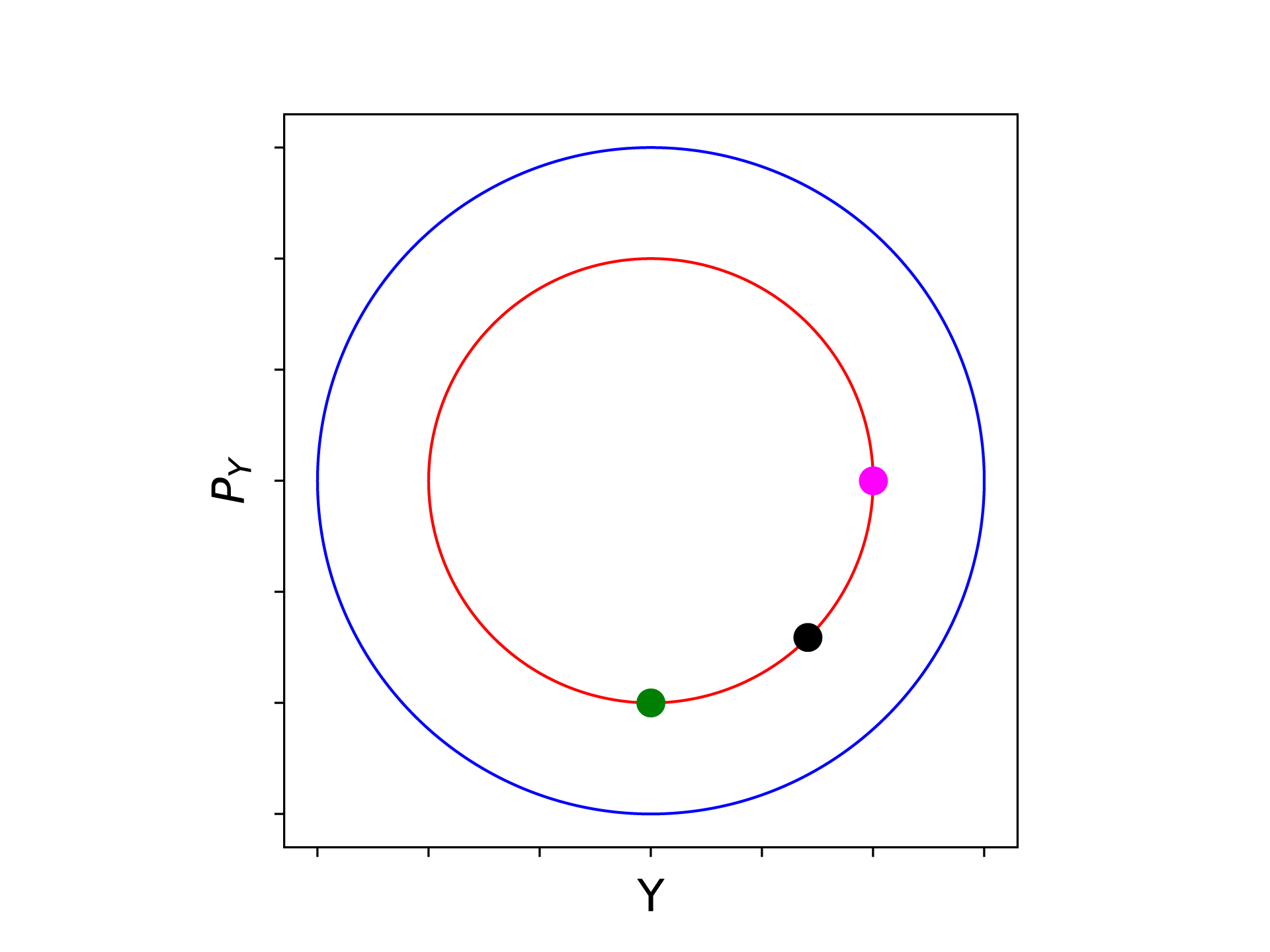}
	\caption{Normalized phase spaces with injected beamlets with $\pi/2$ phase difference. }
	\label{Fig:paintingcond}
\end{figure}

\begin{figure}[tbp]
	\centering
	\includegraphics[width=0.49\linewidth]{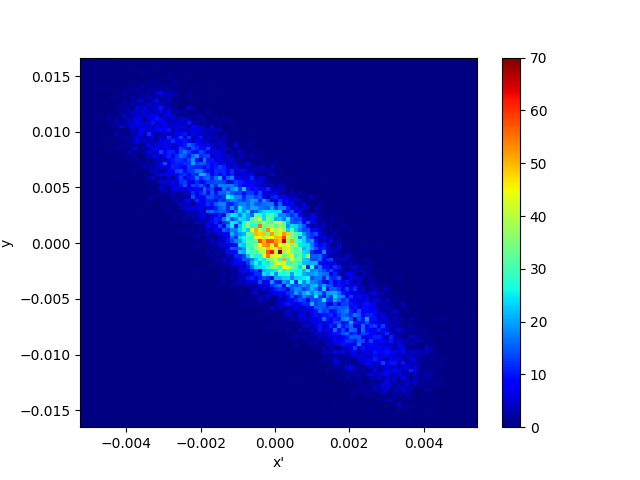}
	\includegraphics[width=0.49\linewidth]{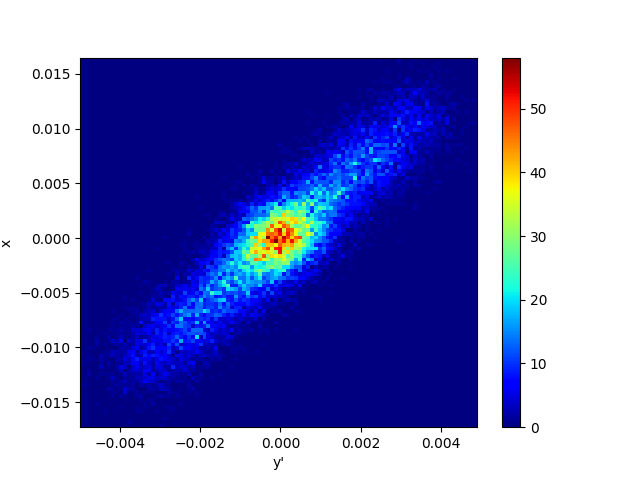}\\
	\includegraphics[width=0.49\linewidth]{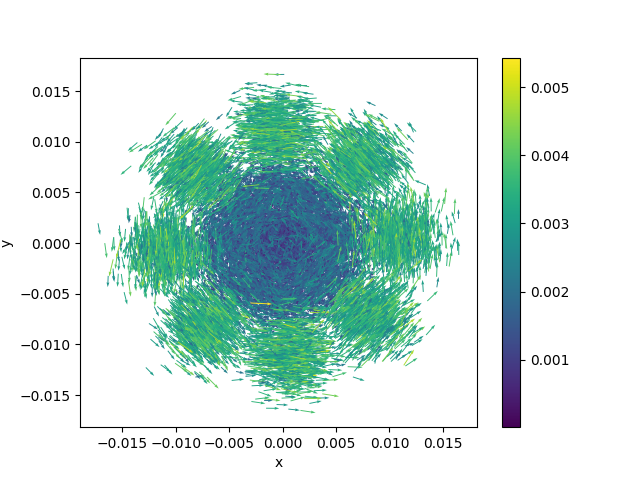}
	\caption{Beamlets formation: $(x',y)$ phase space of 8 beamlets (top left), $(y',x)$ phase space (top right), $(x,y)$ space (bottom).}
	\label{Fig:Multiturnpainting}
\end{figure}

A schematic of the method is shown in Fig.~\ref{Fig:paintingcond}. In the normalized phase space \((X, P_X)\) and \((Y, P_Y)\), with \(X=x/\sqrt{\beta_x}\) and \(P_X=\sqrt{\beta_x}\,x' + (\alpha_x/\sqrt{\beta_x})x\), the color-coded dots represent identical beamlets separated by a phase difference of $\pi/2$ between the two planes. This offset introduces a sense of preferred rotation, enabling the build-up of a beam with dominant non-zero angular momentum, as illustrated in Fig.~\ref{Fig:Multiturnpainting}. Achieving this requires controlled four-dimensional painting and, therefore, both horizontal and vertical kicker magnets. Experiments at the Spallation Neutron Source~(SNS) have shown that such self-consistent beams with small 4D emittance can be painted in practice~\cite{hoover2021computation}.


\section{Preservation of circular modes}
\label{sec:preservationofcircularmodes}

Once a circular-mode beam has been created, two distinct properties must be preserved during transport. First, the eigenemittance separation and the associated canonical angular momentum must be maintained. Second, the beam must retain a round and untilted projection in the physical \((x,y)\) plane. The first requirement concerns the intrinsic four-dimensional structure of the beam, whereas the second concerns how that structure projects onto configuration space. Conservation of angular momentum alone does not necessarily guarantee that the instantaneous \((x,y)\) projection remains round.

In ideal linear symplectic transport without collective effects, the eigenemittances and hence the intrinsic flatness ratio are invariants. In practice, mismatch, nonlinear forces, acceleration conventions, and space charge can modify the rms eigenemittances or change how the eigenmodes project onto the laboratory coordinates.

As derived from the projected covariance matrix and the general Lebedev--Bogacz second moments in Appendix~\ref{subsec:appendixAsecondmoments}, the rms contour in the physical \((x,y)\) plane is
\begin{equation}
    \frac{x^2}{\sigma_x^2}
    -
    \frac{2\widetilde{\alpha}}{\sigma_x\sigma_y}xy
    +
    \frac{y^2}{\sigma_y^2}
    =
    1-\widetilde{\alpha}^{\,2},
    \qquad
    \widetilde{\alpha}
    =
    \frac{
        \epsilon_1\sqrt{\beta_{1x}\beta_{1y}}\cos\nu_1
        +
        \epsilon_2\sqrt{\beta_{2x}\beta_{2y}}\cos\nu_2
    }{
        \sqrt{\epsilon_1\beta_{1x}+\epsilon_2\beta_{2x}}\,
        \sqrt{\epsilon_1\beta_{1y}+\epsilon_2\beta_{2y}}
    }.
    \label{eq:projectedbeamellipse}
\end{equation}
Here, \(\sigma_x\) and \(\sigma_y\) are the rms beam sizes, while
\(\widetilde{\alpha}=\sigma_{xy}/(\sigma_x\sigma_y)\) is the normalized \(x\)-\(y\) correlation parameter. The ratio \(\sigma_x/\sigma_y\) determines the aspect ratio of the projected ellipse, and \(\widetilde{\alpha}\) determines its tilt. A round and untilted projection therefore requires$\sigma_x=\sigma_y,\quad\widetilde{\alpha}=0.
$
The mode-1-dominated limit introduced in Sec.~\ref{subsec:circmodelimittheorysection} is defined by $ \mathcal R=\frac{\epsilon_1}{\epsilon_2}\gg1.$
This terminology refers only to the eigenemittance hierarchy; it does not mean that eigenmode~2 is absent, nor does it refer to the coupling parameter \(u\). In this limit, the contribution proportional to \(\epsilon_2\) in Eq.~\eqref{eq:projectedbeamellipse} can be neglected to leading order, giving
\begin{equation}
    \widetilde{\alpha}\simeq\cos\nu_1,
    \qquad
    \sigma_x^2\simeq\epsilon_1\beta_{1x},
    \qquad
    \sigma_y^2\simeq\epsilon_1\beta_{1y}.
    \label{eq:dominantmodeprojection}
\end{equation}
Consequently, a mode-1-dominated beam has a round and untilted projection when
\begin{equation}
    \beta_{1x}=\beta_{1y},
    \qquad
    \nu_1=\frac{\pi}{2}\pmod{\pi}.
    \label{eq:dominantmodepreservation}
\end{equation}
A difference between \(\beta_{1x}\) and \(\beta_{1y}\) changes the projected aspect ratio, whereas a deviation of \(\nu_1\) from \(\pi/2\) produces an \(x\)-\(y\) correlation and tilts the projected ellipse.

The preservation analysis developed in the following subsections is formulated for linear transverse optics evaluated about the reference momentum. For a beam with finite momentum spread, chromatic focusing makes the normal-mode phase advances, coupled beta functions, and coupling phases momentum dependent. Different momentum slices may therefore satisfy slightly different circular-mode conditions. Although linear symplectic transport preserves the eigenemittances of each momentum slice, their superposition can exhibit phase mixing, weakening the projected vortex correlations and causing departure from a round transverse profile.

Nonlinear forces may likewise distort the phase-space distribution. A rotationally symmetric nonlinear Hamiltonian preserves canonical angular momentum even when the distribution changes shape. General non-axisymmetric nonlinearities, however, need not preserve either the rms angular momentum or the rms eigenemittances. When such nonlinearities are weak and the working point is sufficiently separated from the resonances they drive, their influence is expected to remain perturbative. A quantitative study of chromatic and external nonlinear effects is beyond the scope of the present work.

The preservation problem can therefore be separated into two parts. Section~\ref{subsec:preservingangmom} identifies the lattice conditions required to preserve canonical angular momentum and eigenemittance separation. Section~\ref{subsec:couplingphasepreservation} then examines the conditions required to preserve the coupling phase and hence the circularity of the physical beam projection.

\subsection{Conservation of Angular Momentum}\label{subsec:preservingangmom}
Using the 4D transfer matrix formalism for an uncoupled lattice, the transformation of the phase space vector is  $\vec{z}_{f}=\mathcal{M}\vec{z}_{i}$, where $f$ and $i$ stand for ``final'' and ``initial'' positions. $\mathcal{M}$ is the $4 \times 4$ transfer matrix and $\vec{z}=(x,x',y,y')^{T}$, the uncoupled transfer matrix is 
\begin{equation}
        \mathcal{M}= \begin{pmatrix}
            M & 0 \\
            0 & N
        \end{pmatrix}.
\end{equation}
By defining a special 4D $\mathbf{L}$ matrix
\begin{equation}
    \begin{split}
        \mathbf{L} = \begin{pmatrix}
            0 & S \\
            -S & 0
        \end{pmatrix}, \textnormal{ with } S = \begin{pmatrix}
            0 & 1 \\
            -1 & 0
        \end{pmatrix},
    \end{split}
\end{equation}
the angular momentum defined as $L_{z}=xy'-yx'$ can also be written as $L_{z} = \tfrac{1}{2}\vec{z}^{T}\mathbf{L}\vec{z}$. Conservation of angular momentum, $L_{zf}=L_{zi}$ can also be written in matrix form as $\mathcal{M}^{T}\mathbf{L}\mathcal{M}=\mathbf{L}$. Conservation yields the following constraints on the 2D transfer matrices $M$ and $N$:
\begin{equation}
    \begin{split}
        &  M_{11} N_{21} = M_{21} N_{11}, \\
        &  M_{12} N_{22} =  M_{22} N_{12}, \\
        &  M_{11} N_{22} -M_{21} N_{12} = 1,  \\
        & M_{22} N_{11} - M_{12} N_{21} = 1.
    \end{split}
\end{equation}
Using the first two conditions and substituting them into the last two equations while using the property of matrices $\det(M) = \det(N)=1$, yields $M=N$, which implies equal overall focusing in both planes. In addition, knowing that angular momentum is preserved under rotation, the general form of the matrix $\mathcal{M}$ is
\begin{equation}
    \mathcal{M} = R(\theta)\cdot \begin{pmatrix}
        T & 0 \\
        0 & T
    \end{pmatrix},
    \label{Eq:rotinvariantmatrix}
\end{equation}
where $R(\theta)$ is rotation matrix and $T$ is a $2\times 2$ transfer matrix for both horizontal and vertical planes. Therefore, a transfer map that focuses equally on both $x$ and $y$ planes is rotationally invariant. Consequently, Eq.~\eqref{Eq:rotinvariantmatrix} yields degenerate eigenvalues, i.e., equal transverse tunes $Q_x=Q_y$. This corresponds to the linear difference resonance condition $Q_x-Q_y=0$~\cite{franchi2007emittance} and, in particular, also satisfies the Montague (fourth-order) resonance condition $2Q_x-2Q_y=0$~\cite{montague1968fourth}. To maintain intrinsic flatness efficiently, the degeneracy of the transfer matrix must be broken and the tunes have to split.

A rotationally invariant transfer map preserves canonical angular momentum, but it also produces degenerate transverse eigenvalues. In this limit, the two eigenmode tunes are equal, placing the beam on the linear difference resonance. A practical circular-mode lattice must therefore separate the eigenmode tunes without destroying the rotational symmetry required for angular-momentum conservation. A solenoid provides this separation because its focusing remains axisymmetric while its longitudinal magnetic field gives the two counter-rotating eigenmodes opposite phase-advance contributions.

In the Lebedev--Bogacz parametrization, \(\mu_1(s)\) and \(\mu_2(s)\) are the accumulated betatron phase advances of the two eigenmodes. Their corresponding one-turn tunes are
\[
    Q_j=\frac{1}{2\pi}\oint\frac{d\mu_j}{ds}\,ds,
    \qquad j\in\{1,2\}.
\]
For stable linear on-momentum motion through a lattice containing a longitudinal solenoidal field, transport of the coupled eigenvectors gives~\cite{lebedev2010betatron}
\begin{equation}
    \begin{aligned}
        \frac{d\mu_1}{ds}
        &=
        \frac{1-u}{\beta_{1x}}
        -
        \frac{k}{2}
        \sqrt{\frac{\beta_{1y}}{\beta_{1x}}}
        \sin\nu_1,\\
        \frac{d\mu_2}{ds}
        &=
        \frac{1-u}{\beta_{2y}}
        +
        \frac{k}{2}
        \sqrt{\frac{\beta_{2x}}{\beta_{2y}}}
        \sin\nu_2.
    \end{aligned}
    \label{eq:eigenmodephaseadvancepropagation}
\end{equation}
Here, \(k(s)\) is the signed normalized solenoidal strength, \(u\) is the Lebedev--Bogacz coupling parameter, and \(\nu_{1,2}\) are the relative phases between the horizontal and vertical components of the two eigenmodes. These equations describe the linear reference optics; chromatic, nonlinear, and collective perturbations are not included.

For the ideal circular-mode conditions introduced in Sec.~II B,$u=\frac{1}{2},\nu_1=\nu_2=\frac{\pi}{2},\beta_{1x}=\beta_{1y}=\beta_{2x}=\beta_{2y}\equiv\beta_0,$
Eq.~\eqref{eq:eigenmodephaseadvancepropagation} reduces to
\begin{equation}
    \frac{d\mu_1}{ds}
    =
    \frac{1}{2\beta_0}-\frac{k}{2},
    \qquad
    \frac{d\mu_2}{ds}
    =
    \frac{1}{2\beta_0}+\frac{k}{2}.
    \label{eq:eigenmodephaseadvancecircular}
\end{equation}
The first term is the common axisymmetric focusing contribution. The solenoidal term enters with opposite signs because the two circular eigenmodes rotate with opposite handedness. The solenoid therefore separates the two eigenmode tunes while preserving axisymmetric focusing and canonical angular momentum.

\subsection{Preserving the coupling phase}
\label{subsec:couplingphasepreservation}

Tune separation alone is not sufficient to preserve a circular-mode beam. As shown in Sec.~IV and Appendix B, the coupling phases determine how the eigenmode structure projects onto the physical \((x,y)\) plane. In the mode-1-dominated limit, the projected correlation parameter satisfies \(\widetilde{\alpha}\simeq\cos\nu_1\). Consequently, \(\nu_1=\pi/2\pmod{\pi}\) produces an untilted projection, whereas a deviation from this value produces an \(x\)-\(y\) correlation and exposes part of the intrinsic flatness in configuration space.

It is therefore necessary to determine how the coupling phases evolve through the lattice. In the Lebedev--Bogacz parametrization, \(\nu_j\) is the relative phase between the horizontal and vertical components of eigenmode \(j\). Transport of the coupled eigenvectors through the same linear on-momentum lattice gives~\cite{lebedev2010betatron}
\begin{equation}
    \begin{aligned}
        \frac{d\nu_1}{ds}
        &=
        \frac{1-u}{\beta_{1x}}
        -
        \frac{u}{\beta_{1y}}
        -
        \frac{k}{2}
        \left(
            \sqrt{\frac{\beta_{1y}}{\beta_{1x}}}
            -
            \sqrt{\frac{\beta_{1x}}{\beta_{1y}}}
        \right)
        \sin\nu_1,\\
        \frac{d\nu_2}{ds}
        &=
        \frac{1-u}{\beta_{2y}}
        -
        \frac{u}{\beta_{2x}}
        -
        \frac{k}{2}
        \left(
            \sqrt{\frac{\beta_{2x}}{\beta_{2y}}}
            -
            \sqrt{\frac{\beta_{2y}}{\beta_{2x}}}
        \right)
        \sin\nu_2.
    \end{aligned}
    \label{eq:couplingphasepropagation}
\end{equation}
The first two terms describe the phase slippage caused by unequal horizontal and vertical focusing. The terms proportional to \(k\) describe the additional rotation produced by the solenoidal field.

To make the physical-plane dependence explicit, we introduce the effective horizontal and vertical beta functions
\[
    \beta_x
    \equiv
    2\beta_{1x}
    =
    2\beta_{2x},
    \qquad
    \beta_y
    \equiv
    2\beta_{1y}
    =
    2\beta_{2y},
\]
for the maximally coupled case \(u=1/2\). Equation~\eqref{eq:couplingphasepropagation} then becomes
\begin{equation}
    \begin{aligned}
        \frac{d\nu_1}{ds}
        &=
        \frac{1}{\beta_x}
        -
        \frac{1}{\beta_y}
        -
        \frac{k}{2}
        \left(
            \sqrt{\frac{\beta_y}{\beta_x}}
            -
            \sqrt{\frac{\beta_x}{\beta_y}}
        \right)
        \sin\nu_1,\\
        \frac{d\nu_2}{ds}
        &=
        \frac{1}{\beta_y}
        -
        \frac{1}{\beta_x}
        -
        \frac{k}{2}
        \left(
            \sqrt{\frac{\beta_x}{\beta_y}}
            -
            \sqrt{\frac{\beta_y}{\beta_x}}
        \right)
        \sin\nu_2.
    \end{aligned}
    \label{eq:couplingphasepropagationmaximal}
\end{equation}
This notation avoids confusing the effective physical-plane beta functions with the eigenmode labels.

In an ideal axisymmetric solenoidal channel,
\[
    \beta_x=\beta_y=\beta_r,
    \qquad
    \nu_1=\nu_2=\frac{\pi}{2}\pmod{\pi}.
\]
Both the focusing-difference terms and the solenoidal asymmetry terms in Eq.~\eqref{eq:couplingphasepropagationmaximal} then vanish, so the coupling phases remain constant. This explains why an ideal solenoid preserves the circular projection continuously.

In a quadrupole- or dipole-based section without a solenoidal field, \(k=0\), and the phase evolution reduces to
\begin{equation}
    \frac{d\nu_1}{ds}
    =
    \frac{1}{\beta_x}
    -
    \frac{1}{\beta_y},
    \qquad
    \frac{d\nu_2}{ds}
    =
    \frac{1}{\beta_y}
    -
    \frac{1}{\beta_x}.
    \label{eq:couplingphasepropagationnonsolenoidal}
\end{equation}
The two coupling phases therefore evolve with equal magnitude and opposite sign. They remain constant pointwise only when \(\beta_x=\beta_y\). More generally, the circular-mode condition can be recovered after one lattice period if the integrated phase slippage cancels. This observation motivates the mirror-symmetric alternating-gradient lattice introduced below.

\begin{figure}[tbp]
    \centering
    \includegraphics[width=0.8\linewidth]{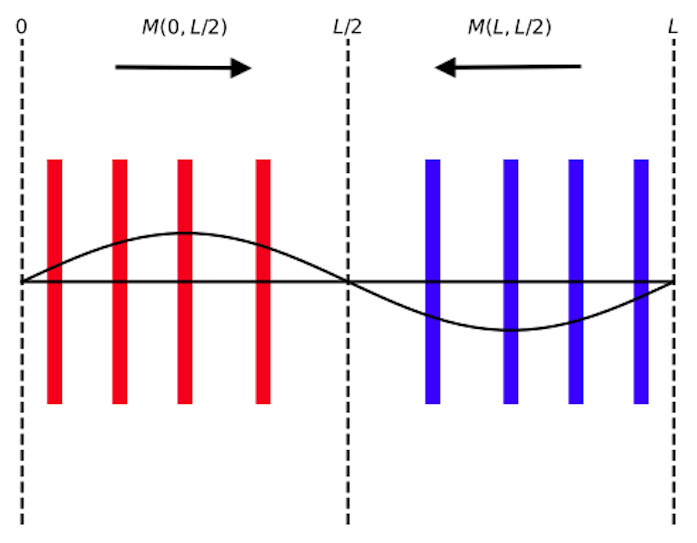}
    \caption{SAG design illustration. The quadrupole elements on the left of $L/2$ line have the same magnitude but opposite polarity of the quadrupole elements on the right of $L/2$ line.}
    \label{fig:mirrorsymmetry}
\end{figure}

The mid-plane mirror symmetry requirement on the $4\times4$ transfer matrix $\mathcal{M}$ for a period of length $L$ is
\begin{equation}
\mathcal{M}(0,L/2)=
\begin{pmatrix}
M & 0\\
0 & N
\end{pmatrix},
\qquad
\mathcal{M}(L,L/2)=
\begin{pmatrix}
N & 0\\
0 & M
\end{pmatrix},
\label{eq:midplanesymmetry}
\end{equation}
where $\mathcal{M}(s_1,s_2)$ denotes the transfer matrix from $s=s_1$ to $s=s_2$. Equation~\eqref{eq:midplanesymmetry} enforces a symmetric-alternating-gradient (SAG) design which is illustrated in Fig.~\ref{fig:mirrorsymmetry}. To conserve coupling phases for quadrupole based lattices, the mid-plane mirror symmetry is enforced on the transfer matrix.
%

\subsection{Lattice Building Blocks for Circular-Mode Beams}\label{subsec:latticebuildingblocks}
\begin{figure*}[tbp]
    \centering
    \includegraphics[width=0.44\linewidth]{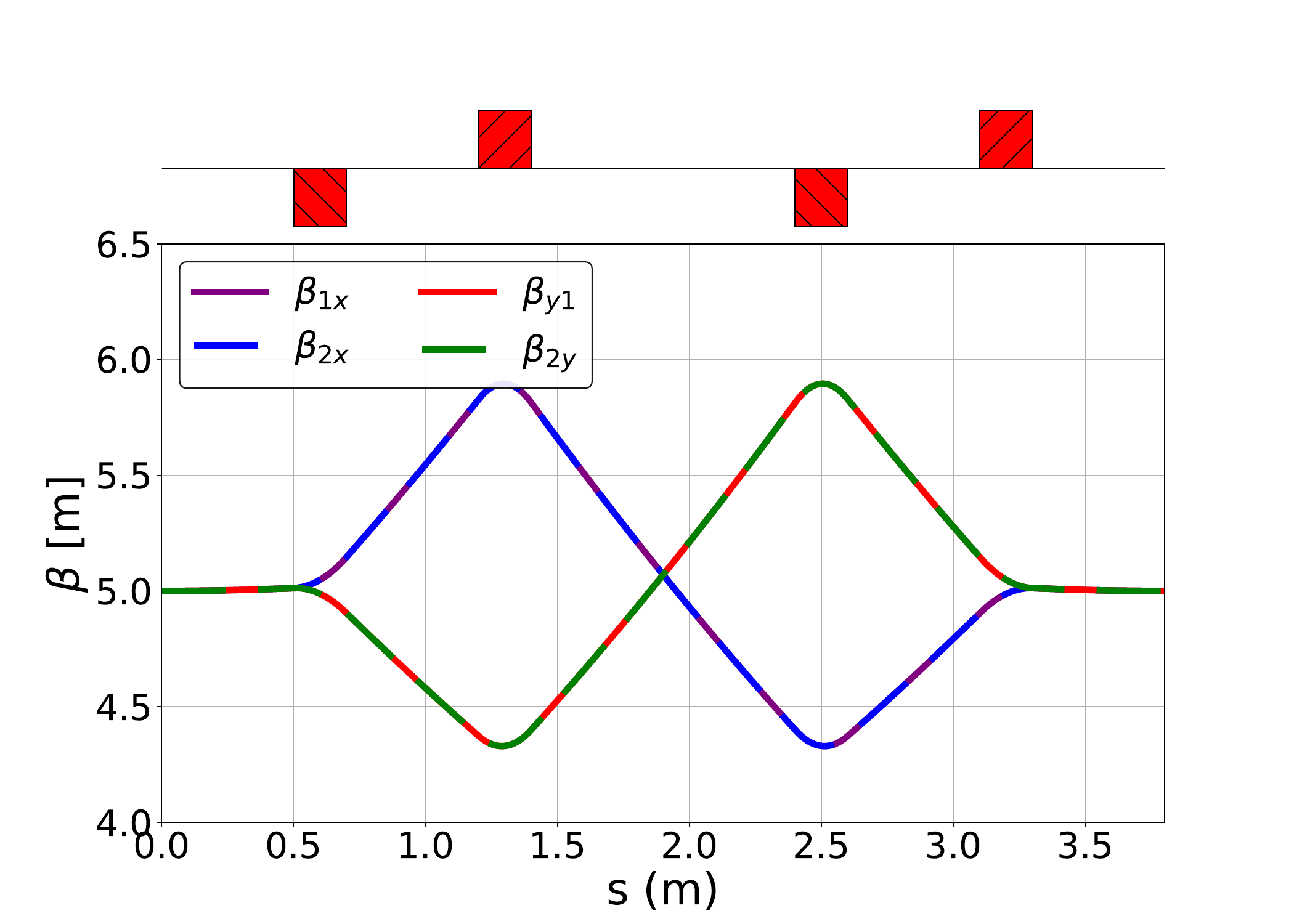}
    \includegraphics[width=0.44\linewidth]{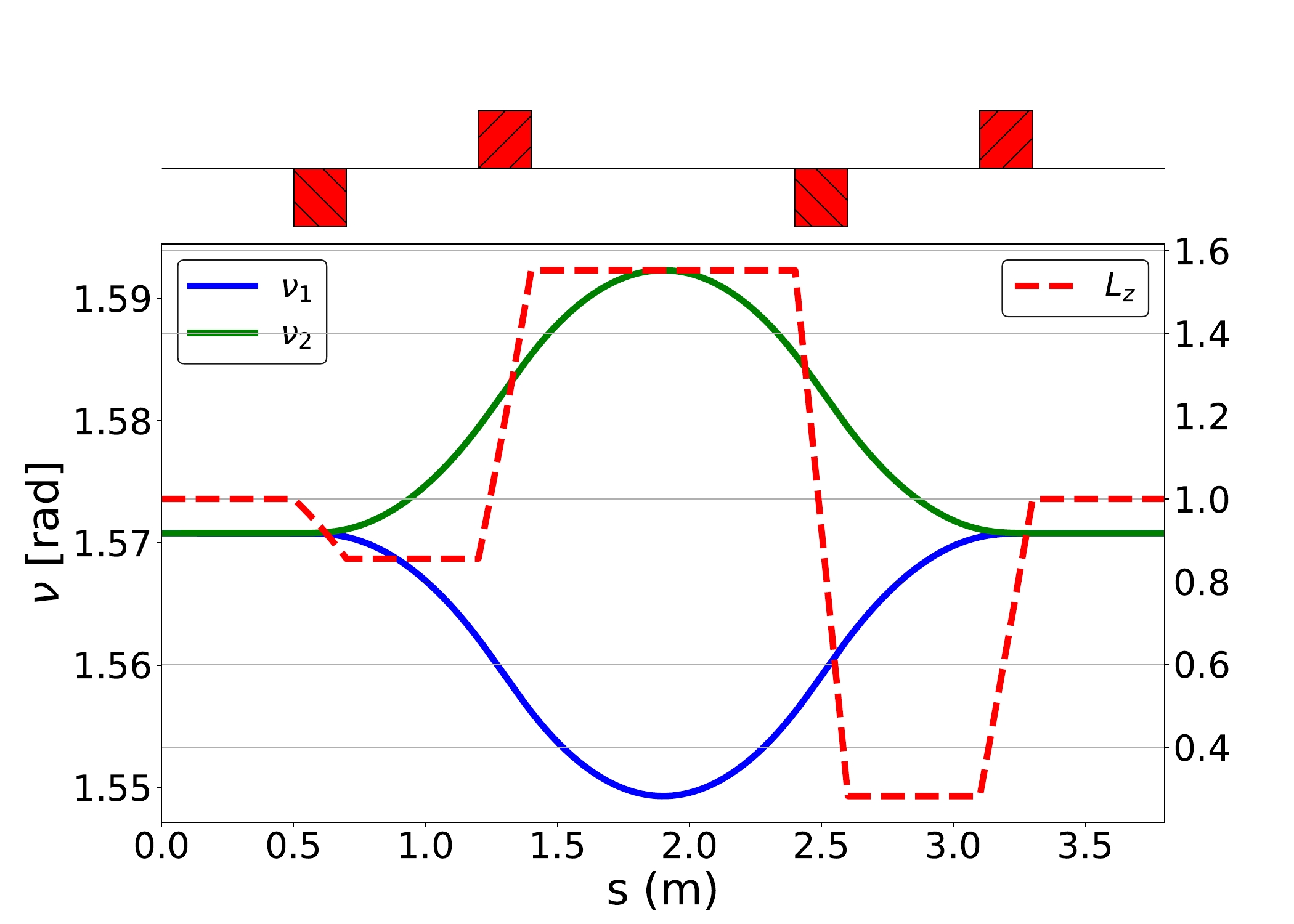}
    \includegraphics[width=0.44\linewidth]{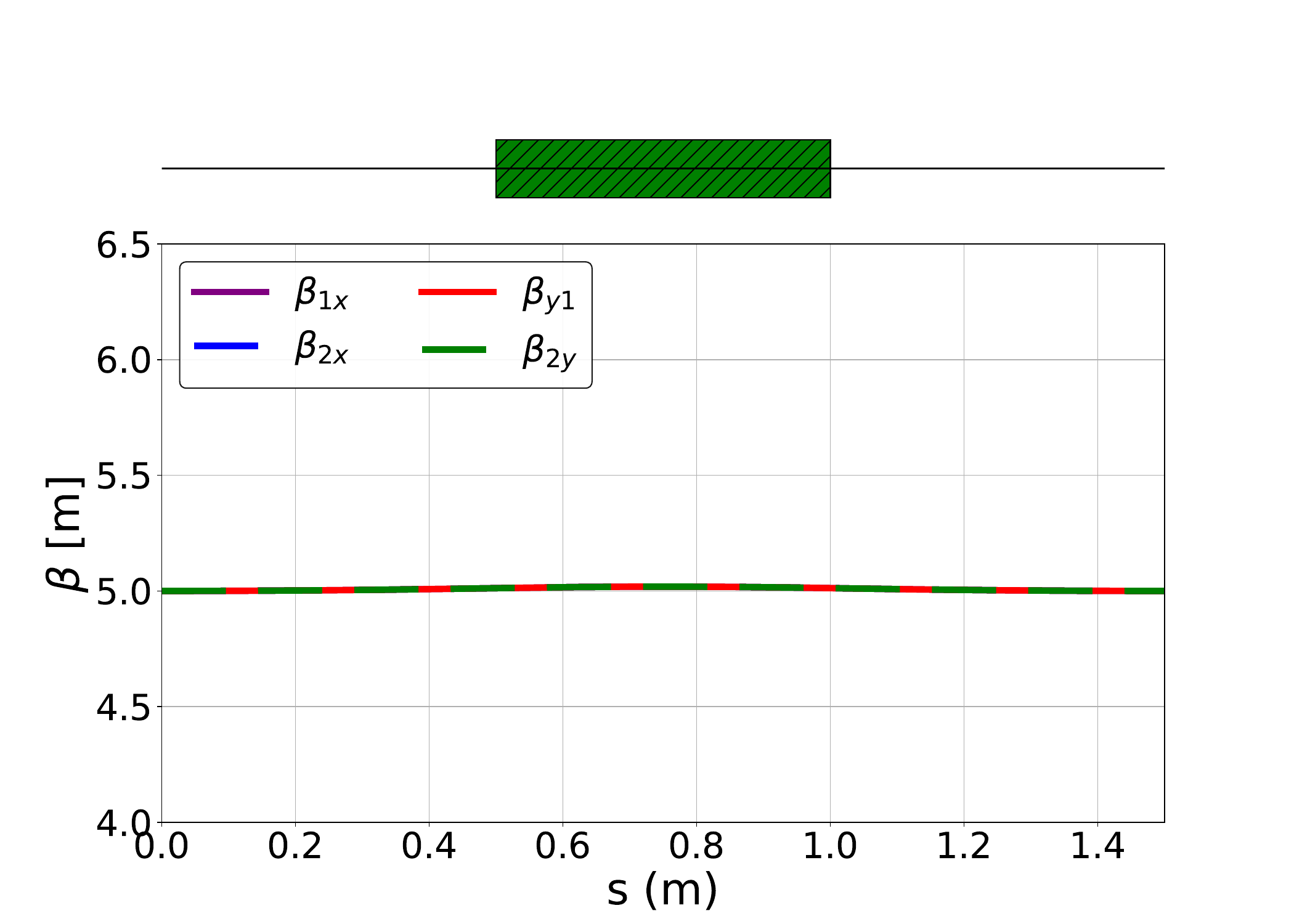}
    \includegraphics[width=0.44\linewidth]{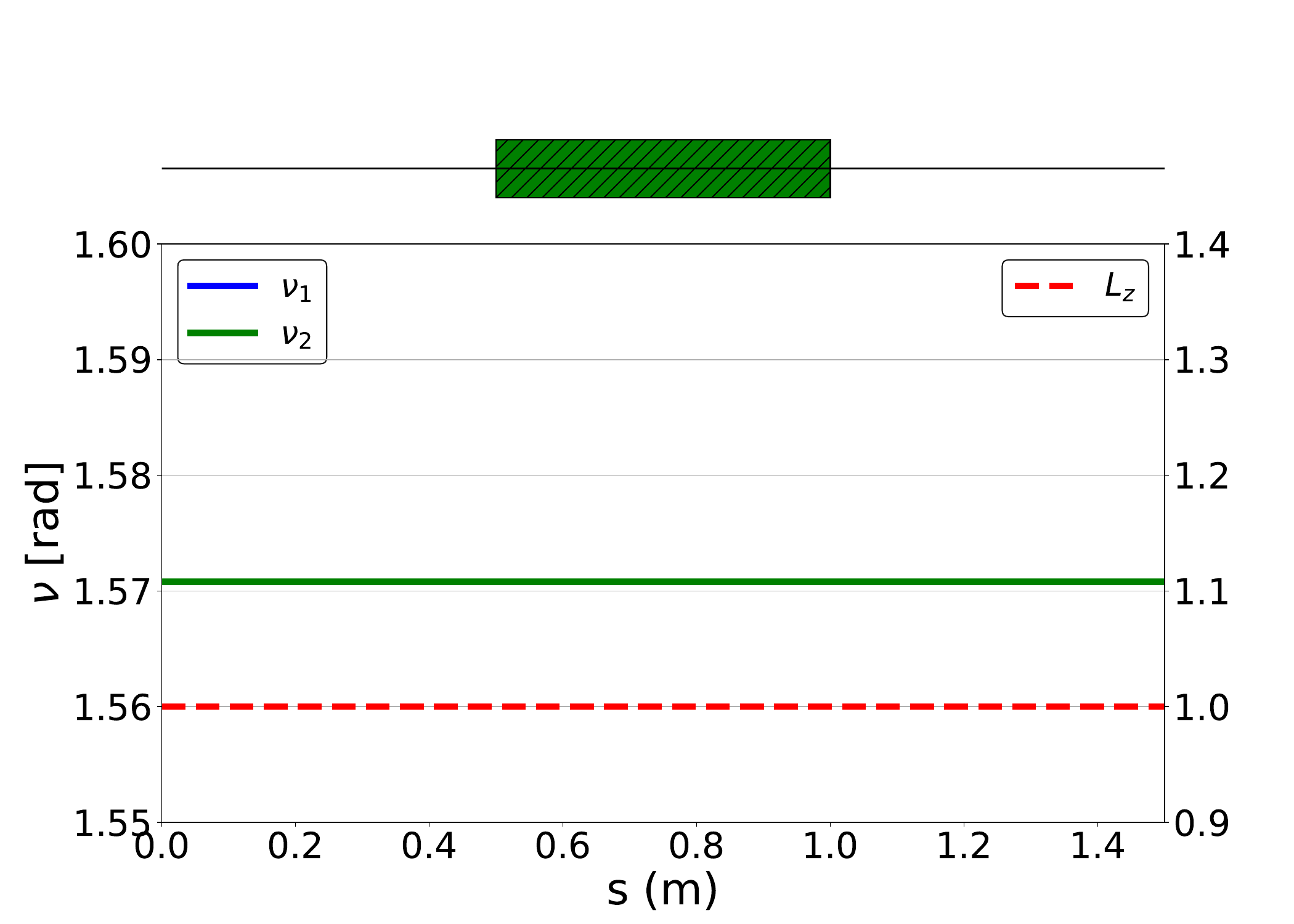}
    \includegraphics[width=0.44\linewidth]{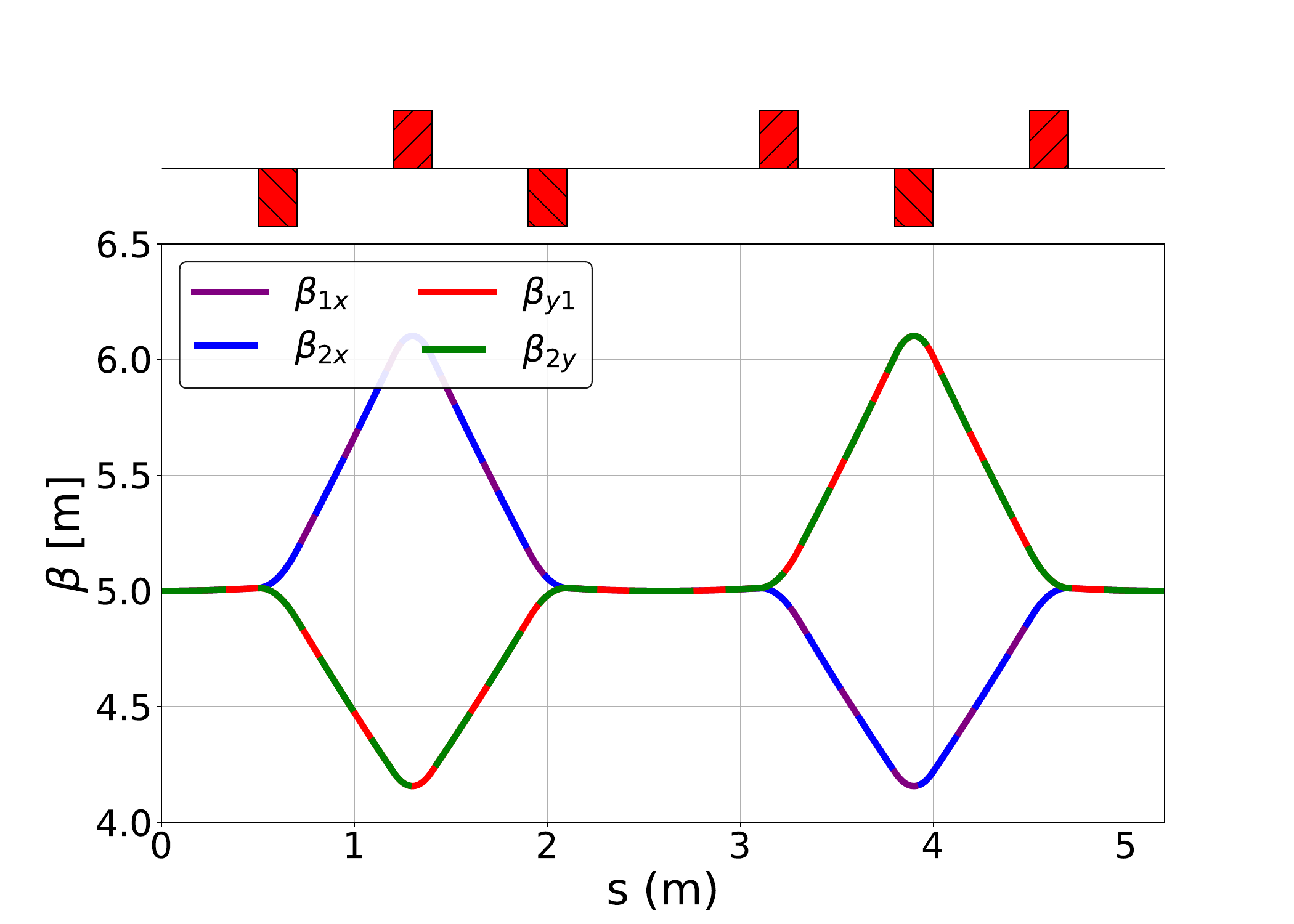}
    \includegraphics[width=0.44\linewidth]{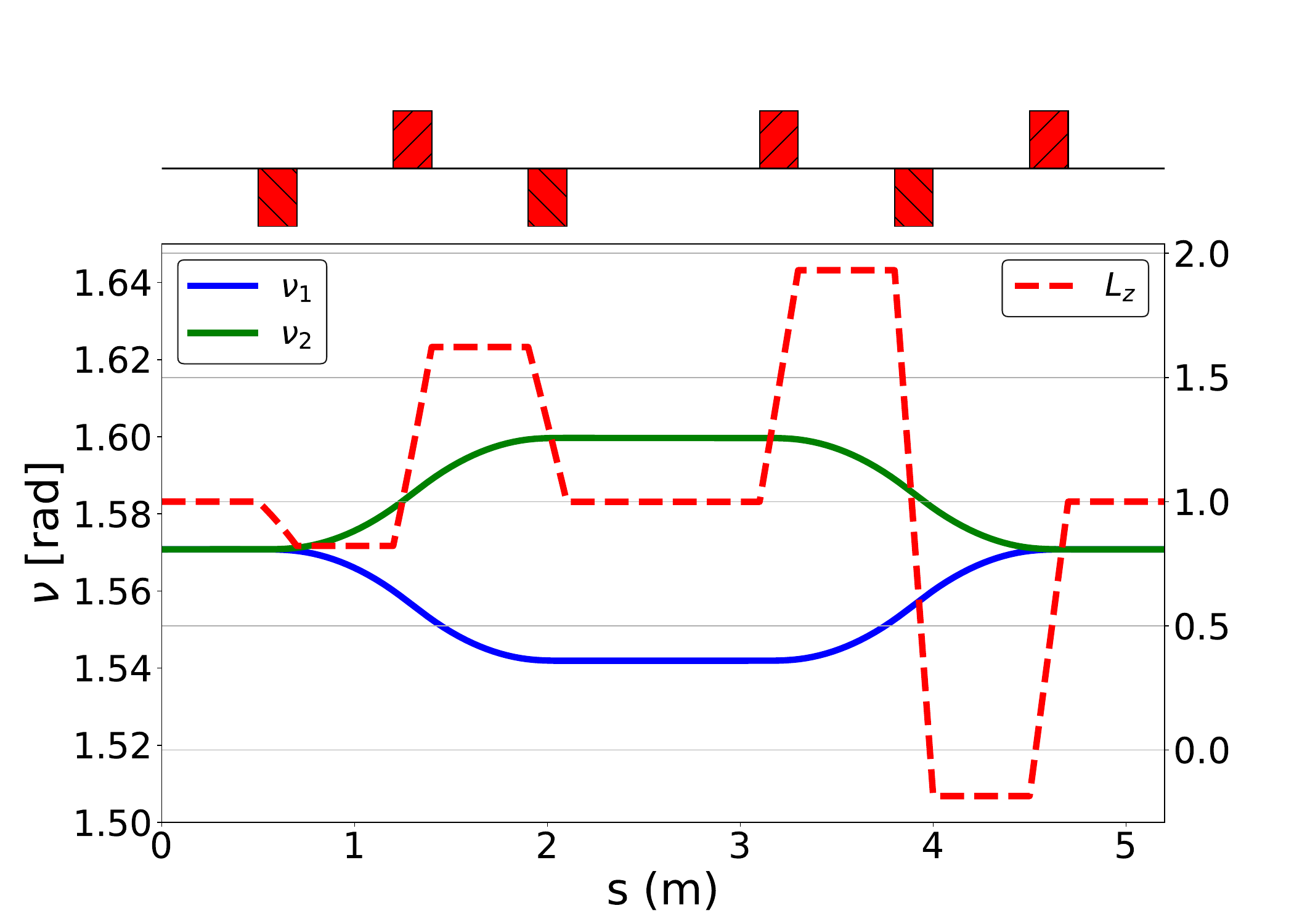}
    \caption{Examples of rotationally invariant periodic cells. Left: coupled beta functions. Right: coupling phases and angular momentum. Red elements denote quadrupoles and green elements denote solenoids. Top: quadrupole-doublet cell. Middle: solenoid cell. Bottom: quadrupole-triplet cell.}
    \label{fig:rotinvcells}
\end{figure*}

\begin{figure*}[tbp]
    \centering
    \includegraphics[width=0.44\linewidth]{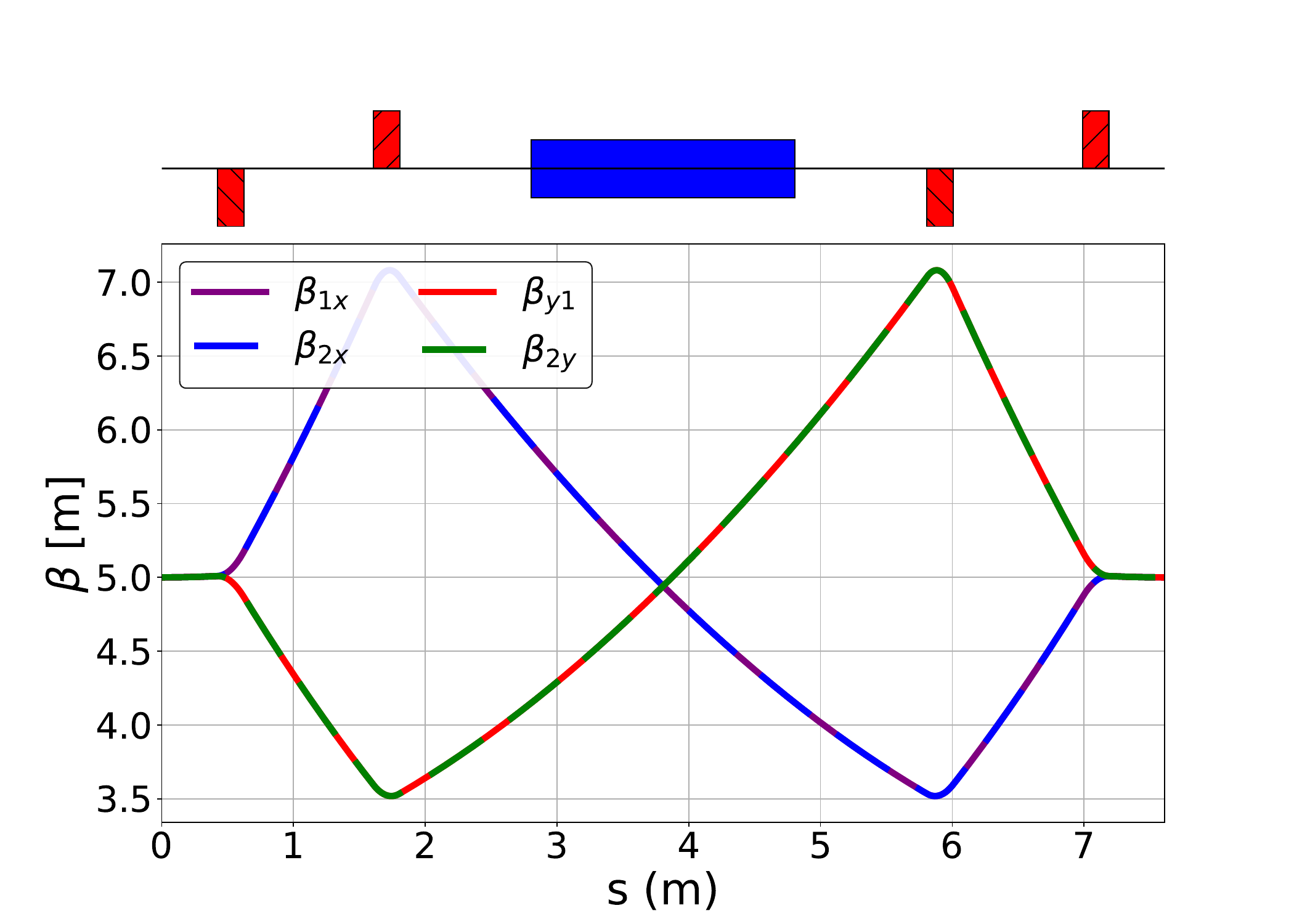}
    \includegraphics[width=0.44\linewidth]{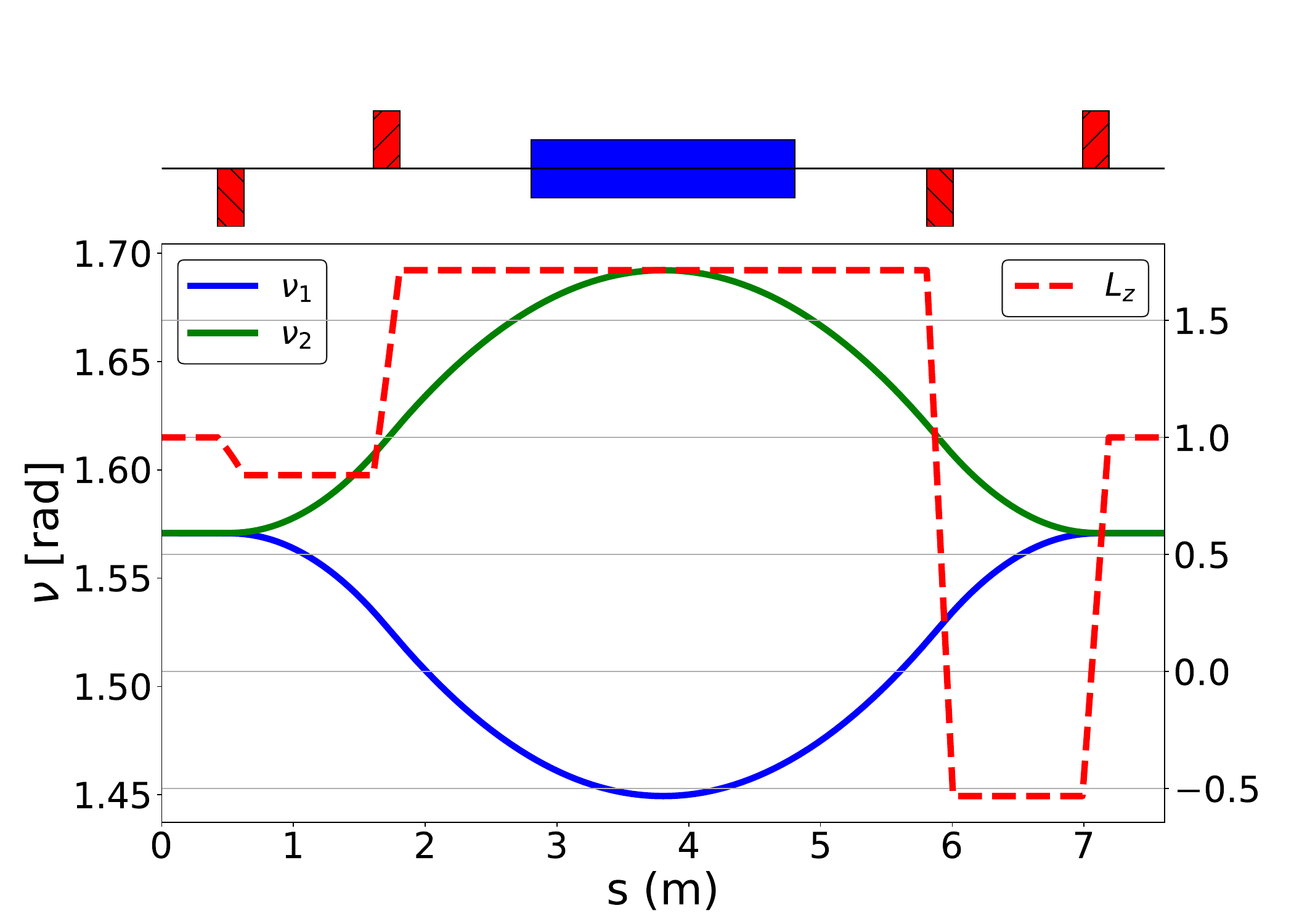}
    \includegraphics[width=0.44\linewidth]{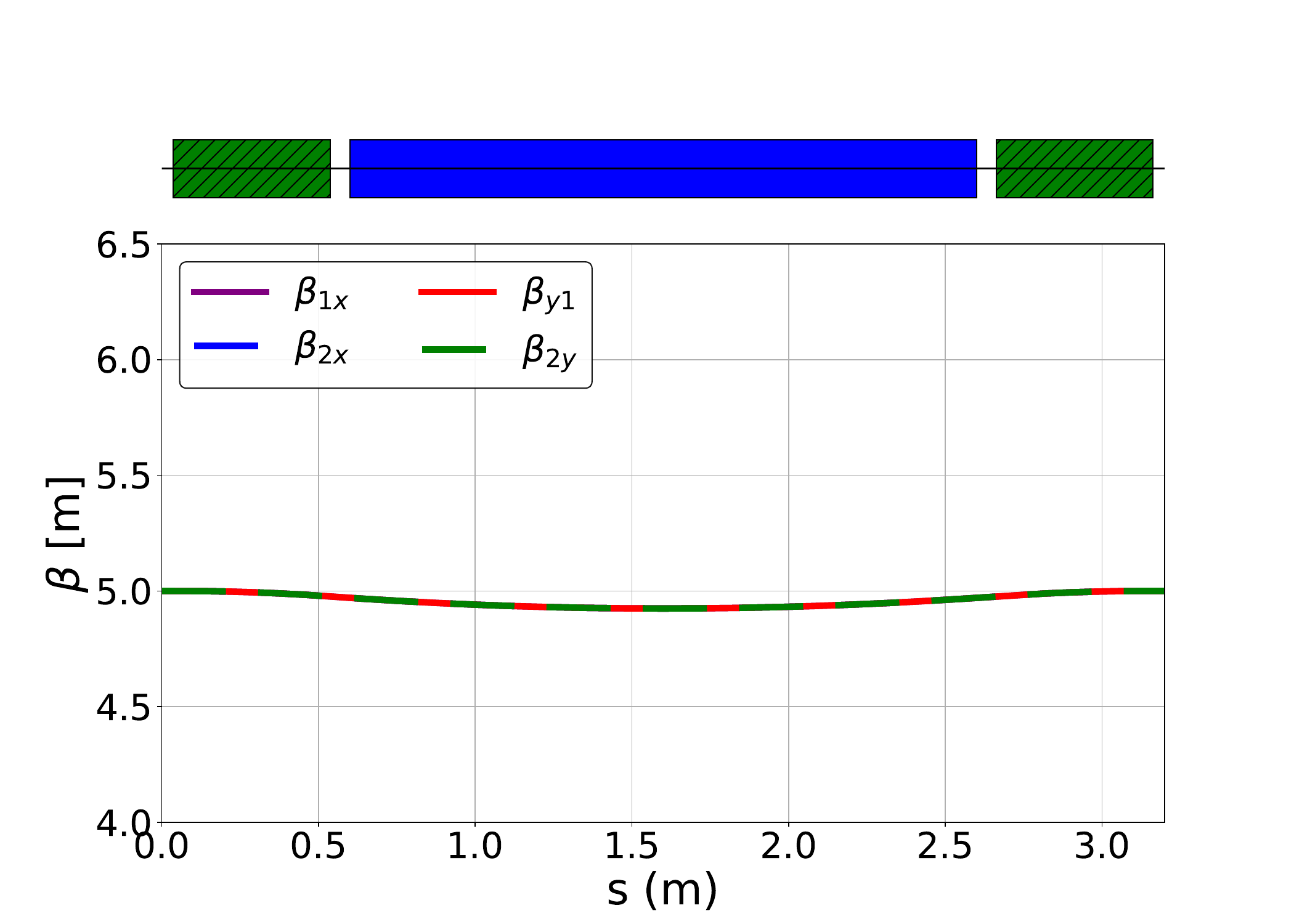}
    \includegraphics[width=0.44\linewidth]{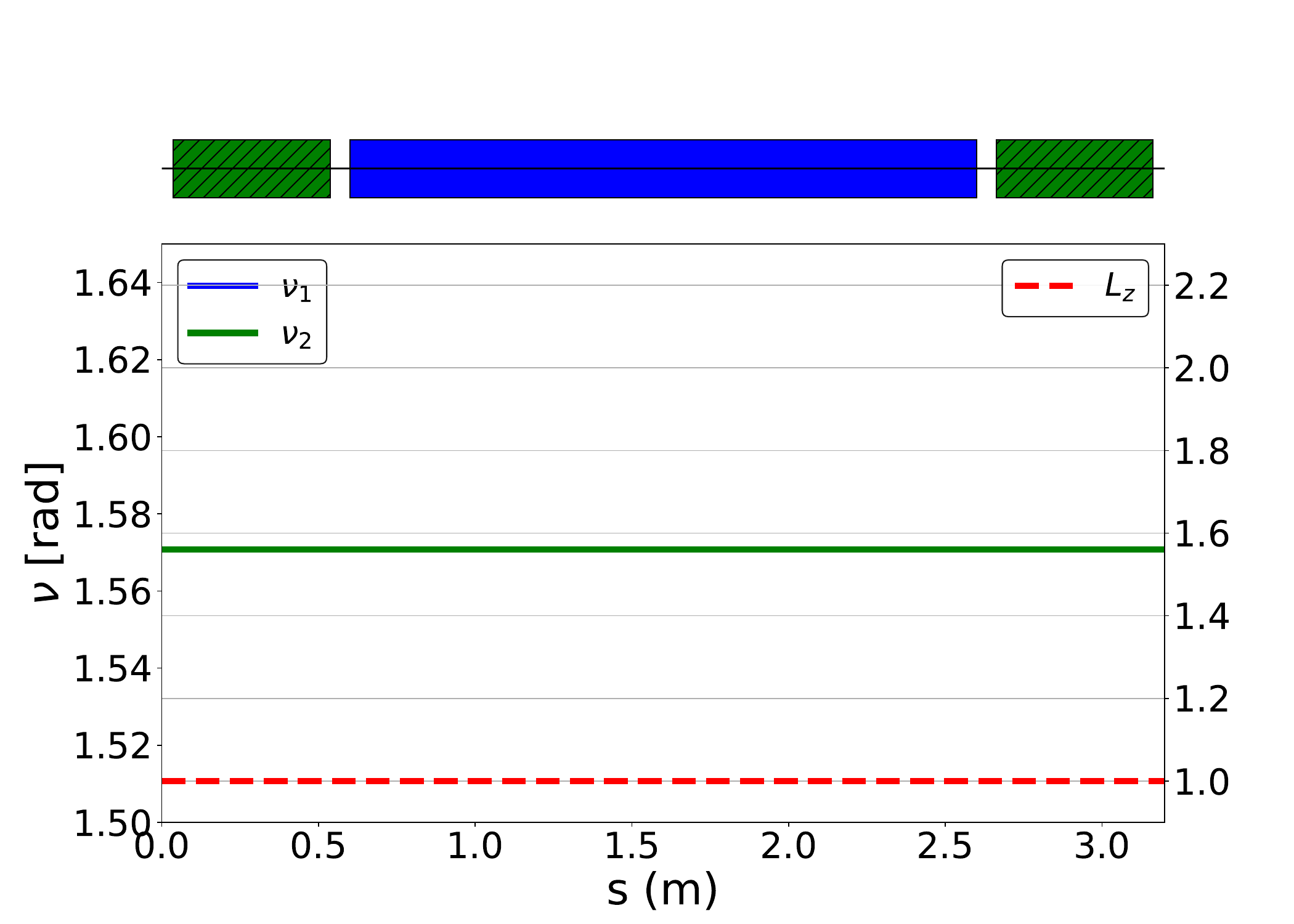}
    \includegraphics[width=0.44\linewidth]{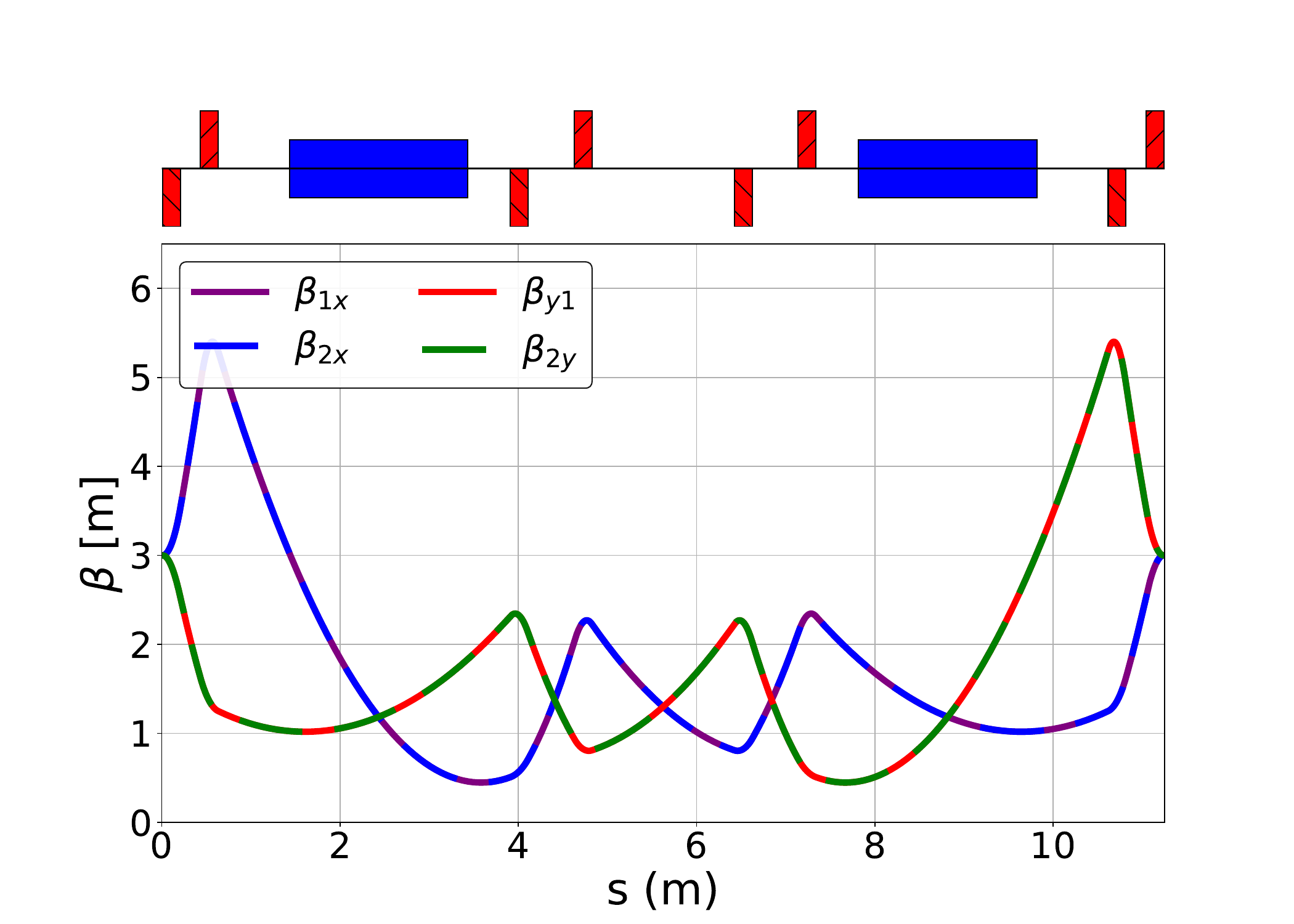}
    \includegraphics[width=0.44\linewidth]{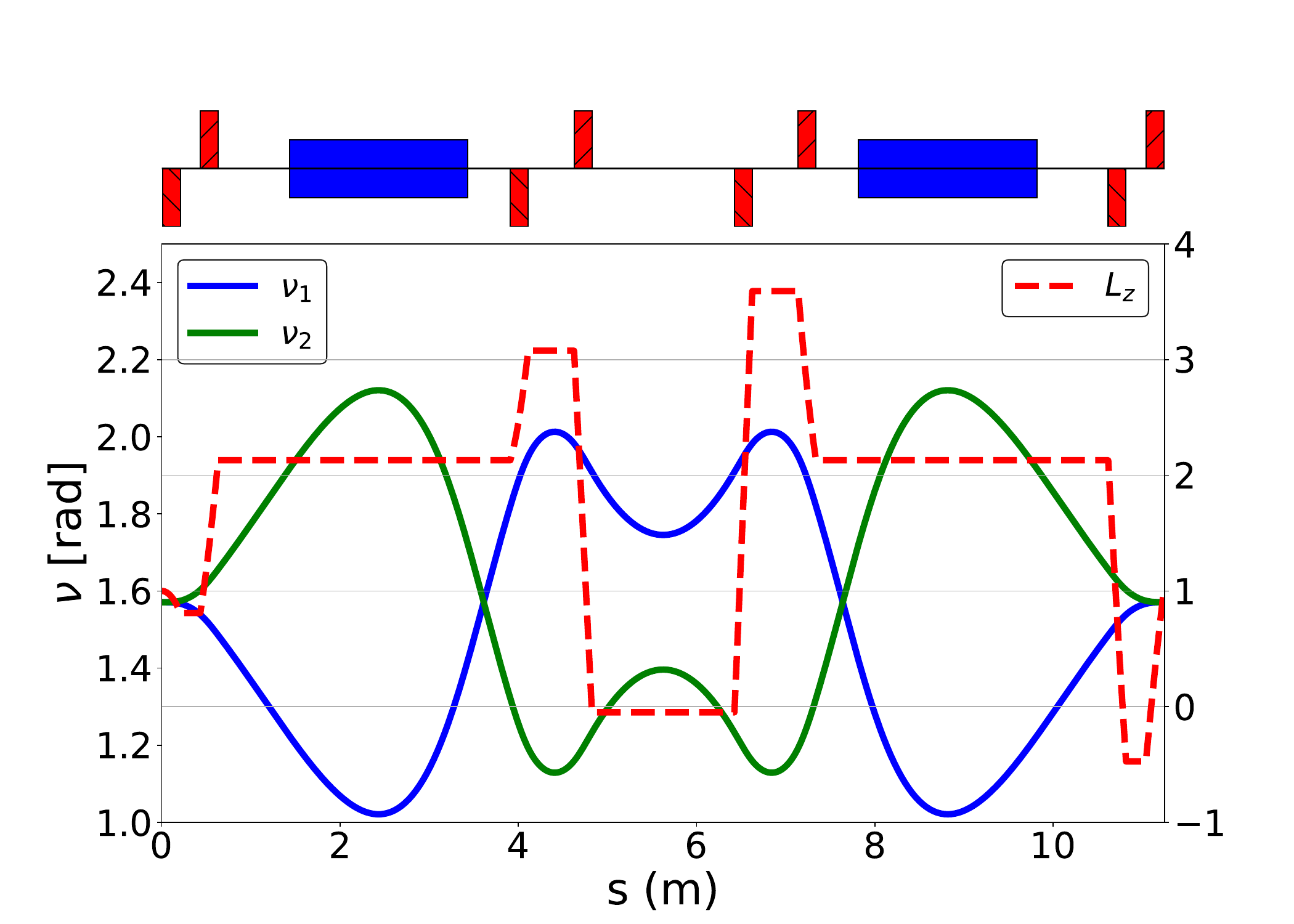}
    \caption{Examples of rotationally invariant periodic cells. Left: coupled beta functions. Right: coupling phases and angular momentum. Red elements denote quadrupoles, blue elements denote indexed dipoles, and green elements denote solenoids. Top: quadrupole-doublet cell. Middle: solenoid cell. Bottom: double-bend achromat cell.}
    \label{fig:rotinvbendingcells}
\end{figure*}

The results of Sections~\ref{subsec:preservingangmom} and ~\ref{subsec:couplingphasepreservation} motivate a set of rotationally invariant lattice building blocks. In the ideal linear lattice, the SAG symmetry restores the circular-mode conditions over one cell and therefore provides discrete conservation of angular momentum and coupling phase from cell to cell. Solenoid-based cells preserve both the coupling phases and angular momentum continuously because of their axisymmetric focusing. However, quadrupole-focusing-based cells must adhere to the SAG design. Conventional dipoles fail to meet the SAG design condition due to a lack of focus in the $y$ plane; instead, indexed dipoles must be used for bending that also have a quadrupole field with an index of one-half, $n=1/2$. Examples of periodic transport cells for circular-mode beams are shown in Fig.~\ref{fig:rotinvcells}, and periodic bending cells are shown in Fig.~\ref{fig:rotinvbendingcells}. Figures~\ref{fig:rotinvcells}
 and~\ref{fig:rotinvbendingcells} show that the angular momentum is periodic over the cell: although $L_z(s)$ varies within the cell, it is restored after one period, $L_z(s+L)=L_z(s)$. Unlike quadrupole-based SAG cells, where $L_z$ is restored only after one full period, solenoids and indexed dipoles with field index $n=1/2$ conserve angular momentum continuously through the element. This conservation is shown in Figs.~\ref{fig:rotinvcells} and~\ref{fig:rotinvbendingcells}, and the solenoids are depicted with a green-colored element. The solenoids are necessary in the design to split the tune and break the eigenvalue degeneracy.

%

\subsection{Identification of Eigenmodes and Measurement of Eigenmode Emittances From Beam Observables}\label{subsec:identificationandmeasurement}
\begin{figure}[tbp]
    \centering
    \includegraphics[width=0.49\linewidth]{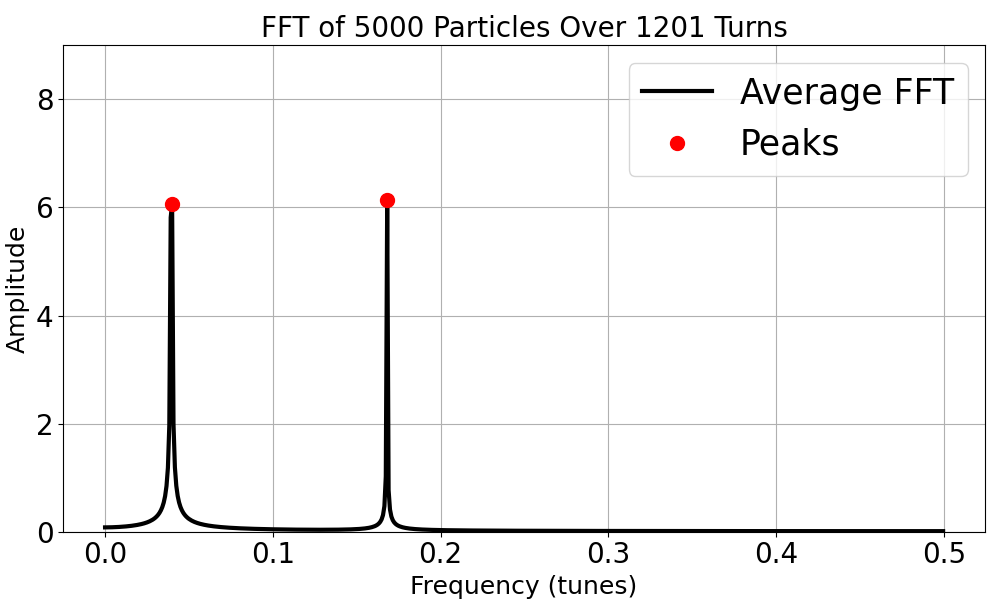}
    \includegraphics[width=0.49\linewidth]{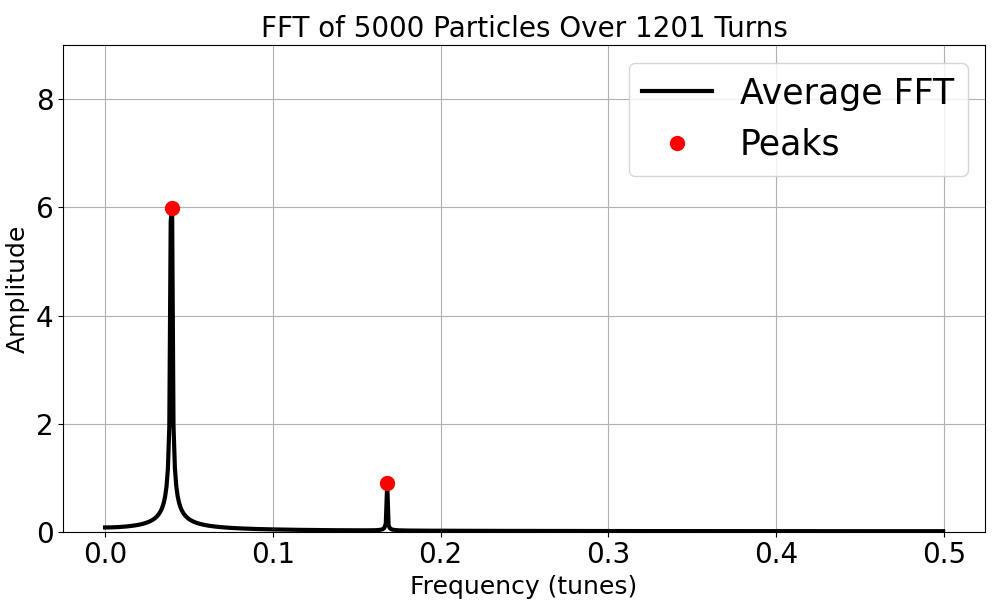}
    \caption{FFT analysis of $u=1/2$ with $\nu_{1,2}=\pi/2$ beam in the $x$ plane. Equal eigenmode emittances (left), and factor of $100$ difference between eigenmode emittances (right). Tracked with ImpactX code in a simple ring design.}
    \label{fig:fftanal}
\end{figure}

\begin{figure}[tbp]
    \centering
    \includegraphics[width=0.8\linewidth]{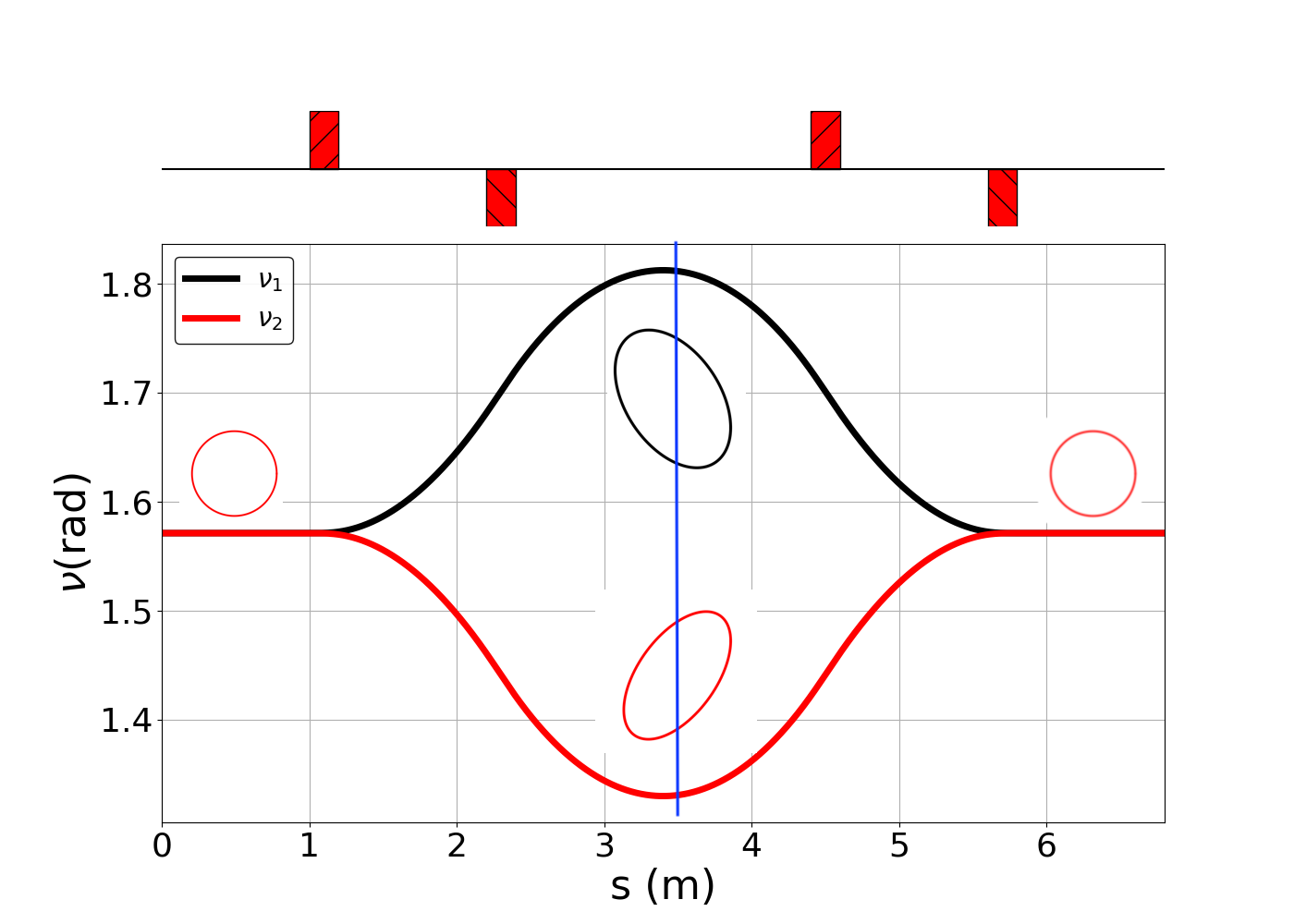}
    \caption{Propagation of coupling phases in a quadrupole channel. The coupling phases $\nu_{1}$ (black) and $\nu_{2}$ (red) evolve with $s$ with opposite sign, producing opposite $(x,y)$-projection tilt angles along the cell; the tilt vanishes at locations where $\nu_{1,2}=\pi/2$.}
    \label{fig:propagationofphaseofcouplingsquad}
\end{figure}

We established in Section~\ref{subsec:preservingangmom} that solenoids are necessary to break the degeneracy of the eigenvalues of the $4\times 4$ transfer matrix. Inclusion of solenoids in the lattice yields a coupled lattice. In~\cite{gilanliogullari2025formalism}, we have shown that the Fourier transformed $(x,y)$ coordinates yield two Fourier peaks in the projected $(x,y)$ plane given by
\begin{equation}
\begin{split}
    X(f) &= \frac{\sqrt{\epsilon_{1}\beta_{1x}}}{2}\delta(f-Q_{1}) + \frac{\sqrt{\epsilon_{2}\beta_{2x}}}{2}e^{i\nu_{2}}\delta(f-Q_{2}),\\
    Y(f) &= \frac{\sqrt{\epsilon_{1}\beta_{1y}}}{2}e^{i\nu_{1}}\delta(f-Q_{1}) + \frac{\sqrt{\epsilon_{2}\beta_{2y}}}{2}\delta(f-Q_{2}).
    \end{split}
    \label{eq:FFTXYcoordinates}
\end{equation}
Here, $X(f)$ and $Y(f)$ are Fourier transformed coordinates, $Q_{1}$ and $Q_{2}$ are the eigenmode tunes, $\beta_{1x},\beta_{2x},\beta_{1y}$ and $\beta_{2y}$ are the betatron functions, $\nu_{1},\nu_{2}$ are the coupling phases, and $\epsilon_{1},\epsilon_{2}$ are the eigenmode emittances. The Fourier amplitudes depend on the eigenmode emittances. FFT analysis of turn-by-turn data from tracking results computed from a simple ring design, including solenoids, using ImpactX simulation~\cite{huebl2022next}, is shown in Fig.~\ref{fig:fftanal}. In Fig.~\ref{fig:fftanal}, the left plot shows a coupled beam with equal eigenmode emittances $\epsilon_{1}\approx\epsilon_{2}$, and the right plot shows a circular-mode beam with $\epsilon_{1}=90\epsilon_{2}$. The two peaks in Fig.~\ref{fig:fftanal} correspond to the tunes of modes 1 and 2, while their relative amplitudes reflect the square-root ratio of the corresponding eigenmode emittances. Computing the relative amplitudes of the Fourier peaks yields the square root of the intrinsic flatness, which allows us to extract the intrinsic flatness ratio of the beam by analyzing the Fourier amplitudes. The Fourier analysis also indicates which mode is dominant; for instance, the amplitude of mode~1 is larger than that of mode~2.
Another way to identify the dominant eigenmode is to measure the $(x,y)$ projection when the coupling phases satisfy $\nu_{j}\neq\tfrac{\pi}{2}$ ($j=1,2$). The tilt angle $\theta$ of the projected ellipse is
\begin{equation}
\tan 2\theta
=\frac{2\sigma_{xy}}{\sigma_x^{2}-\sigma_y^{2}}
=\frac{2\sqrt{\beta_{jx}\beta_{jy}}}{\beta_{jx}-\beta_{jy}}\cos\nu_j,
\label{eq:tiltangle}
\end{equation}
where $j$ denotes the dominant mode (mode~1 or mode~2), and the corresponding optics functions $(\beta_{jx},\beta_{jy},\nu_j)$ are used. In the circular-mode limit, the beam is round and $\nu_{1,2}=\tfrac{\pi}{2}$ (mod $\pi$), so the $(x,y)$ projection is untilted. However, in quadrupole-based channels, the coupling phases are generally $s$-dependent; Fig.~\ref{fig:propagationofphaseofcouplingsquad} shows that the two modes evolve with opposite-sign coupling phases. Consequently, the $(x,y)$ tilt changes sign depending on which mode is dominant, providing an experimental handle to identify the dominant mode from the measured ellipse orientation.
For known optics from the lattice, we can introduce a normalized $(X,Y)$ projection plane as $X=x/\sqrt{\beta_{1}}$ and $Y=y/\sqrt{\beta_{2}}$, where $\beta_{1}$ and $\beta_{2}$ are decoupled plane beta functions. Since only eigenmode~1 dominates the dynamics, the dominant-mode emittance can be calculated from the beam area in the normalized $(X,Y)$ plane and the coupling phase, as
\begin{equation}
    X^{2} + Y^{2} = \epsilon_{1}\sin^{2}\nu_{1}.
\end{equation}

An alternative method to determine the eigenmode emittances is given in Ref.~\cite{gilanliogullari2025formalism}. Consider an uncoupled section of the ring lattice whose linear transverse transfer matrix is block diagonal, $\mathcal{M}=\mathrm{diag}(M,N)$, where $M$ and $N$ are the $2\times2$ transfer matrices in the horizontal $(x,x')$ and vertical $(y,y')$ planes, respectively. Suppose the beam is measured at an initial waist location, yielding rms sizes $\sigma_{xi}$ and $\sigma_{yi}$. After transport through a quadrupole section that generates a nonzero $(x,y)$ tilt, the rms sizes and tilt angle at a downstream location are measured as $\sigma_{xf}$, $\sigma_{yf}$, and $\theta_f$.  The eigenmode emittances are then obtained as
\begin{equation}
\begin{split}
2\epsilon_{1} &= 2\sqrt{\frac{\sigma_{xi}^{2}\sigma_{xf}^{2}}{M_{12}^{2}}-\frac{M_{11}^{2}}{M_{12}^{2}}\sigma_{xi}^{4}}
+ \frac{(\sigma_{xf}^{2}-\sigma_{yf}^{2})\tan2\theta_{f}}{M_{11}N_{12}\frac{\sigma_{xi}}{\sigma_{yi}}-M_{12}N_{11}\frac{\sigma_{yi}}{\sigma_{xi}}}, \\
2\epsilon_{2} &=
\frac{(\sigma_{xf}^{2}-\sigma_{yf}^{2})\tan2\theta_{f}}{M_{11}N_{12}\frac{\sigma_{xi}}{\sigma_{yi}}-M_{12}N_{11}\frac{\sigma_{yi}}{\sigma_{xi}}}
- 2\sqrt{\frac{\sigma_{xi}^{2}\sigma_{xf}^{2}}{M_{12}^{2}}-\frac{M_{11}^{2}}{M_{12}^{2}}\sigma_{xi}^{4}}.
\end{split}
\end{equation}

Overall, methods described in this section enable identification of the dominant mode, extraction of the eigenmode emittances from measurements, and evaluation of the intrinsic flatness ratio.

\section{Space-Charge Effects}\label{sec:spacechargeeffects}
Space charge refers to the beam’s self-generated electromagnetic field. This self-field acts as an additional (generally defocusing) contribution to the external focusing, producing an incoherent tune shift and tune spread. The strength of the space-charge effect is characterized by the perveance parameter,
\begin{equation}
\kappa_{sc}=\frac{2r_{p}}{\beta^{2}\gamma^{3}}\lambda(z),
\end{equation}
where $r_{p}$ is the classical particle radius, $\beta$ and $\gamma$ are the relativistic factors, and $\lambda(z)$ is the line charge density. Space charge scales with the charge density and decreases rapidly with increasing energy. For high-intensity beams at low energy, space-charge effects are typically dominant. In addition, the resulting tune shift and tune spread depend on the transverse charge distribution in the $(x,y)$ plane (through the beam sizes and profile shape). In this manuscript, we use Gaussian distribution profiles for the transverse phase spaces and we distinguish between the tune-shift expressions for uncoupled and coupled lattices. 

For a correlated beam in an uncoupled lattice, it is convenient to describe the transverse distribution in its rotated principal axes $(\eta,\tau)$, defined by
\begin{equation}
(\eta,\tau)^{T}=R(\theta)\,(x,y)^{T},
\label{eq:rotationtoprincipal}
\end{equation}

Here, \(R(\theta)\) is the \(2\times2\) rotation matrix and \(\theta\) is the tilt angle of the projected \((x,y)\) distribution. This rotation diagonalizes the projected transverse covariance matrix, so that \((\eta,\tau)\) are the principal axes of the spatial charge distribution. It is introduced only to simplify the evaluation of the Gaussian space-charge potential and does not represent a transformation to dynamically decoupled particle coordinates.

The space-charge tune shifts with non-zero tilt angle $\theta$ are
\begin{equation}
	\begin{split}
		\Delta Q_{x} &= -\frac{\kappa_{sc}}{4\pi}\oint ds \frac{\beta_{x}\cos^{2}\theta}{\sigma_{\eta}(\sigma_{\eta}+\sigma_{\tau})} + \frac{\beta_{x}\sin^{2}\theta}{\sigma_{\tau}(\sigma_{\eta} + \sigma_{\tau})}, \\
		\Delta Q_{y} &= -\frac{\kappa_{sc}}{4\pi}\oint ds \frac{\beta_{y}\sin^{2}\theta}{\sigma_{\eta}(\sigma_{\eta}+\sigma_{\tau})} + \frac{\beta_{y}\cos^{2}\theta}{\sigma_{\tau}(\sigma_{\eta} + \sigma_{\tau})}.
	\end{split}
\end{equation}
Here, $\sigma_{\eta}$ and $\sigma_{\tau}$ are rms beam sizes in the rotated axes [Eq.~\eqref{eq:principalrmsbeamsizes}]. For aligned projected $(x,y)$ plane distribution, $\theta=0$, the tune-shift formalism for uncoupled beam and lattices are recovered as:
\begin{equation}
    \begin{split}
        \Delta Q_{x}&= -\frac{\kappa_{sc}}{4\pi}\oint ds \frac{\beta_{x}}{\sigma_{x}(\sigma_{x} + \sigma_{y})}, \\
        \Delta Q_{y} &= -\frac{\kappa_{sc}}{4\pi}\oint ds \frac{\beta_{y}}{\sigma_{y}(\sigma_{x} + \sigma_{y})}. 
    \end{split}
\end{equation}
Here, for $\theta=0$ there is no rotation of axes, $\eta=x$ and $\tau=y$. For circular-mode beams, the goal is to maintain round projection through the lattice by minimizing the oscillation of the coupling phase, $\nu_{1}\approx\tfrac{\pi}{2}$. To leading order, the tune shifts for a circular-mode beam and an uncorrelated round beam are equivalent
\begin{equation}
    \Delta Q_{x,y} = -\frac{\kappa_{sc}}{8\pi}\langle \frac{\beta_{c}}{\sigma_{b}^{2}} \rangle_{s},
    \label{eq:tuneshiftround}
\end{equation}
where $\sigma_{b}\equiv\sigma_{x}=\sigma_{y}$ for a round beam and $\beta_c$ denotes the effective round-beam beta function, related to the circular-mode projected beta functions by $\beta_c=2\beta_{1x}=2\beta_{1y}$ in the convention used below. 

\begin{figure}[tbp]
    \centering
    \includegraphics[width=0.47\linewidth]{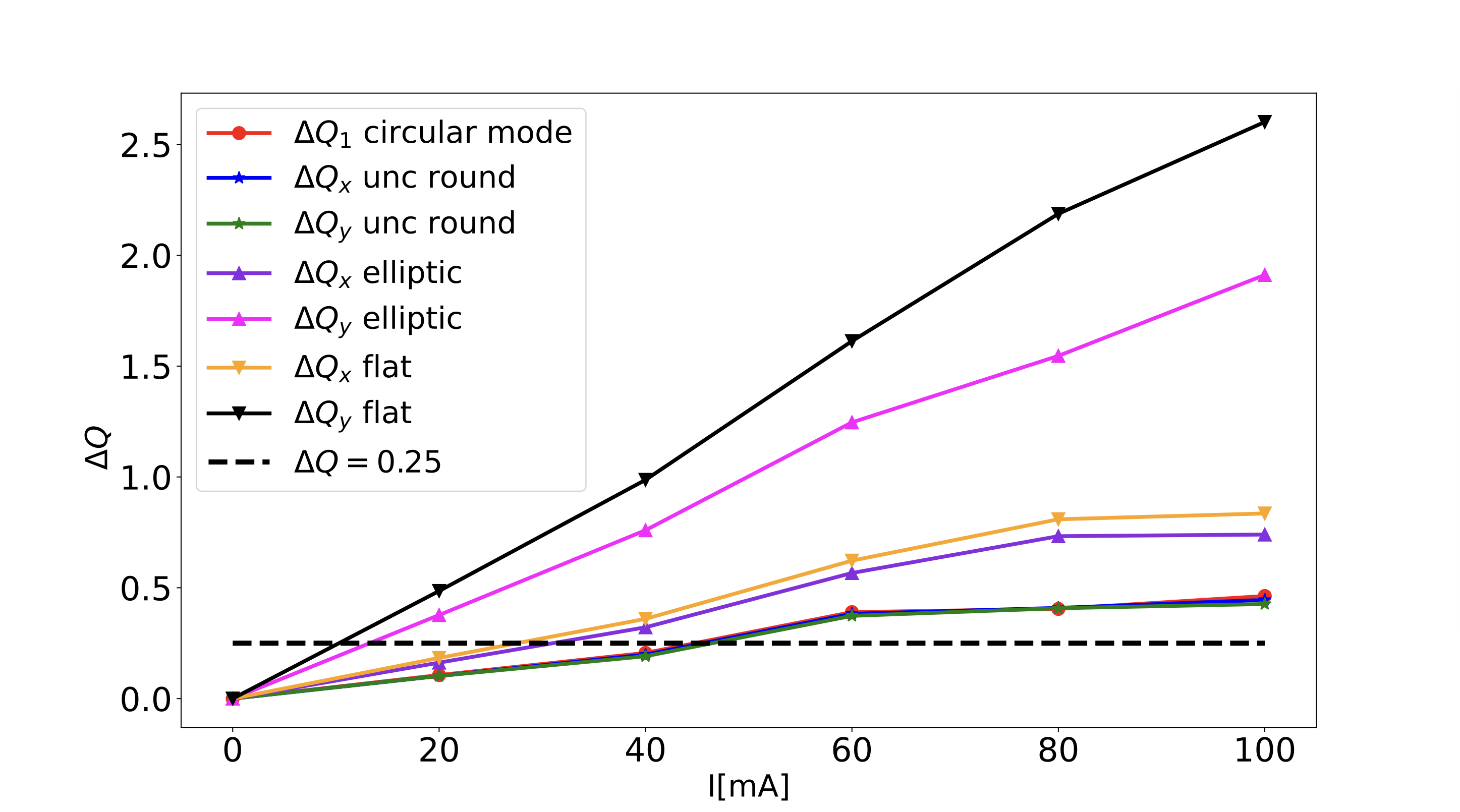}
    \includegraphics[width=0.47\linewidth]{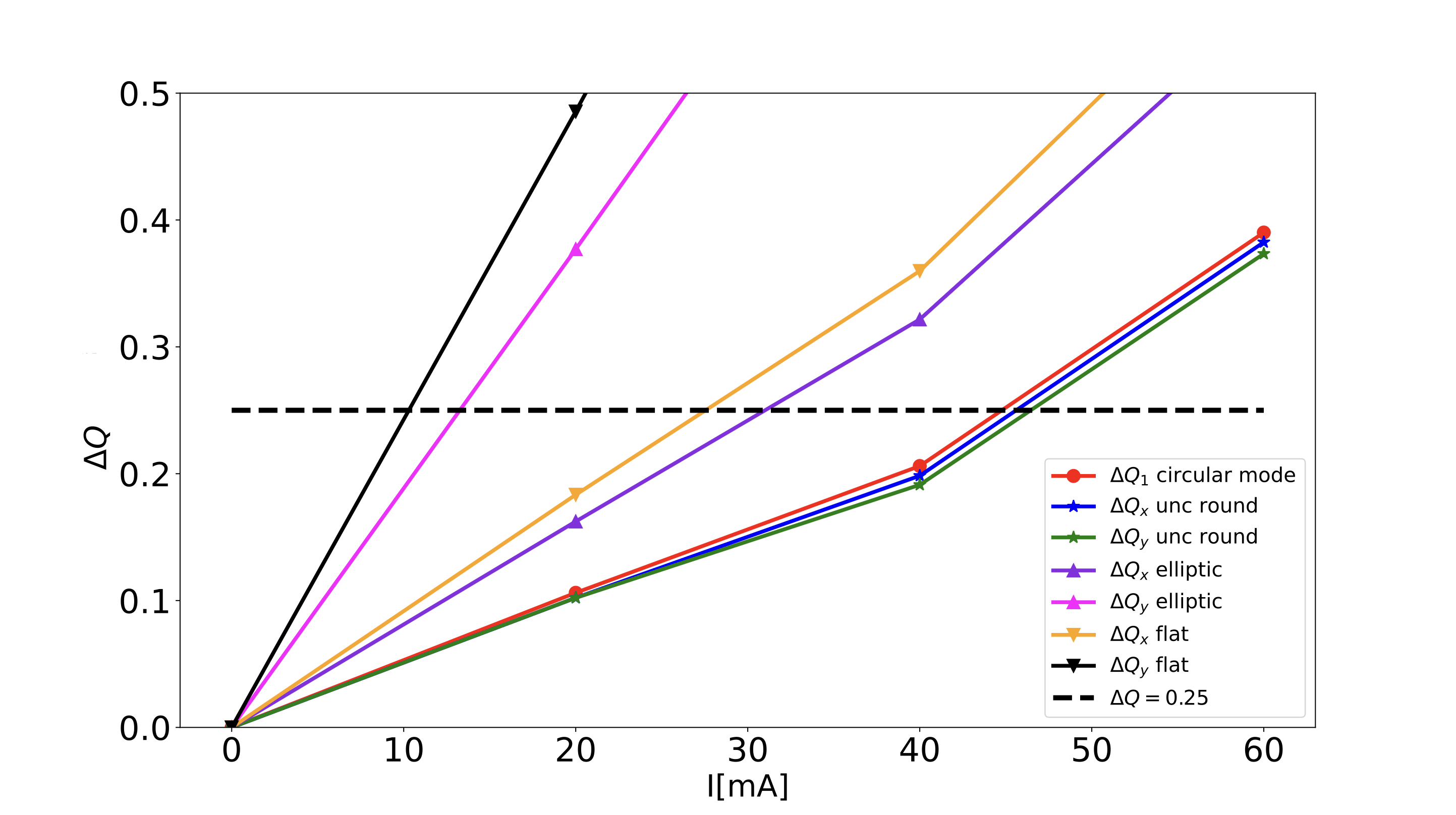}
    \caption{Tune shifts for different beam shapes with $E_{k}=100$\,MeV energy with protons. Elliptic cross section beam, $\epsilon_{x}=10\epsilon_{y}$, flat beam $\epsilon_{x}=100\epsilon_{y}$, $\epsilon_{1}=100\epsilon_{2}$ for circular-mode beam. Tune shifts for different current values (left), tune shifts for current values between $I\in (0.0,60.0)$\,mA (right). The dotted line marks a tune-shift value of $\Delta Q = 0.25$, which illustrates a reference point for comparison. }
    \label{fig:tuneshiftfordifferentbeamshapes}
\end{figure}

Figure~\ref{fig:tuneshiftfordifferentbeamshapes} compares the incoherent tune shifts obtained from TRACK~\cite{TRACKweb} simulations for elliptic, uncorrelated round, flat, and circular-mode beams. Each marker represents a tune shift extracted from the tracking calculation. For the same intensity and projected rms beam size, the results for the circular-mode and uncorrelated round beams nearly overlap, providing a numerical consistency check of the leading-order equivalence predicted by Eq.~\eqref{eq:tuneshiftround}. By contrast, the physically flat beam with the same intrinsic emittance ratio exhibits a substantially larger vertical tune shift. Thus, maintaining a round projected density allows the circular-mode beam to retain the favorable leading-order space-charge tune shift of an uncorrelated round beam while preserving its intrinsic eigenemittance asymmetry.

For a coupled beam in a coupled lattice, the transverse axes can still be rotated to align with the principal axes $(\eta,\tau)$, as defined in Eq.~\eqref{eq:rotationtoprincipal}. The space-charge tune shifts for coupled lattices are
\begin{equation}
    \begin{split}
        \Delta Q_{1}&= -\frac{\kappa_{sc}}{4\pi}\oint ds\frac{1}{\sigma_{\eta}\sigma_{\tau}(\sigma_{\eta}+\sigma_{\tau})}((\beta_{1x}\sigma_{\tau}+\beta_{1y}\sigma_{\eta})\cos^{2}\theta \\
        &+(\beta_{1x}\sigma_{\eta}+\sigma_{\tau}\beta_{1y})\sin^{2}\theta + (\sigma_{\tau}-\sigma_{\eta})\sqrt{\beta_{1x}\beta_{1y}}\cos\nu_{1}\sin2\theta), \\
        \Delta Q_{2} &=  -\frac{\kappa_{sc}}{4\pi}\oint ds\frac{1}{\sigma_{\eta}\sigma_{\tau}(\sigma_{\eta}+\sigma_{\tau})}((\beta_{2x}\sigma_{\tau}+\beta_{2y}\sigma_{\eta})\cos^{2}\theta \\
        &+(\beta_{2x}\sigma_{\eta}+\sigma_{\tau}\beta_{2y})\sin^{2}\theta + (\sigma_{\tau}-\sigma_{\eta})\sqrt{\beta_{2x}\beta_{2y}}\cos\nu_{2}\sin2\theta).
    \end{split}
    \label{Eq:tuneshiftproven}
\end{equation}
Here, $\Delta Q_{1,2}$ are the eigenmode tune shifts. In the circular-mode limit, Eq.~\eqref{Eq:tuneshiftproven} reduces to Eq.~\eqref{eq:tuneshiftround}, yielding equal tune shifts for an uncorrelated round beam and the circular-mode beam.

In addition to the tune shift, space charge also generates tune spread. For a beam with a transverse Gaussian density, the self-field is nonlinear, so particles at different transverse amplitudes experience different focusing strengths. This amplitude-dependent incoherent tune shift produces a tune spread across the beam. The space charge tune spread formalism for uncoupled beams in uncoupled lattices is given in~\cite{sen2023density}. In the same reference, the space charge is treated as a perturbative effect and the Hamiltonian is 
\begin{equation}
    \mathcal{H} = Q_{x}J_{x} + Q_{y}J_{y} + V_{sc}(\Psi_{x},J_{x},\Psi_{y},J_{y}),
    \label{eq:perturbedhamiltonianunc}
\end{equation}
where $V_{sc}$ is the space charge potential for a Gaussian distribution given in Eq.~\eqref{eq:spacechargepotentialgauss}. The amplitude dependent tune shift is
\begin{equation}
    \Delta Q_{x,y}= \langle \frac{\partial V_{sc}}{\partial J_{x,y}} \rangle_{\psi_{x},\psi_{y},\psi_{z},s}.
    \label{eq:tunespreaddifferential}
\end{equation}

Here, the brackets denote averaging over the betatron and synchrotron phases and around the lattice, while \(J_x\) and \(J_y\) are the transverse Courant--Snyder actions. This averaging is not performed by treating \((\eta,\tau)\) as dynamically decoupled coordinates. For an uncoupled lattice, the unperturbed particle motion is decoupled in the horizontal and vertical Courant--Snyder variables, whereas for the coupled lattice considered below it is decoupled in the normal-mode variables \((J_1,\psi_1)\) and \((J_2,\psi_2)\) introduced in Sec.~\ref{subsec:theoryeigenmodedesc}. The space-charge potential can therefore be evaluated in the principal-axis representation of the transverse density and subsequently expressed in terms of the appropriate dynamical actions and phases before carrying out the perturbative average.

\begin{figure}[tbp]
    \centering
    \includegraphics[width=0.49\linewidth]{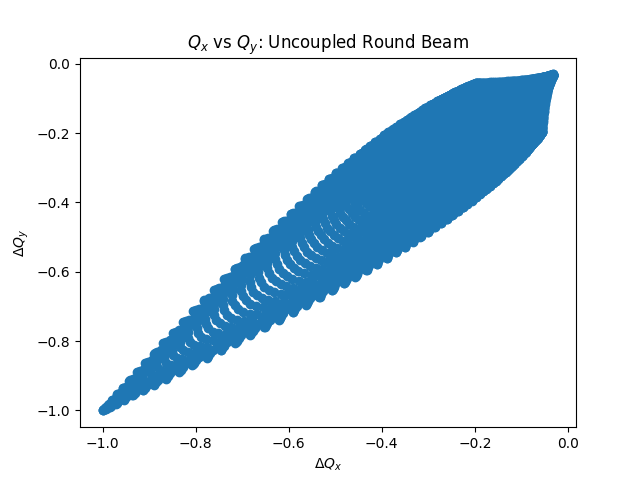}
    \includegraphics[width=0.49\linewidth]{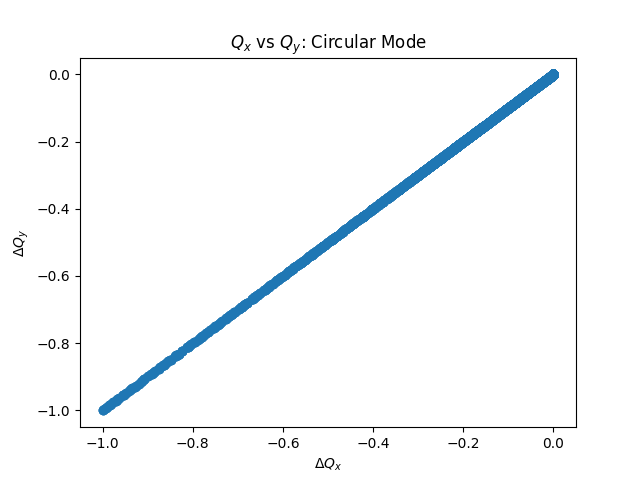}
    \caption{The footprint of the dimensionless tune-shift contribution defined by the integrals in Eq.~\eqref{eq:uncoupledroundbeamspread} and Eq.~\eqref{eq:spreadcircmodeuncbeam}. Uncorrelated round beam footprint is shown in the left figure and circular-mode beam footprint is shown in the right figure. The space-charge prefactor, $C_{SC}$, is removed to highlight the dependence on transverse amplitudes and optics.}
    \label{fig:unctunespreadcircmoderoundbeam}
\end{figure}

In an uncoupled lattice, single-particle motion may be written in Courant--Snyder form~\cite{courant1958theory}, $x=\sqrt{2J_x\beta_x}\cos\psi_x$, $y=\sqrt{2J_y\beta_y}\cos\psi_y$, and $z=\sqrt{2J_z\beta_z}\cos\psi_z$. Defining the normalized amplitudes relative to the rms emittances as $a_j=\sqrt{2J_j/\epsilon_j}$ for $j\in\{x,y,z\}$, the coordinates can be written as
$x=a_x\sigma_x\cos\psi_x$, $y=a_y\sigma_y\cos\psi_y$, and $z=a_z\sigma_z\cos\psi_z$,
where $\sigma_j=\sqrt{\beta_j\epsilon_j}$. For a round, long bunch with $\sigma_{x,y}\ll\sigma_{z}$, the incoherent tune spreads are
{\small
\begin{equation}
\begin{split}
\Delta Q_{x}(a_{x},a_{y},a_{z})&= -C_{SC}\int_{0}^{1}dt \,\frac{\beta_{r}}{\sigma_{b}^{2}}
\left[ I_{0}\!\left(\frac{a_{x}^{2}t}{2}\right) - I_{1}\!\left(\frac{a_{x}^{2}t}{2}\right) \right]
I_{0}\!\left(\frac{a_{y}^{2}t}{2}\right) I_{0}\!\left(\frac{a_{z}^{2}t}{2}\right)\exp\left(-\frac{t}{4}(a_{x}^{2}+a_{y}^{2}+a_{z}^{2}) \right), \\
\Delta Q_{y}(a_{x},a_{y},a_{z})&= -C_{SC}\int_{0}^{1}dt \,\frac{\beta_{r}}{\sigma_{b}^{2}}
\left[ I_{0}\!\left(\frac{a_{y}^{2}t}{2}\right) - I_{1}\!\left(\frac{a_{y}^{2}t}{2}\right) \right]
I_{0}\!\left(\frac{a_{x}^{2}t}{2}\right) I_{0}\!\left(\frac{a_{z}^{2}t}{2}\right)\exp\left(-\frac{t}{4}(a_{x}^{2}+a_{y}^{2}+a_{z}^{2}) \right).
\end{split}
\label{eq:uncoupledroundbeamspread}
\end{equation}
}
Here, $I_{0}$ and $I_{1}$ are modified Bessel functions, and $C_{SC}=\tfrac{N_{p}r_{p}}{\sqrt{\pi}\beta^{3}\gamma^{2}}$ is the space-charge coefficient. In the ideal single-mode circular limit, or to leading order for a strongly mode-1-dominated beam, the projected transverse actions satisfy $J_x\simeq J_y$. In circular-mode limit, the tune spreads in $(x,y)$ plane become equal yielding
{\small
\begin{equation}
    \begin{split}
        \Delta Q_{x,y}(a_{b},a_{b},a_{z})&= -C_{SC}\int_{0}^{1}dt \,\frac{\beta_{r}}{\sigma_{b}^{2}}
\left[ I_{0}\!\left(\frac{a_{b}^{2}t}{2}\right) - I_{1}\!\left(\frac{a_{b}^{2}t}{2}\right) \right]
I_{0}\!\left(\frac{a_{b}^{2}t}{2}\right) I_{0}\!\left(\frac{a_{z}^{2}t}{2}\right)\exp\left(-\frac{t}{4}(2a_{b}^{2}+a_{z}^{2}) \right),
    \end{split}
    \label{eq:spreadcircmodeuncbeam}
\end{equation}
}
where $a_{b}\equiv a_{x}=a_{y}$ for a circular-mode beam. The dimensionless tune-footprint contributions obtained from Eqs.~\eqref{eq:uncoupledroundbeamspread} and~\eqref{eq:spreadcircmodeuncbeam} are shown in Fig.~\ref{fig:unctunespreadcircmoderoundbeam}. The space-charge prefactor \(C_{\mathrm{SC}}\) is omitted to emphasize the dependence on the transverse amplitudes. For an uncorrelated round beam, the horizontal and vertical actions vary independently, producing the conventional two-dimensional necktie-shaped footprint. In the circular-mode limit considered here, the projected transverse amplitudes are constrained by \(a_x=a_y=a_b\), and the corresponding tune shifts satisfy \(\Delta Q_x=\Delta Q_y\). Consequently, the footprint collapses onto a one-dimensional diagonal line. This result describes the tune spread in the projected \((Q_x,Q_y)\) coordinates of an uncoupled lattice; the tune spread in the eigenmode coordinates of a coupled lattice is considered next.

To compute the coupled lattice tune spread (see Appendix~\ref{sec:appendixD}), we follow the method described in~\cite{sen2023density}. The Hamiltonian perturbed with the space charge potential, similar to Eq.~\eqref{eq:perturbedhamiltonianunc}, is
\begin{equation}
    \mathcal{H} = Q_{1}J_{1} + Q_{2}J_{2} + V_{sc}(J_{1},\Psi_{1},J_{2},\Psi_{2}).
\end{equation}
Here, $J_{1}$ and $J_{2}$ are the eigenmode actions, $Q_{1}$ and $Q_{2}$ are the eigenmode tunes. The amplitude-dependent tune shift is computed analogously to Eq.~\eqref{eq:tunespreaddifferential}; however, in this case it is expressed in terms of eigenmodes rather than the uncoupled planar variables. The coordinates are parametrized using Lebedev-Bogacz parametrization~\cite{lebedev2010betatron}. Introducing particle amplitudes $a_{j}=\sqrt{2J_{j}/\epsilon_{j}}$ for $j\in\left\{1,2 \right\}$, the coordinates are
\begin{equation}
    \begin{split}
        x&= a_{1}\sigma_{1x}\cos\psi_{1} + a_{2}\sigma_{2x}\cos(\psi_{2}-\nu_{2}), \\
        y&= a_{1}\sigma_{1y}\cos(\psi_{1}-\nu_{1}) + a_{2}\sigma_{2y}\cos\psi_{2}.
    \end{split}
\end{equation}
Here, $\sigma_{ij}=\sqrt{\epsilon_{i}\beta_{ij}}$ is rms mode beam sizes for $i\in \{1,2 \}$ and $j\in\{x,y\}$, and $\psi_{1,2}$ are betatron phases. For a round, long bunch the tune spread yields
\begin{equation}
	\begin{split}
		\Delta Q_{1}&= -C_{sc}\int_{0}^{1}dt\frac{\beta_{1}}{\sigma_{b}^{2}}\left(I_{0}(2\tilde{a}_{1}\tilde{a}_{2}t) - \frac{\tilde{a}_{2}}{\tilde{a}_{1}}I_{1}(2t\tilde{a}_{1}\tilde{a}_{2}) \right)I_{0}(\frac{a_{z}^{2}t}{2})\exp\left(-\frac{t}{2}(\tilde{a}_{1}^{2} + \tilde{a}_{2}^{2}+2a_{z}^{2})\right), \\
		\Delta Q_{2}&= -C_{sc}\int_{0}^{1}dt\frac{\beta_{2}}{\sigma_{b}^{2}}\left(I_{0}(2t\tilde{a}_{1}\tilde{a}_{2}) - \frac{\tilde{a}_{1}}{\tilde{a}_{2}}I_{1}(2t\tilde{a}_{1}\tilde{a}_{2}) \right)I_{0}(\frac{a_{z}^{2}t}{2})\exp\left(-\frac{t}{2}(\tilde{a}_{1}^{2} + \tilde{a}_{2}^{2}+2a_{z}^{2})\right).
	\end{split}
    \label{eq:circmodetunespreadcoupled}
\end{equation}
Here, $\tilde{a}_{1}=a_{1}\sigma_{1}/\sigma_{b}$ and $\tilde{a}_{2}=a_{2}\sigma_{2}/\sigma_{b}$, as defined in Appendix~\ref{sec:appendixD}. Figure~\ref{fig:coupledspreadcircmode} compares the resulting footprints in the eigenmode tune plane for an uncorrelated round beam with equal emittances and for a circular-mode beam. In the equal-emittance case, the two eigenmode amplitudes contribute comparably, producing the conventional necktie-shaped footprint. For the circular-mode beam, the large intrinsic flatness ratio makes mode~1 dominant and suppresses the contribution associated with mode~2. The resulting amplitude dependence is therefore asymmetric, and the footprint occupies only the corresponding one-sided portion of the necktie structure. The purpose of Fig.~\ref{fig:coupledspreadcircmode} is to identify this characteristic change in topology rather than to predict the exact footprint boundary for a particular accelerator.
\begin{figure}[tbp]
	\centering
	\includegraphics[width=0.49\linewidth]{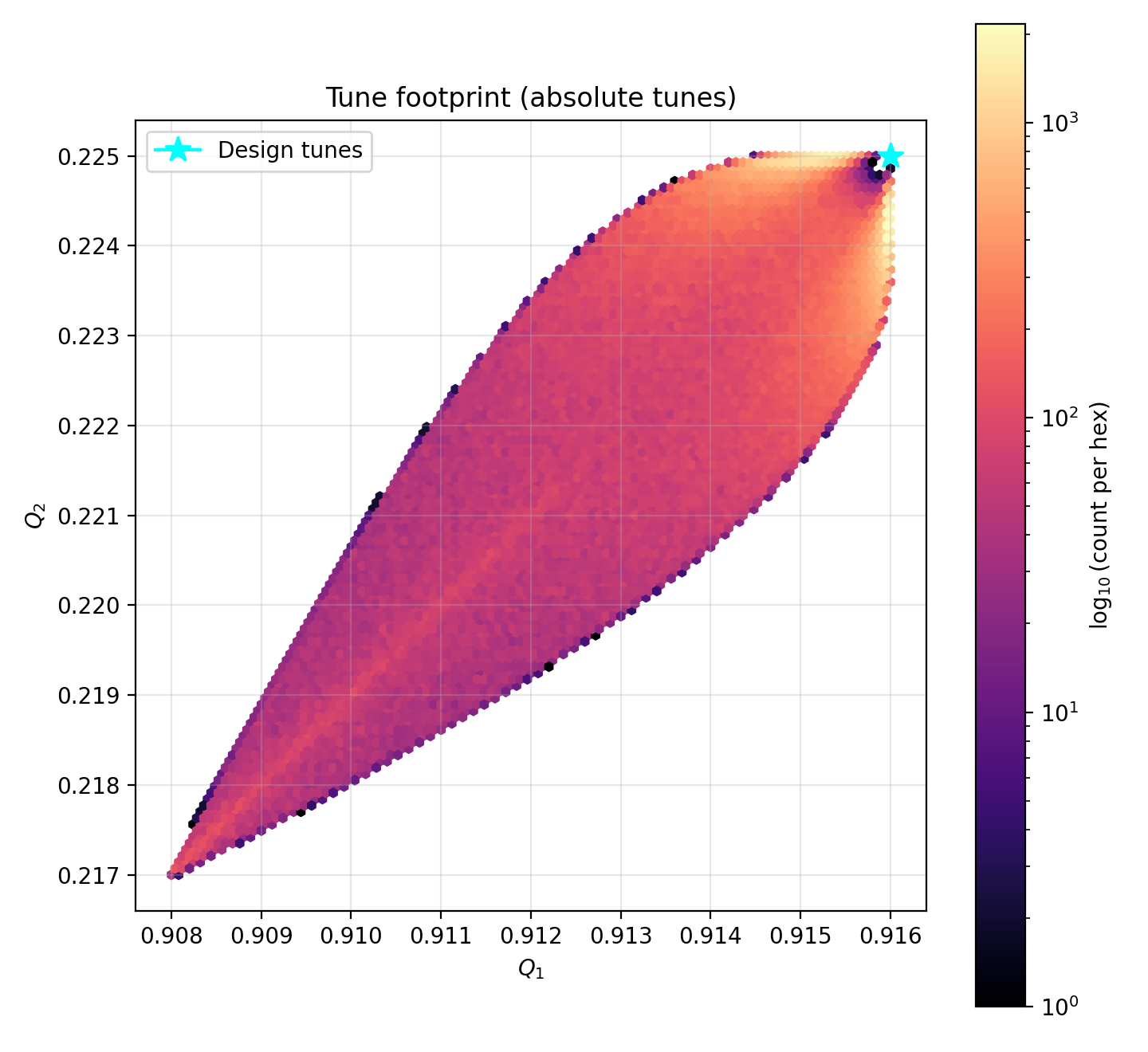}
	\includegraphics[width=0.49\linewidth]{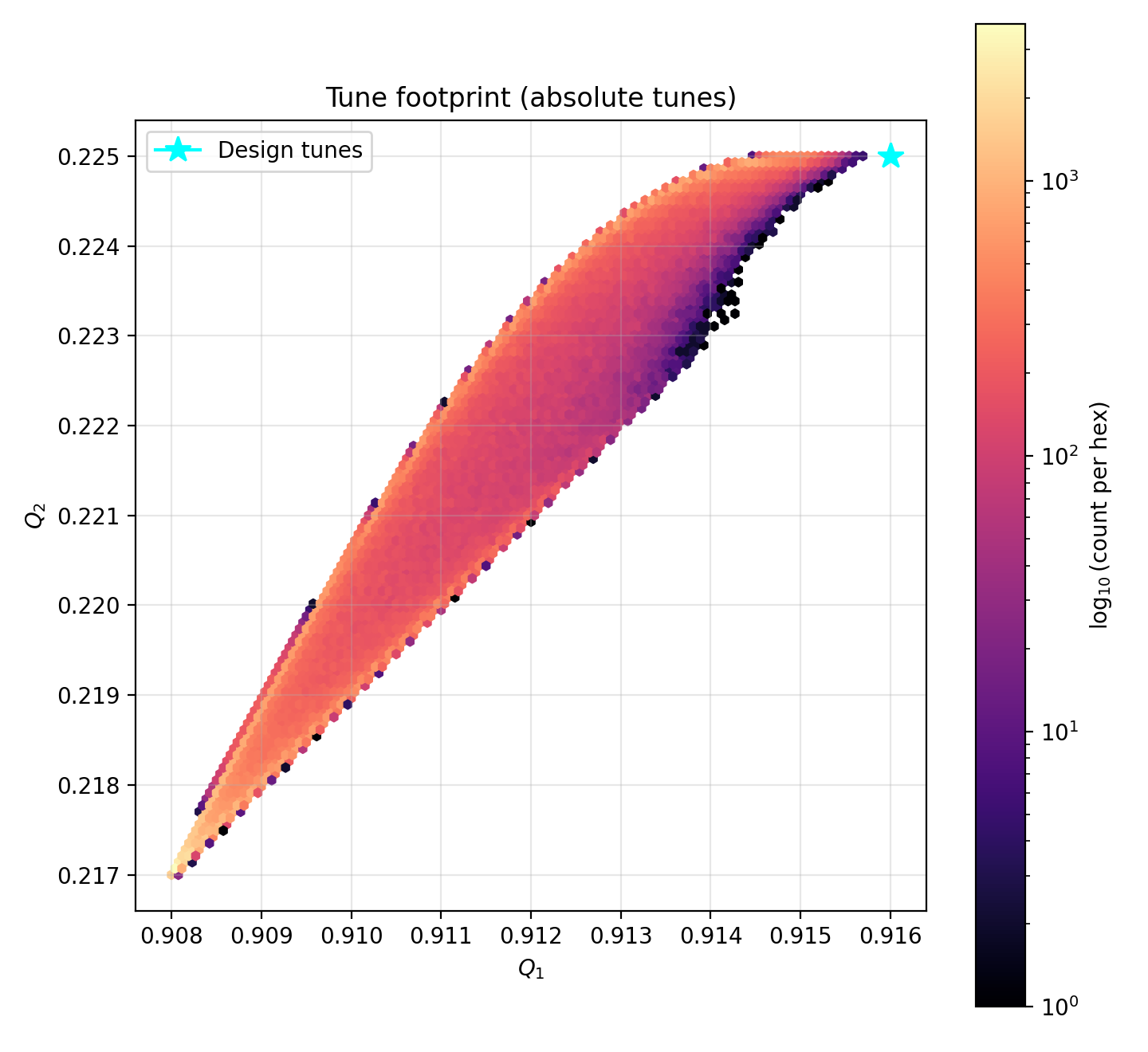}
	\caption{Tune-spread footprints for a round beam with equal emittances (left) and a circular-mode beam (right).}
	\label{fig:coupledspreadcircmode}
\end{figure}

For the matched examples and amplitude ranges considered here, the space-charge-induced tune footprint of the circular-mode beam is reduced relative to the corresponding uncorrelated round-beam case. A smaller tune spread improves resonance control by reducing overlap with low-order resonance lines.

\section{Transport Lines and Rings}\label{sec:Results}
In this section, we demonstrate conservation of angular momentum in the rotationally invariant cell designs introduced in Sec.~\ref{subsec:latticebuildingblocks} using particle-in-cell (PIC) simulations. We employ two PIC codes, TRACK~\cite{TRACKweb} and ImpactX~\cite{huebl2022next}, and study space-charge effects during acceleration while maintaining circular modes. The eigenmode emittances are extracted from the 4D covariance matrix $\Sigma$ via
\begin{equation}
\begin{split}
\epsilon_{1} &= \frac{1}{2}\sqrt{-\mathrm{Tr}\!\left[(\Sigma \mathbf{S})^{2}\right] + \sqrt{\mathrm{Tr}\!\left[(\Sigma \mathbf{S})^{2}\right]^{2} -16\,\epsilon_{4D}^{2}}}, \\
\epsilon_{2} &= \frac{1}{2}\sqrt{-\mathrm{Tr}\!\left[(\Sigma \mathbf{S})^{2}\right] - \sqrt{\mathrm{Tr}\!\left[(\Sigma \mathbf{S})^{2}\right]^{2} -16\,\epsilon_{4D}^{2}}}.
\end{split}
\end{equation}
Here, $\mathrm{Tr}$ is the trace of a matrix, and $\mathbf{S}$ is $4\times 4$ symplectic unit matrix defined in Eq.~\eqref{eq:symplecticunitmatrix4b4}. In the case of acceleration, the eigenmode emittances shown in the figures are normalized eigenmode emittances, $\epsilon_{1,2,N}=\beta\gamma\epsilon_{1,2}$, and normalized angular momentum, $L_{zN}=\beta\gamma L_{z}$ (see Appendix~\ref{sec:appendixC}). 

All ring simulations use six-dimensional particle tracking with full three-dimensional PIC space charge. Acceleration and longitudinal focusing are provided by five short-RF cavities in each acceleration section, operating at the first harmonic. The RF parameters are matched to the longitudinal beam parameters, and the momentum compaction of the lattice is included. The particles therefore execute synchrotron motion during acceleration, although the longitudinal phase-space evolution is not shown because the present study focuses on transverse circular-mode preservation.

The external quadrupoles, dipoles, and solenoids are represented by ideal linear hard-edge maps evaluated about the reference momentum. Consequently, synchrotron motion does not directly modulate the transverse normal-mode optics through chromaticity in the present model. Its influence enters through the evolving longitudinal bunch density and the corresponding three-dimensional space-charge field. Thus, the effects of synchrotron motion on the transverse rms angular momentum and eigenemittances are included in the reported tracking results.

\subsection{Regular Quadrupole-Based Doublet Transport Channel}
\label{subsec:quadchannels}
\begin{figure}[tbp]
    \centering
    \includegraphics[width=0.328\linewidth]{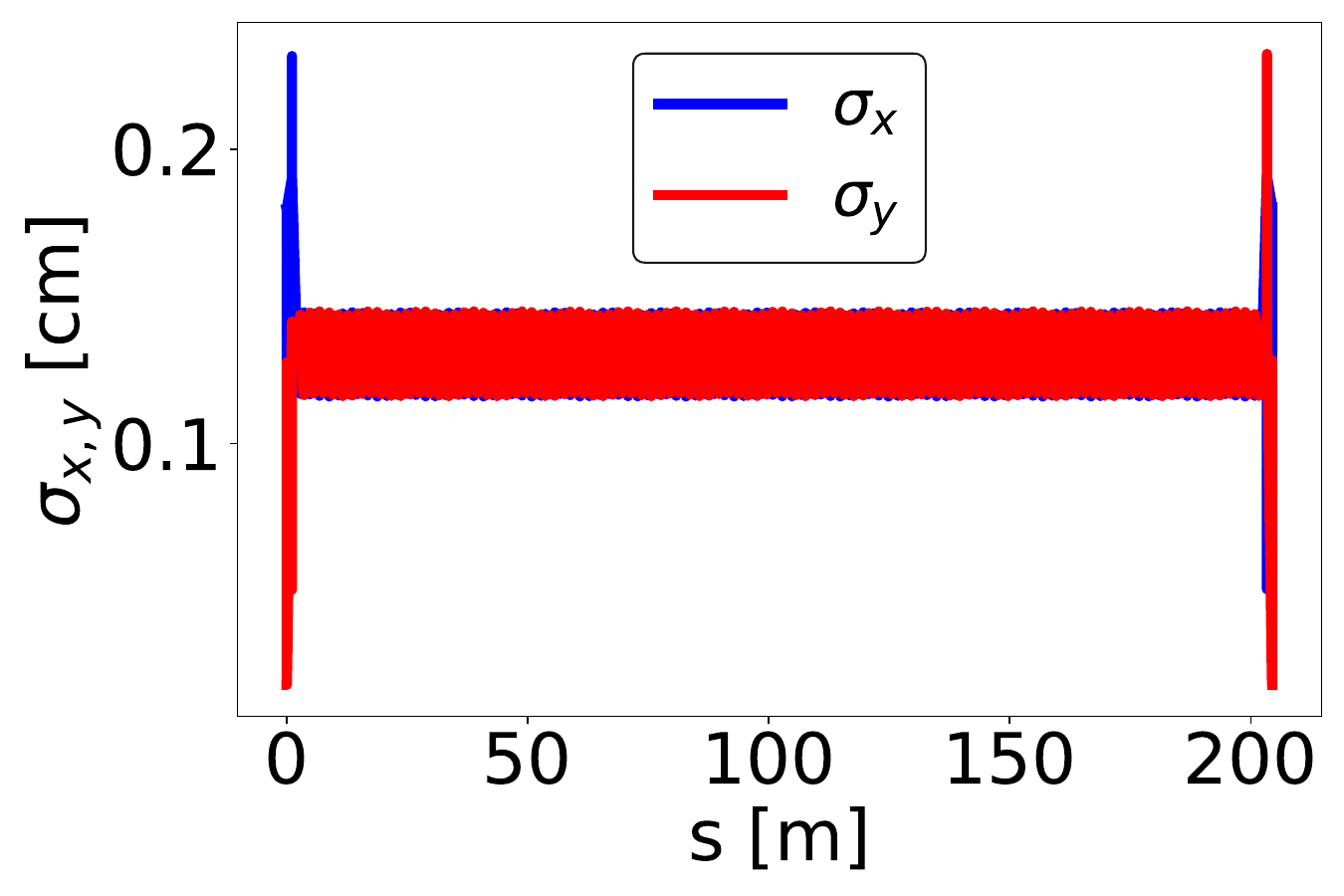}
    \includegraphics[width=0.328\linewidth]{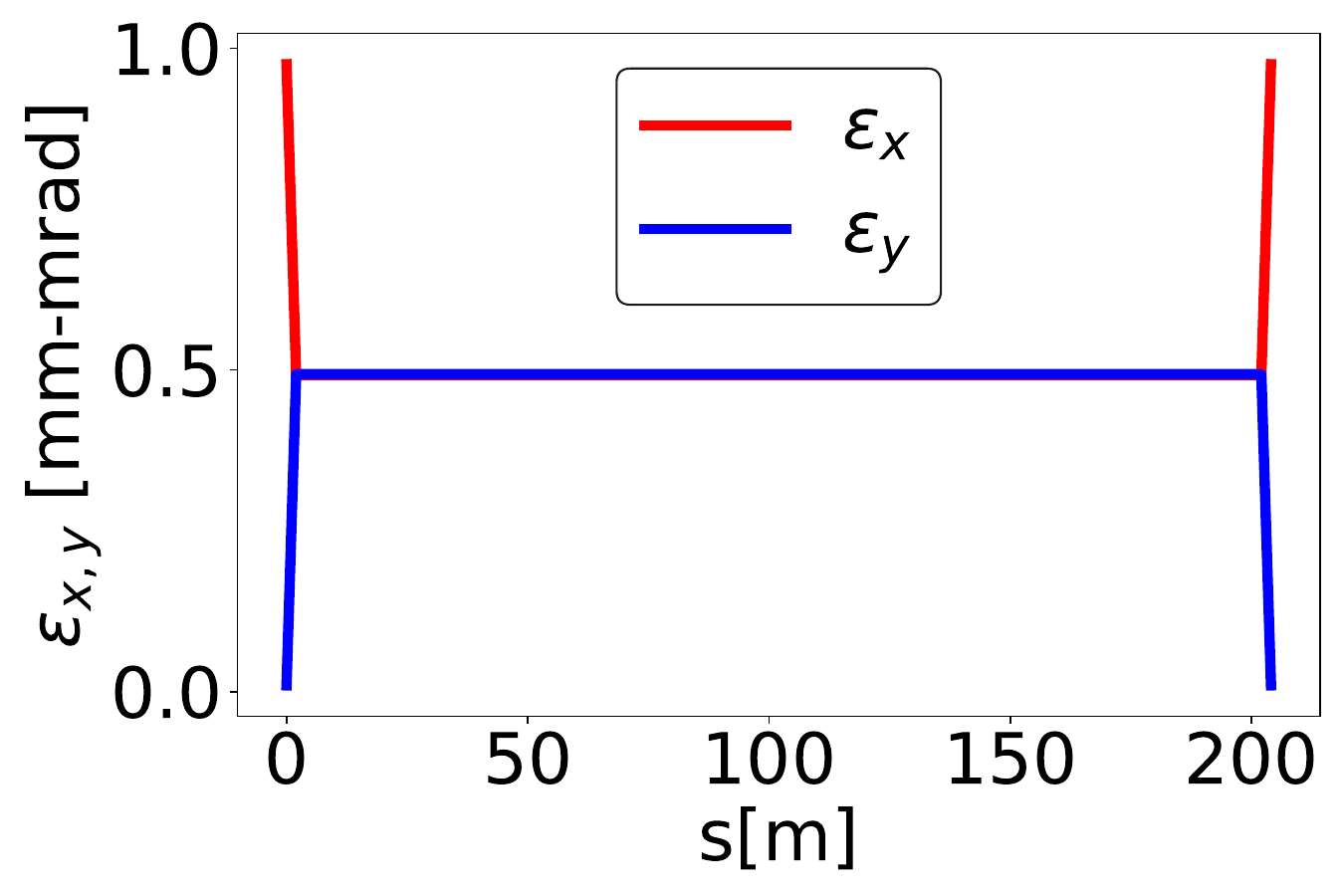}
    \includegraphics[width=0.328\linewidth]{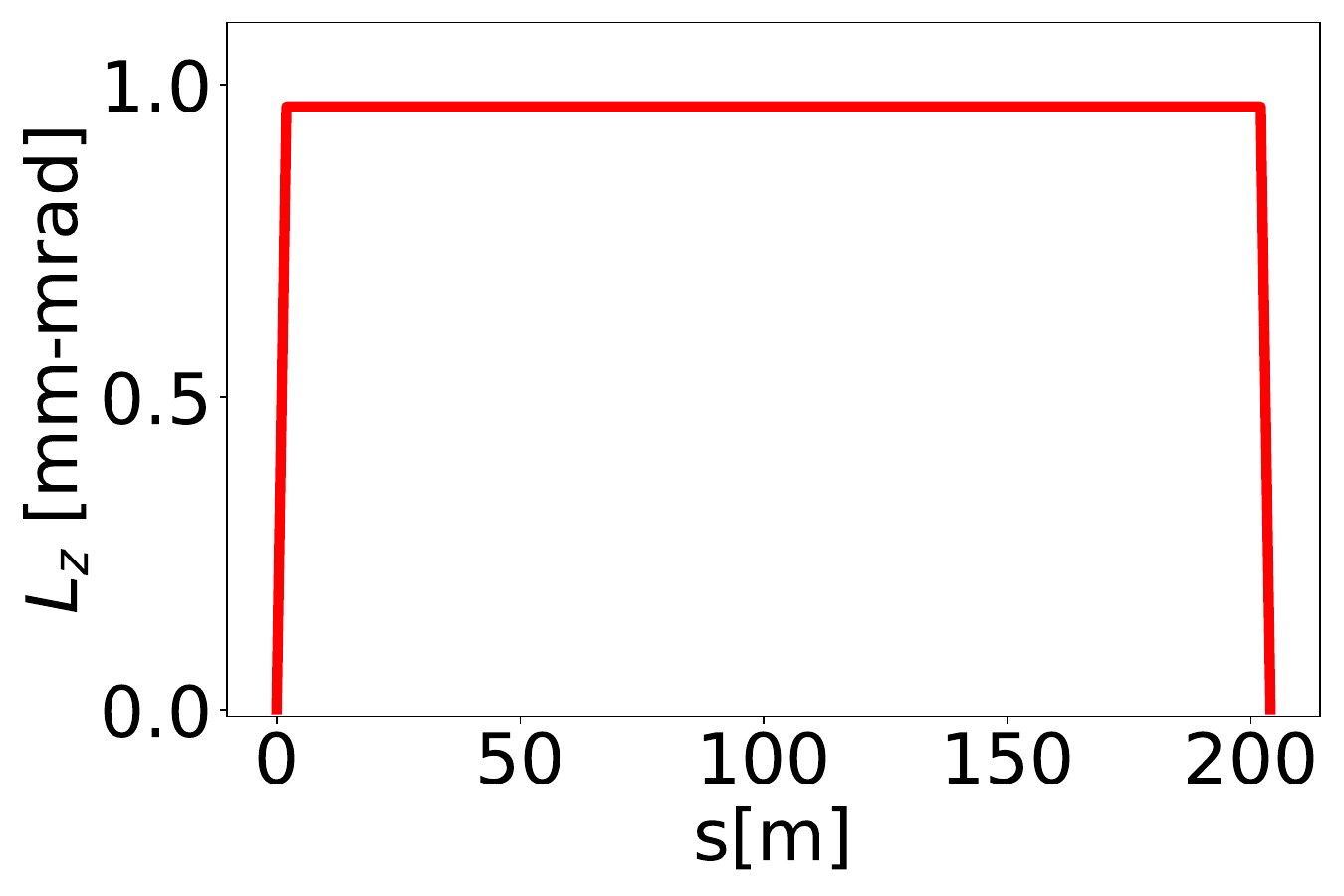}
    \caption{Tracking results of a FODO transport without space charge. Derbenev's adapters are placed at the beginning and at the end of the transport. The rms envelopes are shown in the left plot. The rms apparent emittances, $\epsilon_{x}$ and $\epsilon_{y}$, shown in the middle plot. The rms angular momentum is shown in the right plot.}
    \label{fig:angmomconservation}
\end{figure}

In this section, we demonstrate simple FODO-based transport with circular-mode beams. The FODO structure of the periodic cell is given in the top-left plot in Fig.~\ref{fig:rotinvcells}. The results of the simulations in the transport channel without space charge are shown in Fig.~\ref{fig:angmomconservation}. Derbenev's adapters are added to the transport, initially transforming a flat proton beam into a circular-mode beam and, at the end, transforming a circular-mode beam into a flat beam. The conversion from a flat beam to a round beam is evident from the rms apparent emittances, $\epsilon_{x}$ and $\epsilon_{y}$, where $\epsilon_{y}$ was smaller than $\epsilon_{x}$; after transformation, they become equal. Figure~\ref{fig:angmomconservation} illustrates that rotationally invariant cell designs conserve angular momentum, hence the circular-mode beam.

\begin{figure}[tbp]
    \centering
    \includegraphics[width=0.49\linewidth]{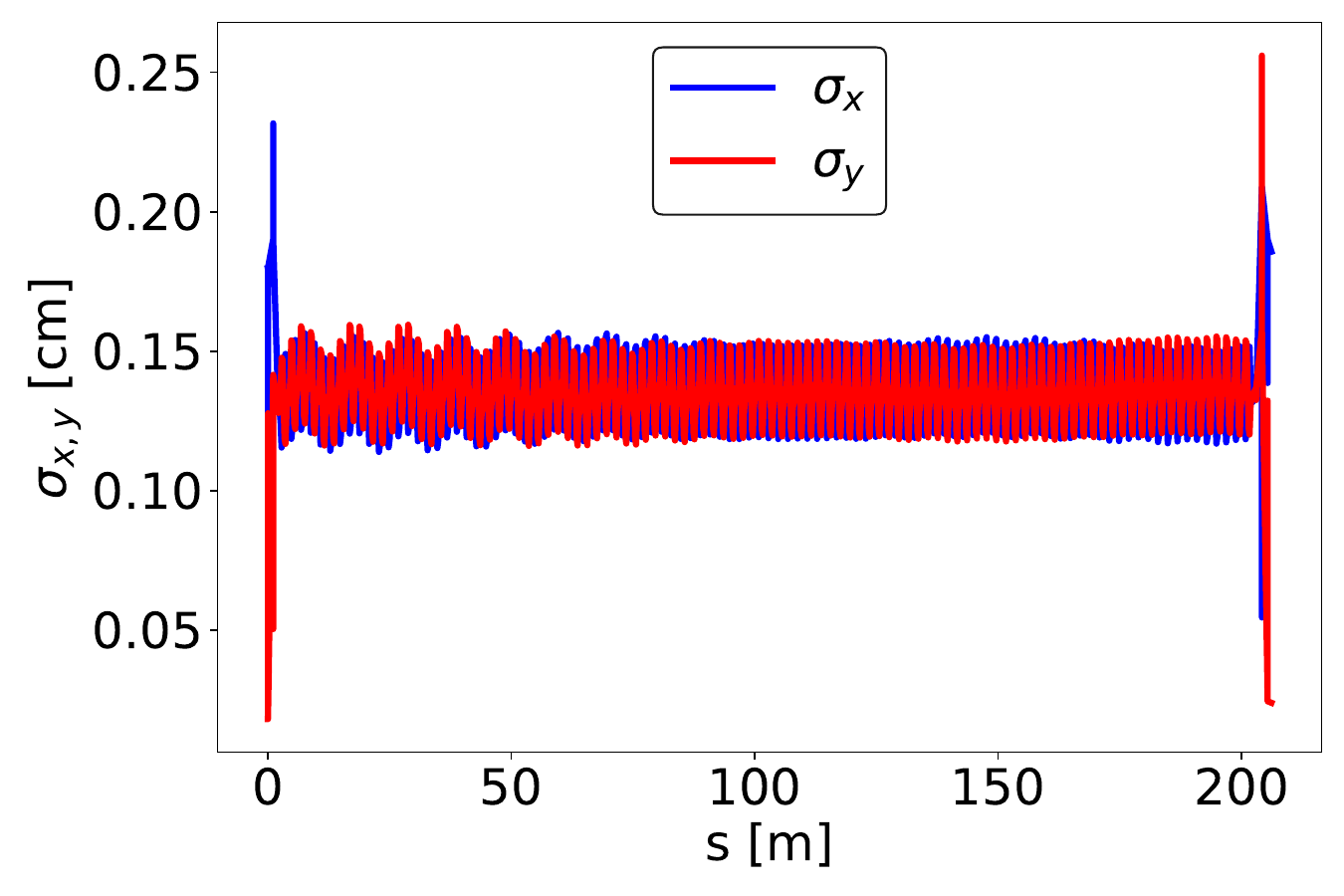}
    \includegraphics[width=0.49\linewidth]{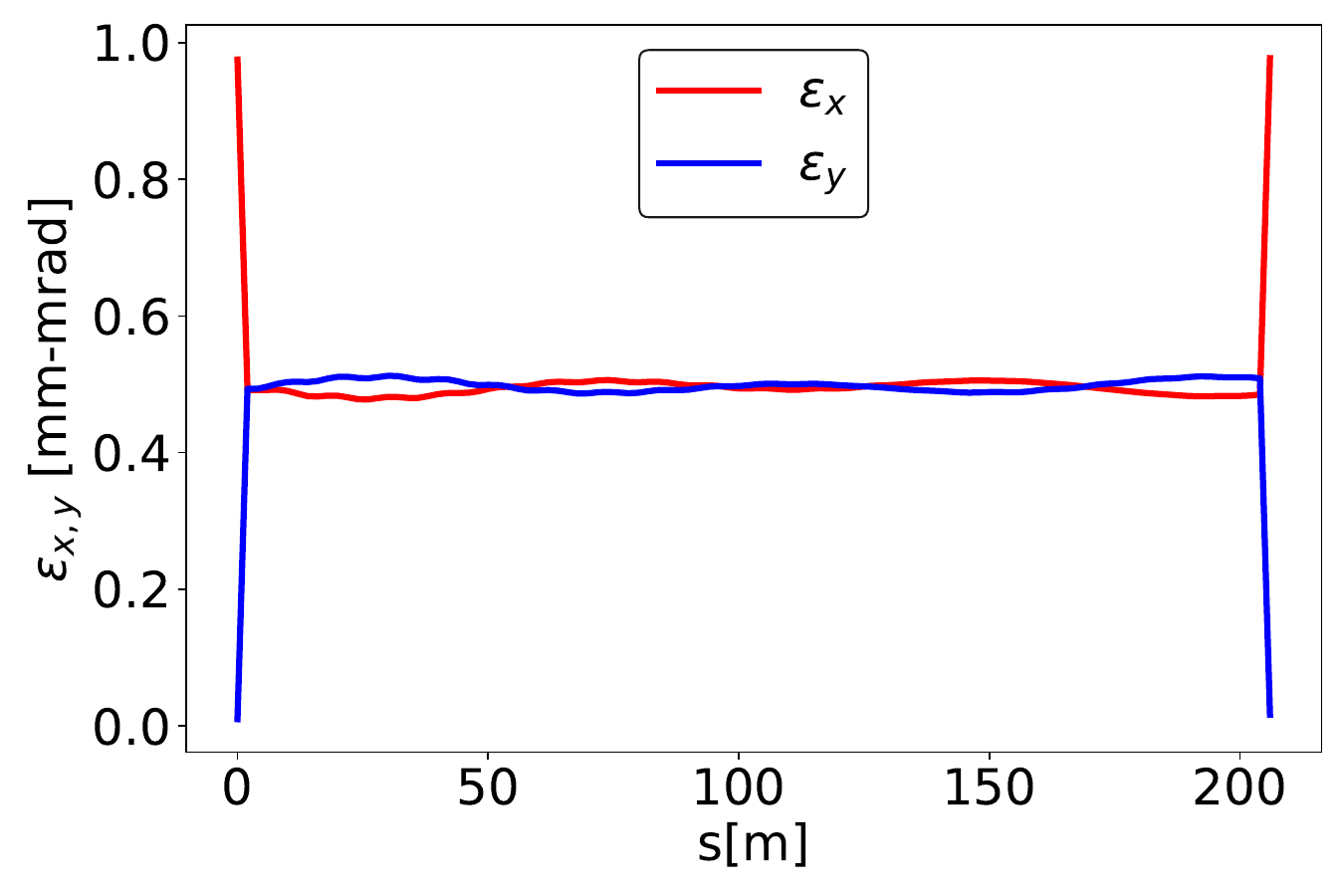}
    \includegraphics[width=0.95\linewidth]{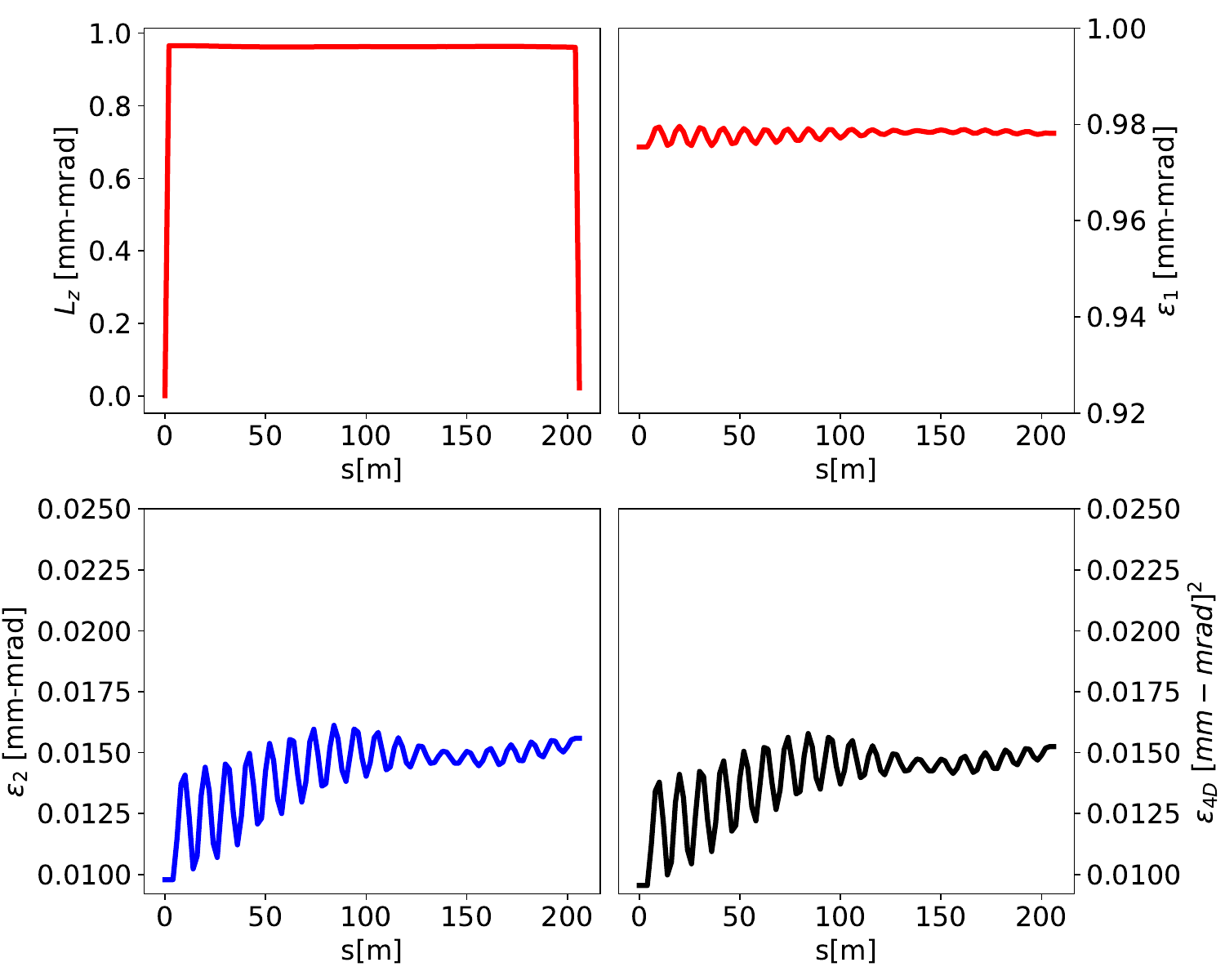}
    \caption{Tracking results of a FODO transport with space charge. Proton beam with $E_{\mathrm{kin}}=200.0$\,MeV and current $I=500.0$\,mA tracked with TRACK simulation. The rms beam sizes are shown in top left figure. The rms apparent emittances, $\epsilon_{x}$ and $\epsilon_{y}$, are given in top right figure. The rms angular momentum (middle-left), rms eigenmode~1 emittance (middle-right), rms eigenmode~2 emittance (bottom-left), and rms 4D eigenmode emittance (bottom-right) are shown.}
    \label{fig:quadchannelwithspacecharge}
\end{figure}

\begin{figure}[tbp]
    \centering
    \includegraphics[width=0.49\linewidth]{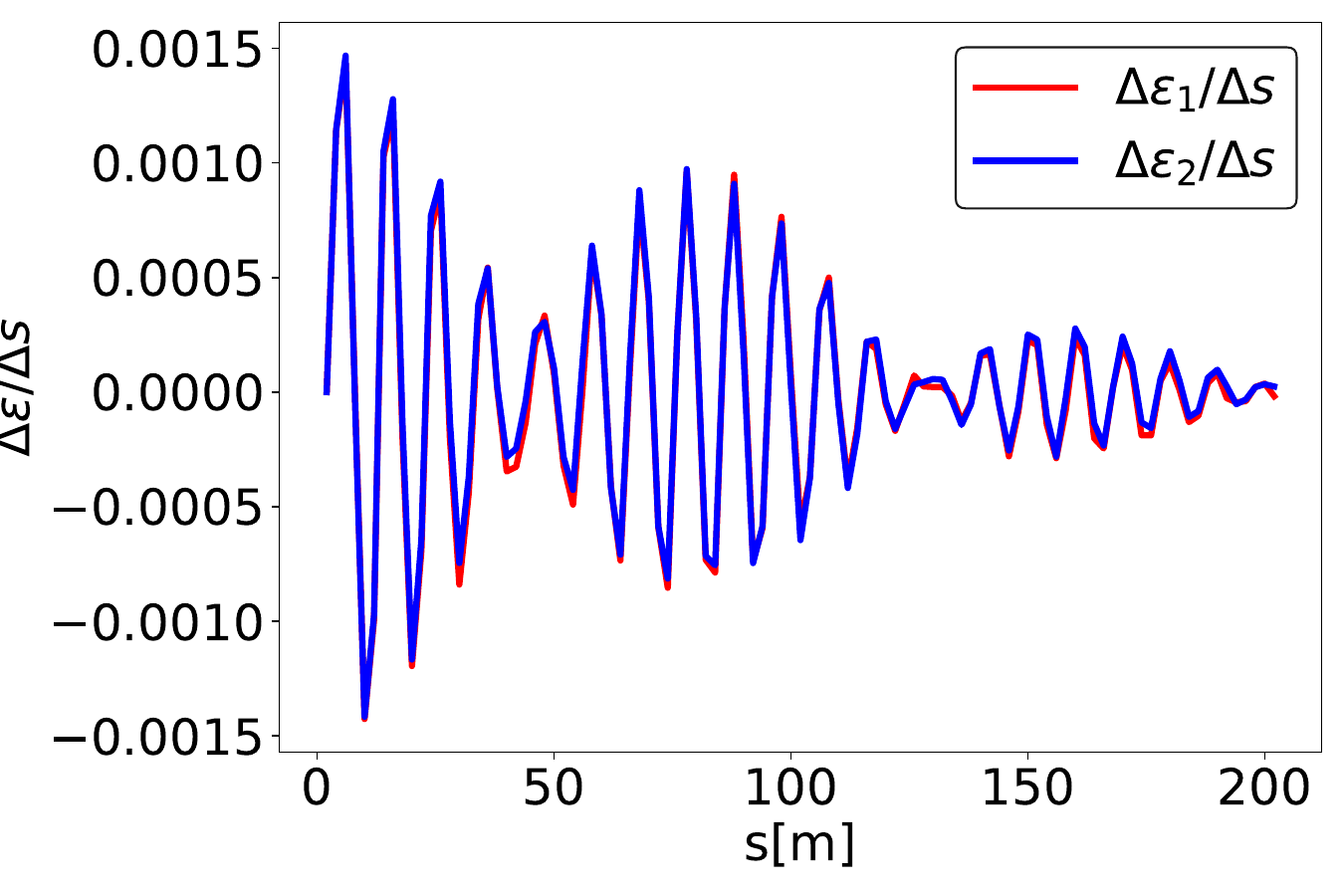}
    \includegraphics[width=0.49\linewidth]{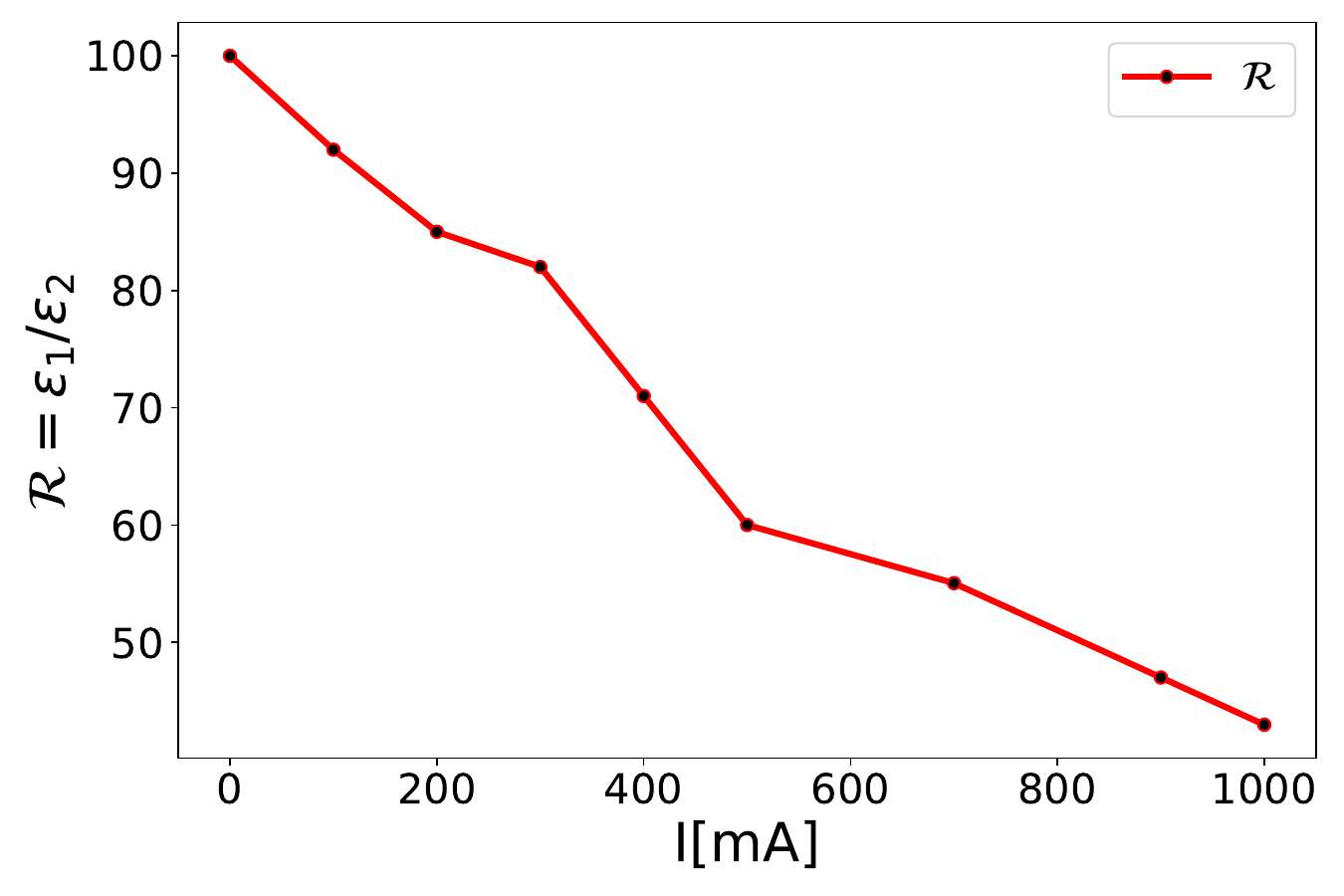}
    \caption{Left plot: relative rms eigenmode emittance growth rates in the FODO-based transport channel given in Fig.~\ref{fig:quadchannelwithspacecharge}. Right plot: intrinsic flatness ratios recorded at the end of the channel as a function of the beam current.}
    \label{fig:eigenmodegrowthratesandratio}
\end{figure}

After confirming the conservation of the circular mode through the FODO-based transport in Fig.~\ref{fig:angmomconservation}, we track a 200-MeV proton beam with 500-mA current in the same transport. The tracking results are shown in Fig.~\ref{fig:quadchannelwithspacecharge} with computed rms eigenmode emittances and angular momentum. The eigenmode~1 emittance barely increases, and the eigenmode~2 emittance grows the most. The transport is also matched to the space charge using methods described in~\cite{xiao2024periodic}. Since the eigenmode~2 emittance is smaller, it experiences greater space-charge-induced emittance growth compared to the larger eigenmode~1 emittance. The relative eigenmode emittance growth rates are shown in Fig.~\ref{fig:eigenmodegrowthratesandratio}, where both eigenmode~1 and eigenmode~2 rms emittance growth rates are equal to each other for beam current $I=500$\,mA. The dominant mode sets the dynamics with respect to the space-charge effect, leading to equal relative growth rates for both modes even though mode~2 is smaller. The initial flatness ratio for the transport was $\mathcal{R}=100$ and for different beam currents the final ratios are also shown in Fig.~\ref{fig:eigenmodegrowthratesandratio}. For an intense beam with $700$\,mA the final ratio after the transport is $55.0$. 

\subsection{Solenoid-Based Booster Ring Design}\label{subsec:solenoidbooster}
\begin{figure}[tbp]
    \centering
    \includegraphics[width=0.49\linewidth]{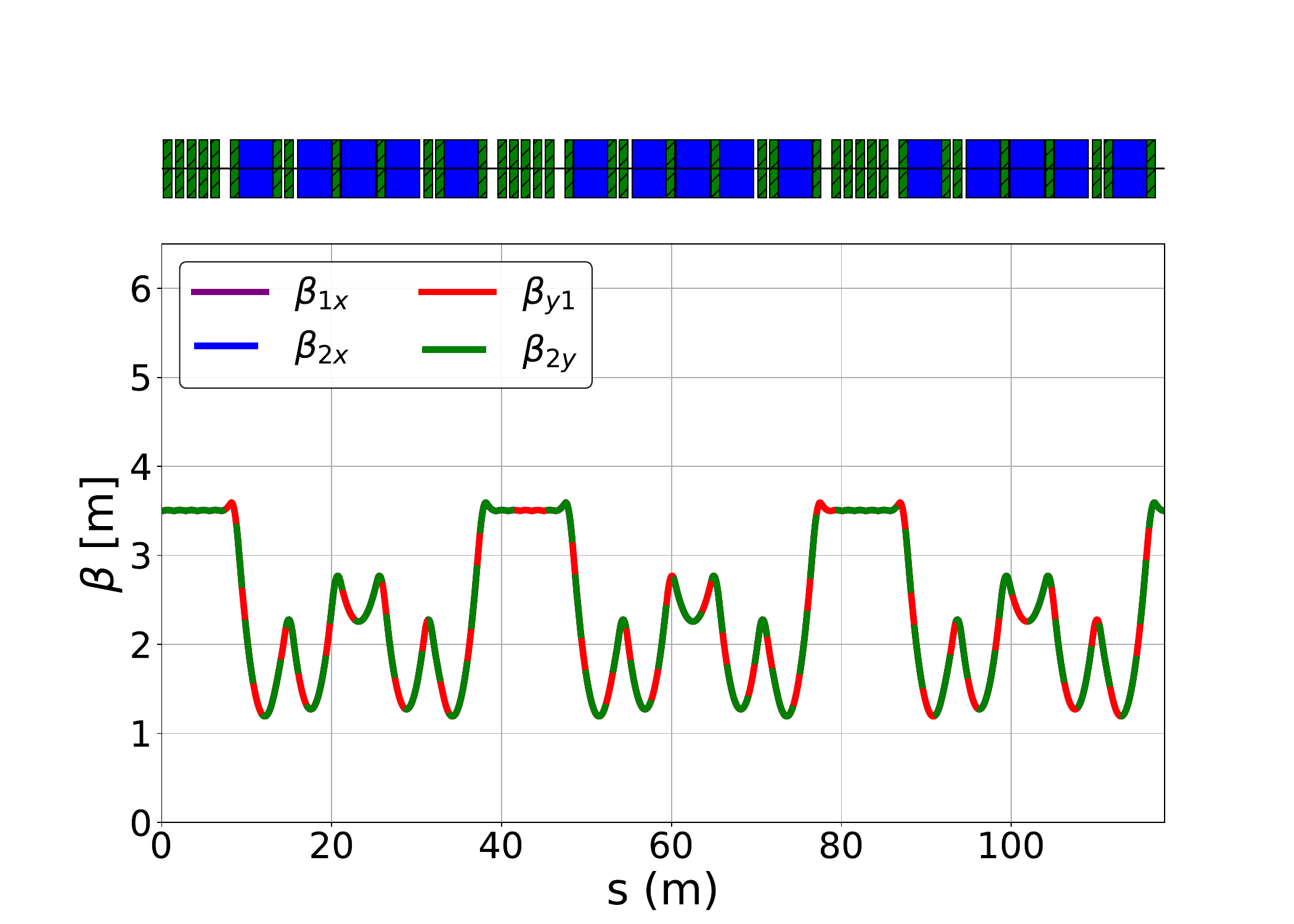}
    \includegraphics[width=0.49\linewidth]{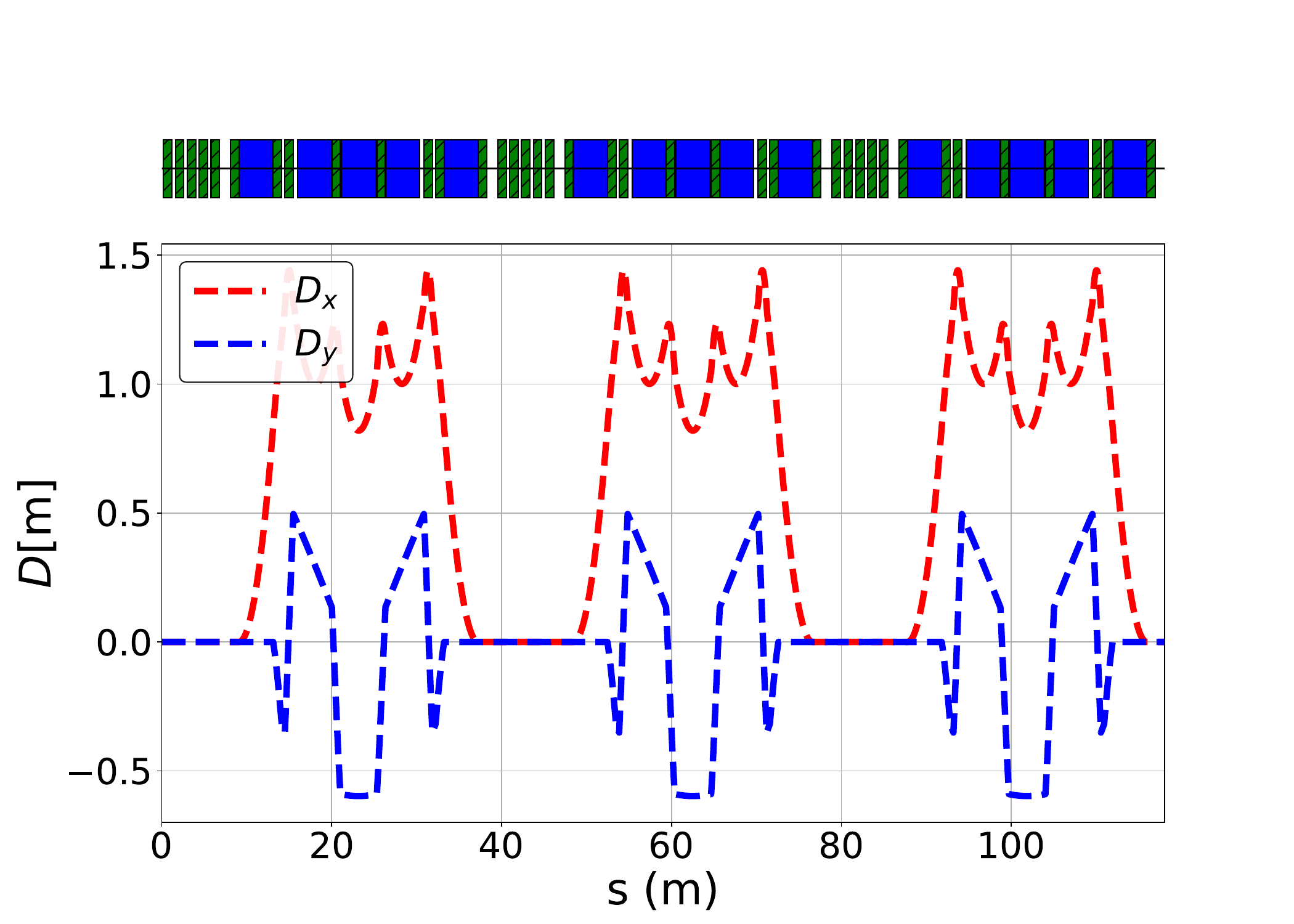}
    \includegraphics[width=0.49\linewidth]{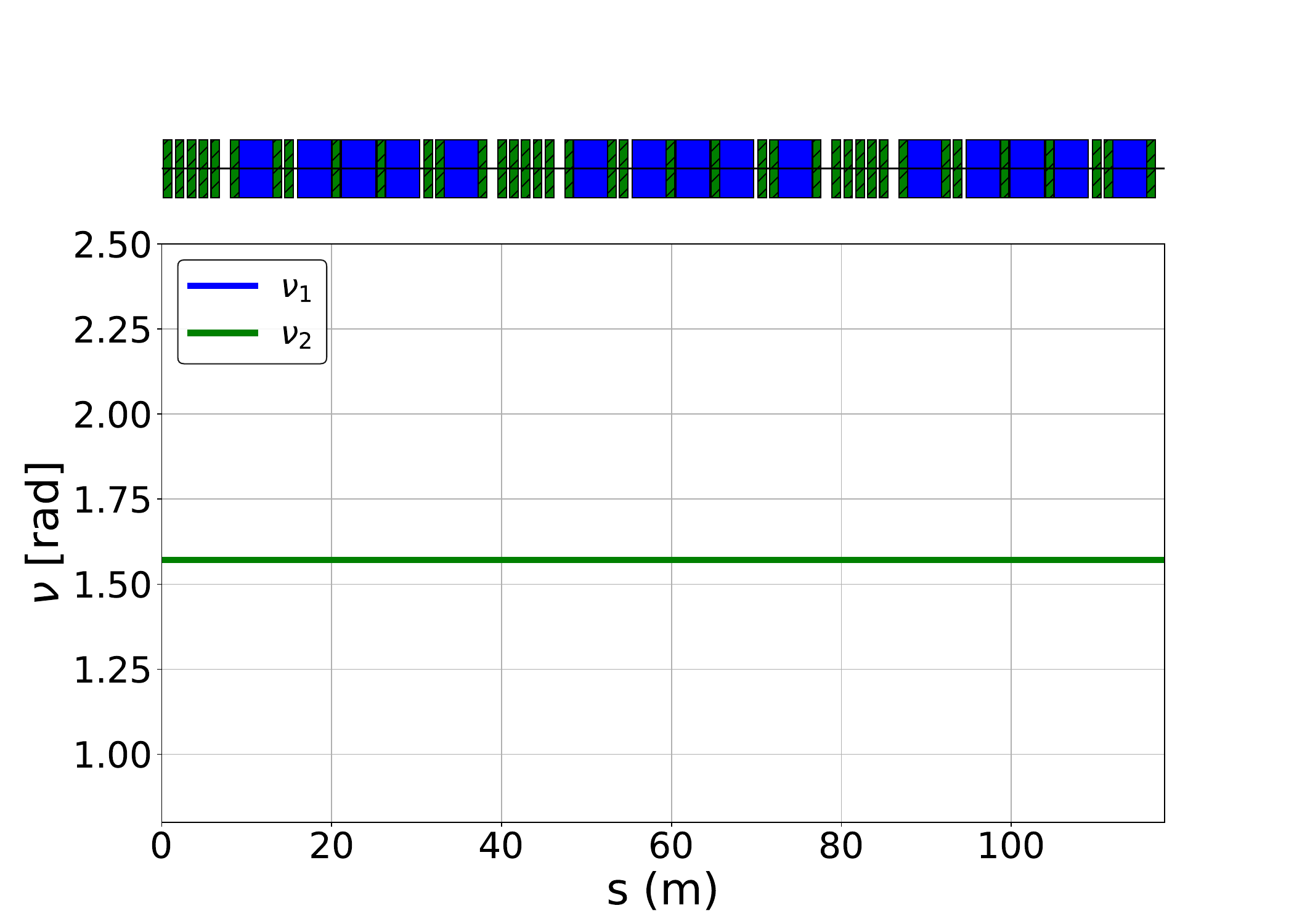}
    \includegraphics[width=0.49\linewidth]{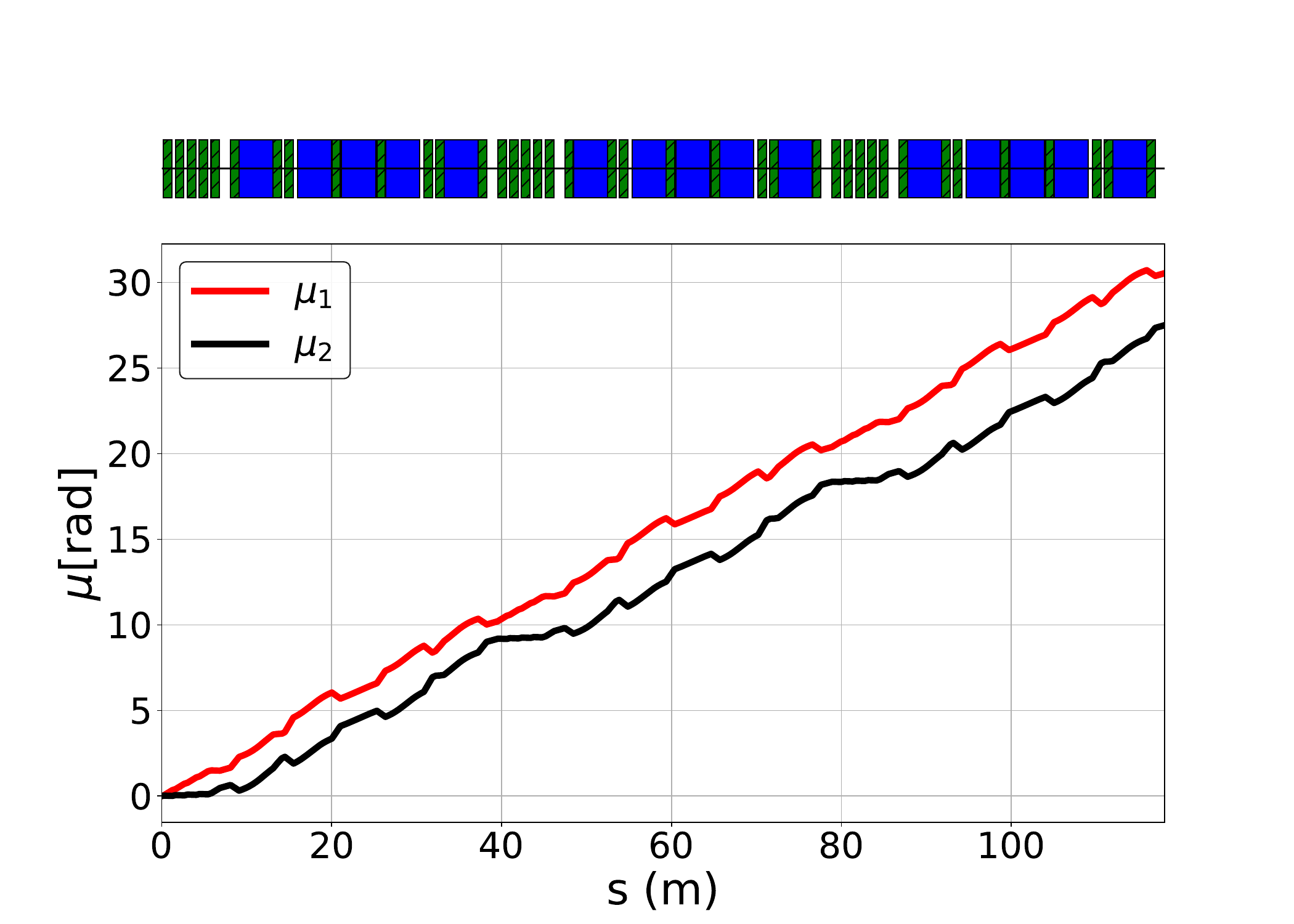}
    \caption{Solenoid indexed dipole ring: coupled $\beta$ functions are shown in the top-left plot. Dispersions $(D_x, D_y)$ are shown in the top-right plot. Coupling phases $(\nu_1,\nu_2)$ are shown in the bottom-left plot. Phase advances $(\mu_1,\mu_2)$ are shown in the bottom-right plot.}
    \label{fig:solenoidringoptics}
\end{figure}

\begin{figure}[tbp]
    \centering
    \includegraphics[width=0.49\linewidth]{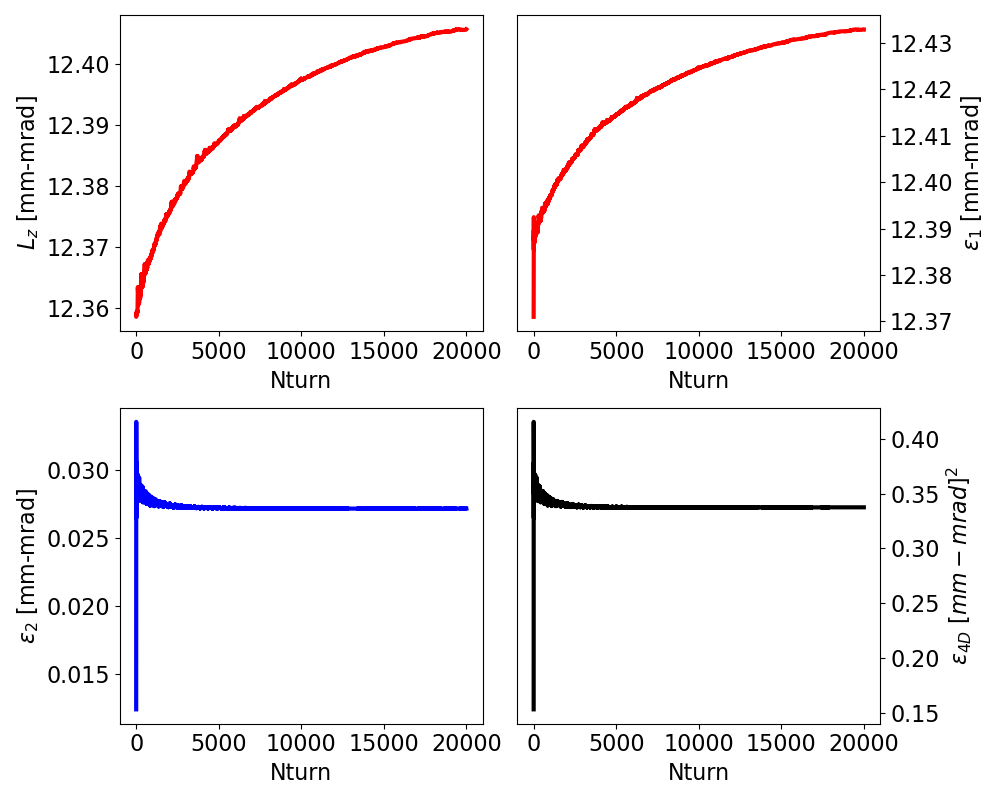}
    \includegraphics[width=0.49\linewidth]{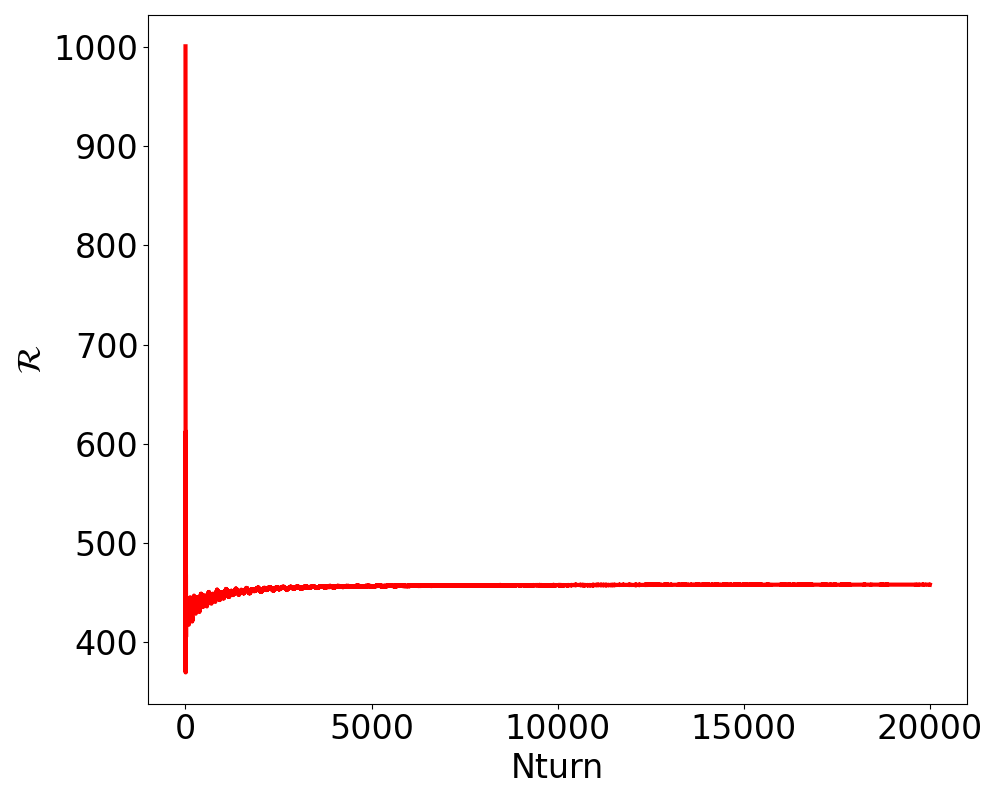}
    \caption{Particle tracking results with $Q_{\mathrm{bunch}}=60\mathrm{\,nC}$. Eigenmode parameters are given in the left plot, where the top-left plot is the normalized angular momentum $L_{ZN}=\beta\gamma L_z$, the top-right plot is the normalized eigenmode~1 emittance, the bottom-left plot is the normalized eigenmode~2 emittance, and the bottom-right plot is the normalized $\epsilon_{4D}$ emittance. The right plot is the eigenmode emittance ratio $\mathcal{R}=\epsilon_1/\epsilon_2$. ImpactX code is used for tracking with 20000 turns.}
    \label{fig:trackingresultssolenoid}
\end{figure}

The solenoid focusing structure will enable the highest performance for circular modes, due to equal focusing in both planes, thereby maintaining a coupling phase of $\pi/2$ everywhere in the lattice. To obtain an axisymmetric transfer map, we employed a lattice consisting of solenoids and indexed dipoles. The coupled optics functions of the design are shown in Fig.~\ref{fig:solenoidringoptics}, where all the $\beta$ functions are equal throughout the ring. With equal $\beta$ functions, the coupling phases $\nu_{1,2}$ do not change; hence, the angular momentum is conserved everywhere in the ring. Achromat conditions are used for suppressing dispersion, and straight section solenoids are used for tune splitting. We used a 200.0-MeV proton beam and accelerated it to 3.2\,GeV. The initial $200$-\,MeV proton bunch has an rms bunch length of $\sigma_z=2.65$\,m and an rms energy spread of $\sigma_{\Delta E}=0.579$\,MeV, corresponding to $\sigma_\delta \approx 1.59\times 10^{-3}$. The bunch has $60$\,nC charge which corresponds to eigenmode tune shifts: $\Delta Q_{1,2} = -0.15$, where working points are $Q_{1}=4.86$ and $Q_2=4.37$. The particle tracking results are shown in Fig.~\ref{fig:trackingresultssolenoid}. 

The tracking results in Fig.~\ref{fig:trackingresultssolenoid} show that eigenmode~2 emittance heats up quickly and settles down at a stable value, whereas the eigenmode~1 emittance and angular momentum increase gradually. The dynamic behavior originates from Eq.~\eqref{eq:magnetizedbeameigenmodeemitcomputation}, where angular momentum is the invariant of the system and eigenmode emittances are deduced from the angular momentum. Equation~\eqref{eq:magnetizedbeameigenmodeemitcomputation} is
\begin{equation}
    \begin{split}
        \epsilon_1 &= \frac{L_z}{2} + \sqrt{\frac{L_z^2}{4} + \epsilon_{4D}}\, , \qquad \epsilon_2 = -\frac{L_z}{2} + \sqrt{\frac{L_z^2}{4} + \epsilon_{4D}}\,.
    \end{split}
    \label{eq:eigenmodeemittancesfromangmom}
\end{equation}
The beam is angular-momentum-dominated, $L_z\gg\epsilon_{4D}$, therefore the eigenmode emittances are expressed as
\begin{equation}
    \epsilon_1 = L_z + \frac{\epsilon_{4D}}{L_z}, \qquad \epsilon_2 = \frac{\epsilon_{4D}}{L_z}.
    \label{eq:approximatedeigenmodedependenceonangmom}
\end{equation}
As the space-charge effect is ``turned on'' at the beginning of the simulation, $\epsilon_{4D}$ grows from the self-generated potential energy and reaches its equilibrium value. The solenoid ring exhibits both local and global axial symmetry, thus preserving the circularity of the mode throughout. Since the beam is round everywhere, the self-generated space-charge potential does not violate angular momentum conservation. Hence, most of the emittance growth and loss of the intrinsic flatness ratio arise from the increase in $\epsilon_{4D}$ due to the mismatch. As a result of the increase in $\epsilon_{4D}$, both $\epsilon_1$ and $\epsilon_2$ increase and stabilize quickly, as is evident from Eq.~\eqref{eq:approximatedeigenmodedependenceonangmom}. Due to the nature of the solenoid-based ring design, it is most favorable for hosting circular-mode beams.

\subsection{Quadrupole-Based Focusing Booster Ring Design}\label{subsec:quadbooster}

\begin{figure}[tbp]
    \centering
    \includegraphics[width=0.49\linewidth]{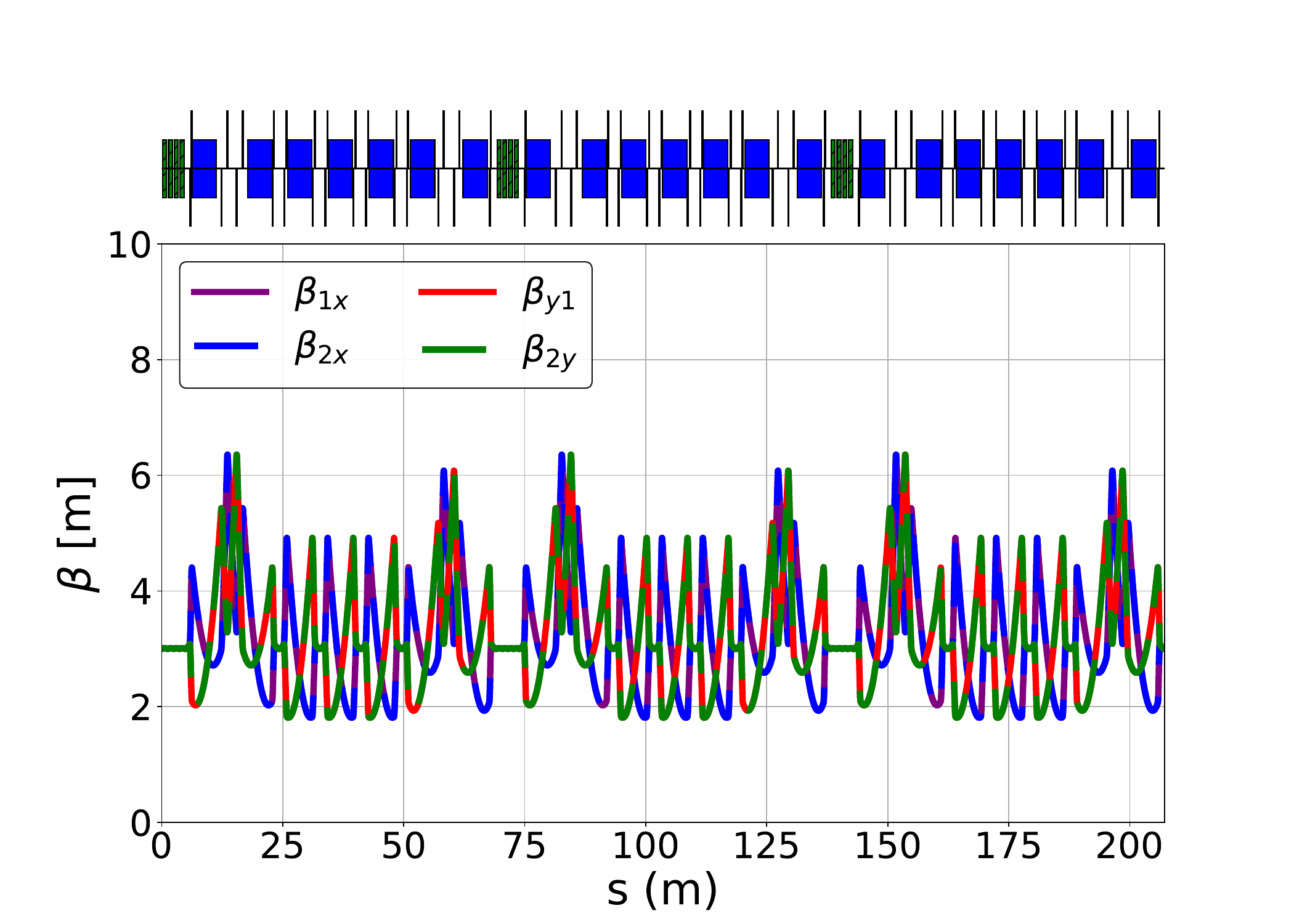}
    \includegraphics[width=0.49\linewidth]{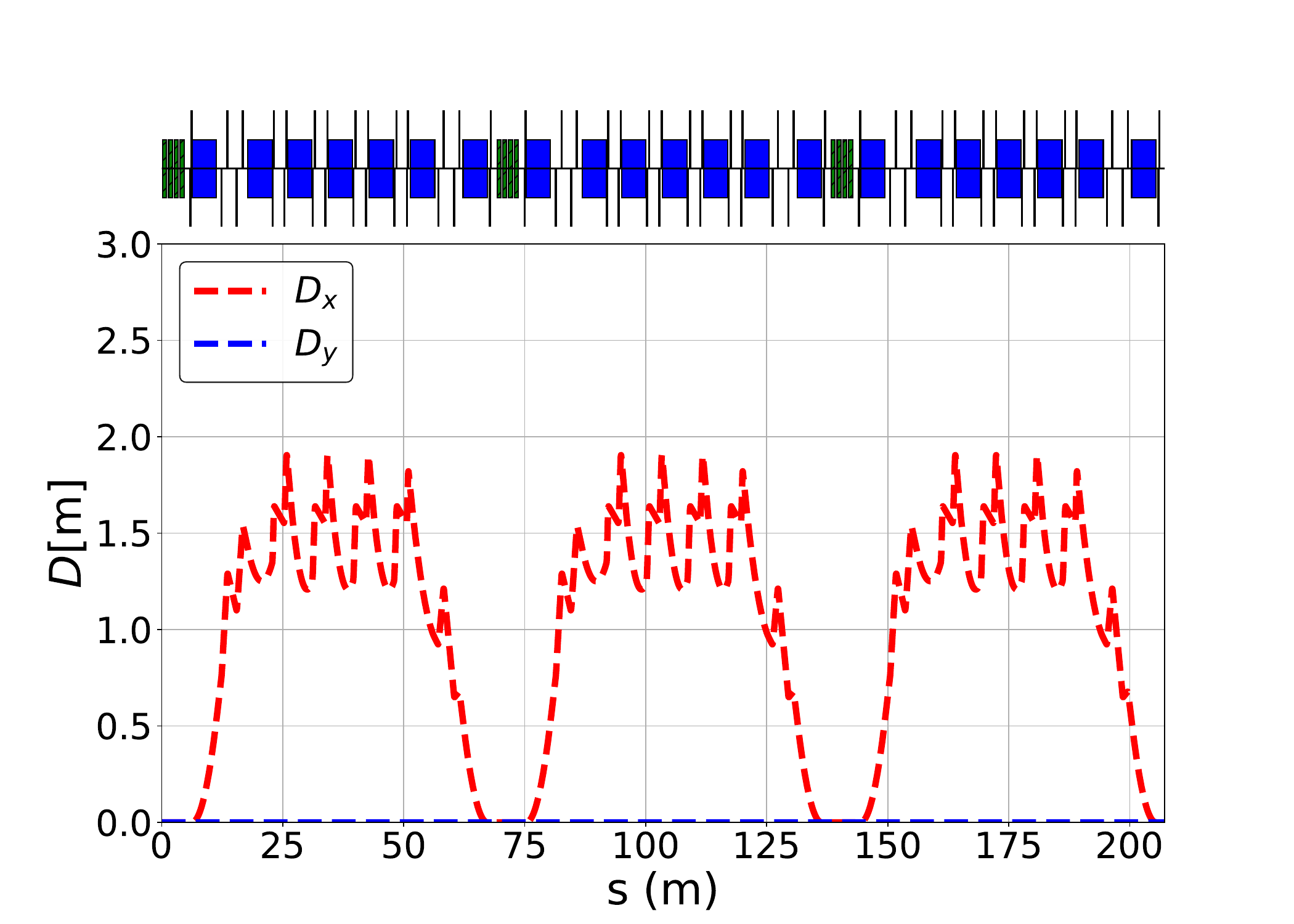}
    \includegraphics[width=0.49\linewidth]{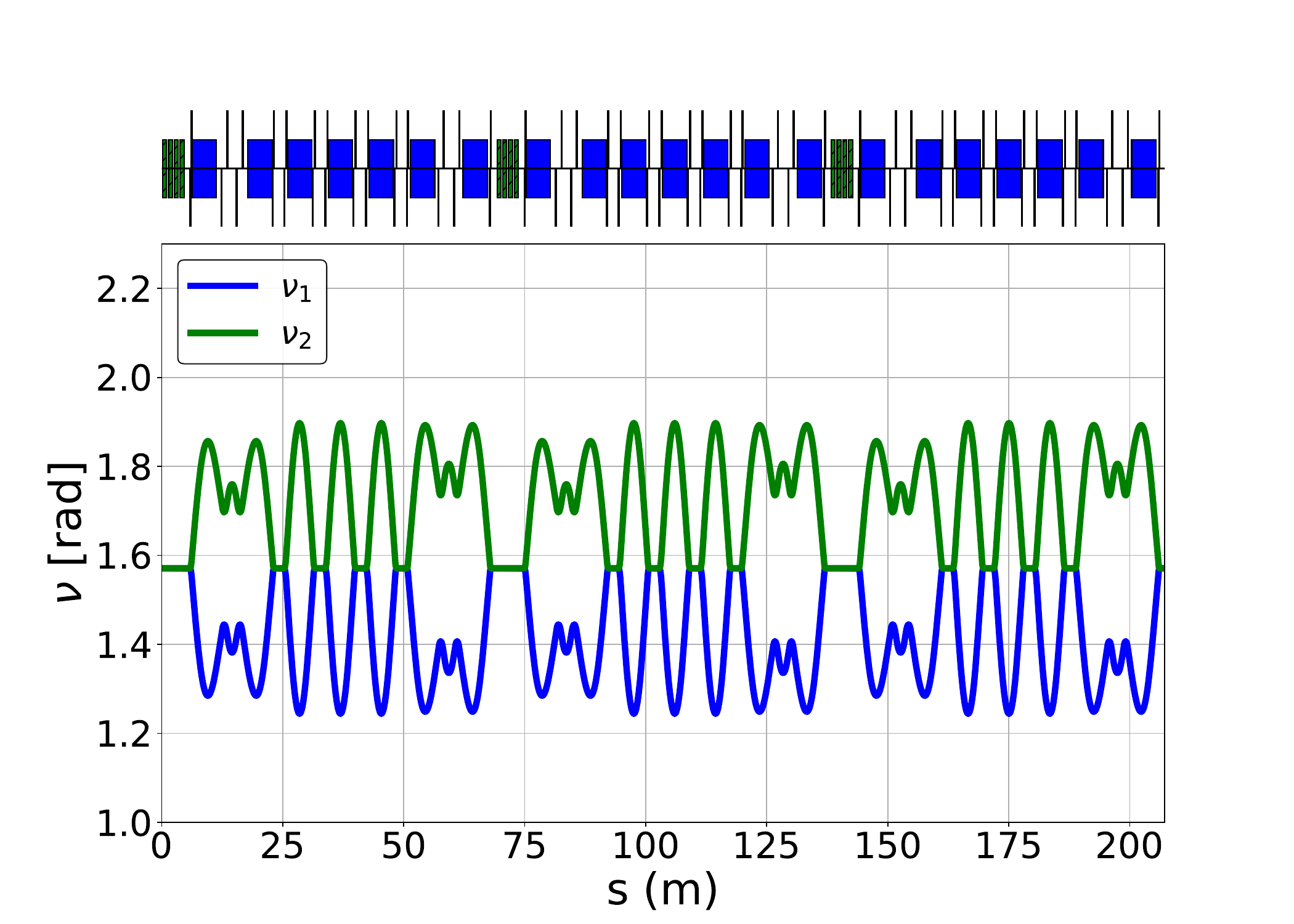}
    \includegraphics[width=0.49\linewidth]{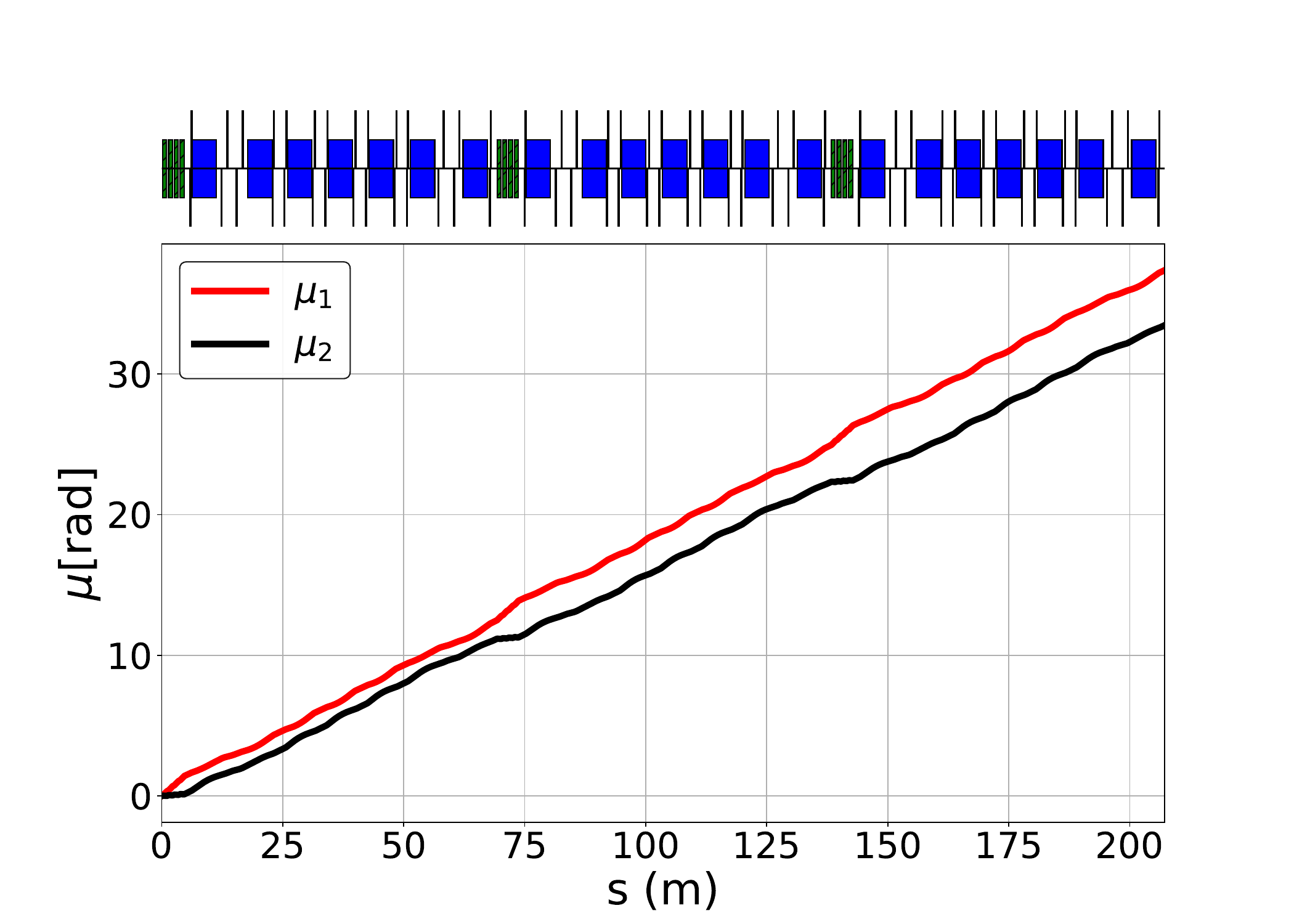}
    \caption{Quadrupole-based booster ring: coupled $\beta$ functions are shown in the top-left plot, dispersions are shown in the top-right plot, the coupling phases are shown in the bottom-left plot, and phase advances are shown in the bottom-right plot.}
    \label{fig:quadboosteroptics}
\end{figure}

\begin{figure}[tbp]
    \centering
    \includegraphics[width=0.49\linewidth]{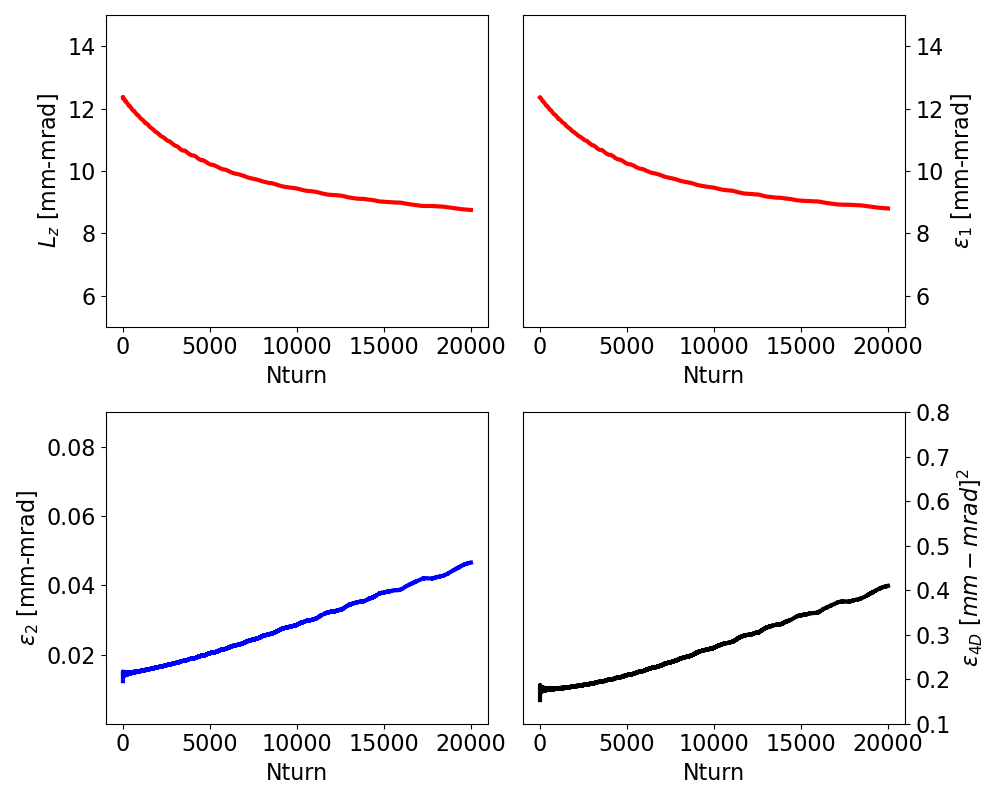}
    \includegraphics[width=0.49\linewidth]{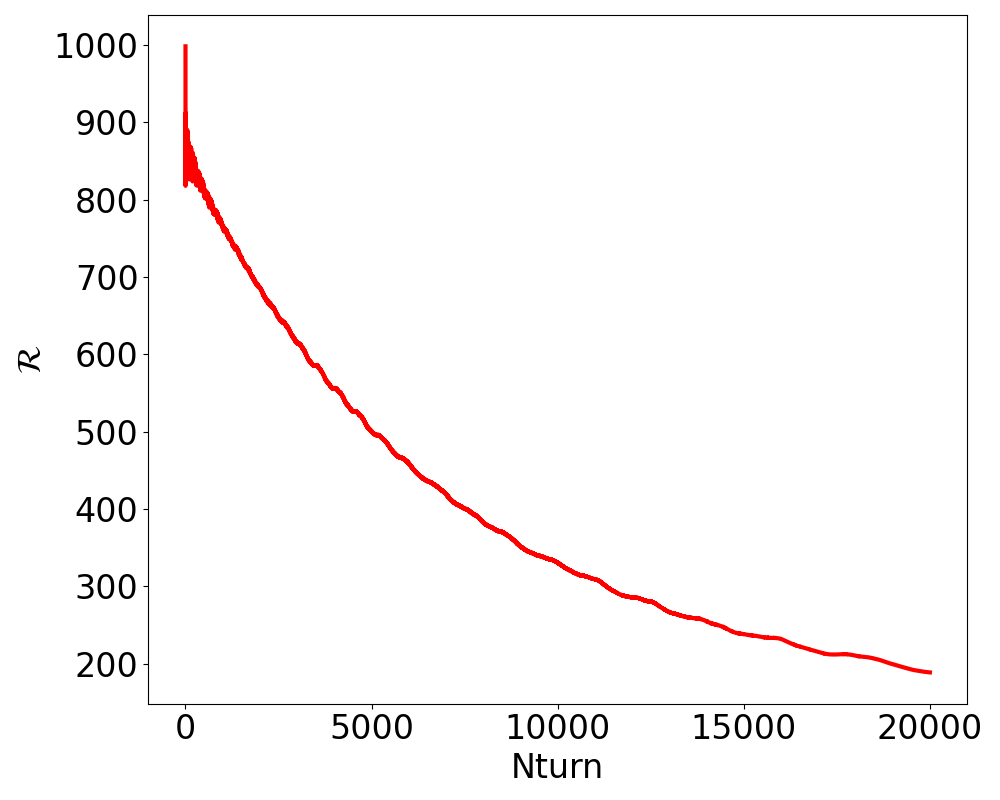}
    \caption{Particle tracking results with $Q_{\mathrm{bunch}}=55\mathrm{\,nC}$. Eigenmode parameters are given in the left plot, where the top-left plot is the normalized angular momentum $L_{ZN}=\beta\gamma L_z$, the top-right plot is the normalized eigenmode~1 emittance, the bottom-left plot is the normalized eigenmode~2 emittance, and the bottom-right plot is the normalized $\epsilon_{4D}$ emittance. The right plot shows the eigenmode emittance ratio $\mathcal{R}=\epsilon_1/\epsilon_2$. ImpactX code is used for tracking with 20000 turns.}
    \label{fig:quadboostersimresults}
\end{figure}

The second booster design is a quadrupole-based ring, in which the main focusing is provided by quadrupoles, while solenoids are placed in the straight sections to establish a well-defined eigenmode separation. The design is illustrated in Fig.~\ref{fig:quadboosteroptics}. The initial proton energy is $200.0$\,MeV, the same as in the solenoid-based booster. In this case, a bunch charge of $55$\,nC is used, resulting in tune shifts of $\Delta Q_{1,2}=-0.110$. The working point is $Q_1=5.94$ and $Q_2=5.32$. The tracking results are shown in Fig.~\ref{fig:quadboostersimresults}, yielding a final flatness ratio of $\mathcal{R}=200$. The beam is initialized with an intrinsic flatness ratio of $\mathcal{R}=1000$. Compared with the solenoid-based ring, the quadrupole booster design therefore leads to a stronger degradation of intrinsic flatness.

This loss of intrinsic flatness is a consequence of the mirror-symmetric design criterion described in Section~\ref{subsec:couplingphasepreservation}. In the ideal lattice, this symmetry yields discrete conservation of canonical angular momentum from cell to cell, rather than continuous conservation along the orbit. This discrete conservation follows from the imposed mid-plane symmetry: as illustrated in Fig.~\ref{fig:mirrorsymmetry}, the quadrupoles on one side of the symmetry plane are matched by corresponding quadrupoles on the other side with equal strength and opposite polarity. However, the direct space-charge force does not respect this symmetry, since it depends on the evolving beam distribution and on the longitudinal coordinate $s$. As the beam propagates through the quadrupole channel, the space-charge potential acquires an azimuthal-angle dependence, which leads to progressive demagnetization over the cell. Demagnetization refers to the reduction of the beam's rms canonical angular momentum, $L_z$, by analogy with magnetized beams (Sec.~\ref{subsec:magnetizedbeams}), whose defining property is a nonzero canonical angular momentum. According to Eq.~\eqref{eq:approximatedeigenmodedependenceonangmom}, 
a decrease in angular momentum reduces the separation between the eigenmode emittances, decreasing $\epsilon_1$ and increasing $\epsilon_2$, thereby lowering the intrinsic flatness ratio.

As a result of this demagnetization effect, the quadrupole booster performs worse in terms of maintaining intrinsic flatness than the solenoid-based booster. However, it remains a viable option when solenoidal focusing is impractical, provided the required intrinsic flatness ratio is moderate.

\subsection{Quadrupole-Based Synchrotron}\label{subsec:synch}

\begin{figure}[tbp]
    \centering
    \includegraphics[width=0.49\linewidth]{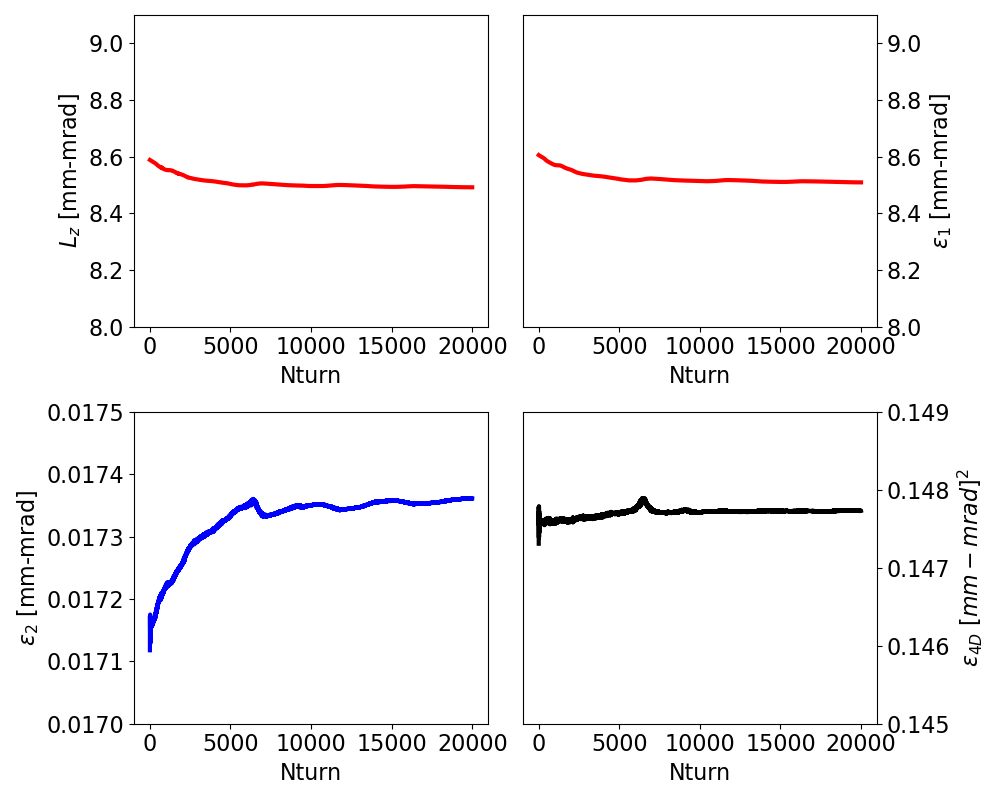}
    \includegraphics[width=0.49\linewidth]{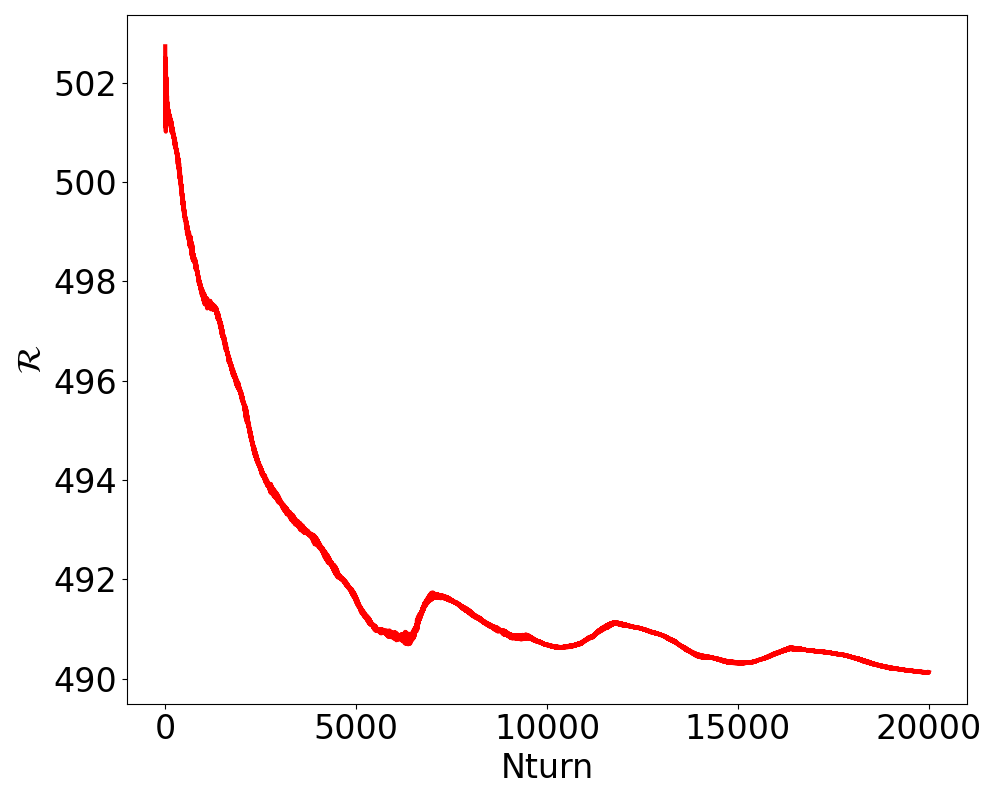}
    \includegraphics[width=0.49\linewidth]{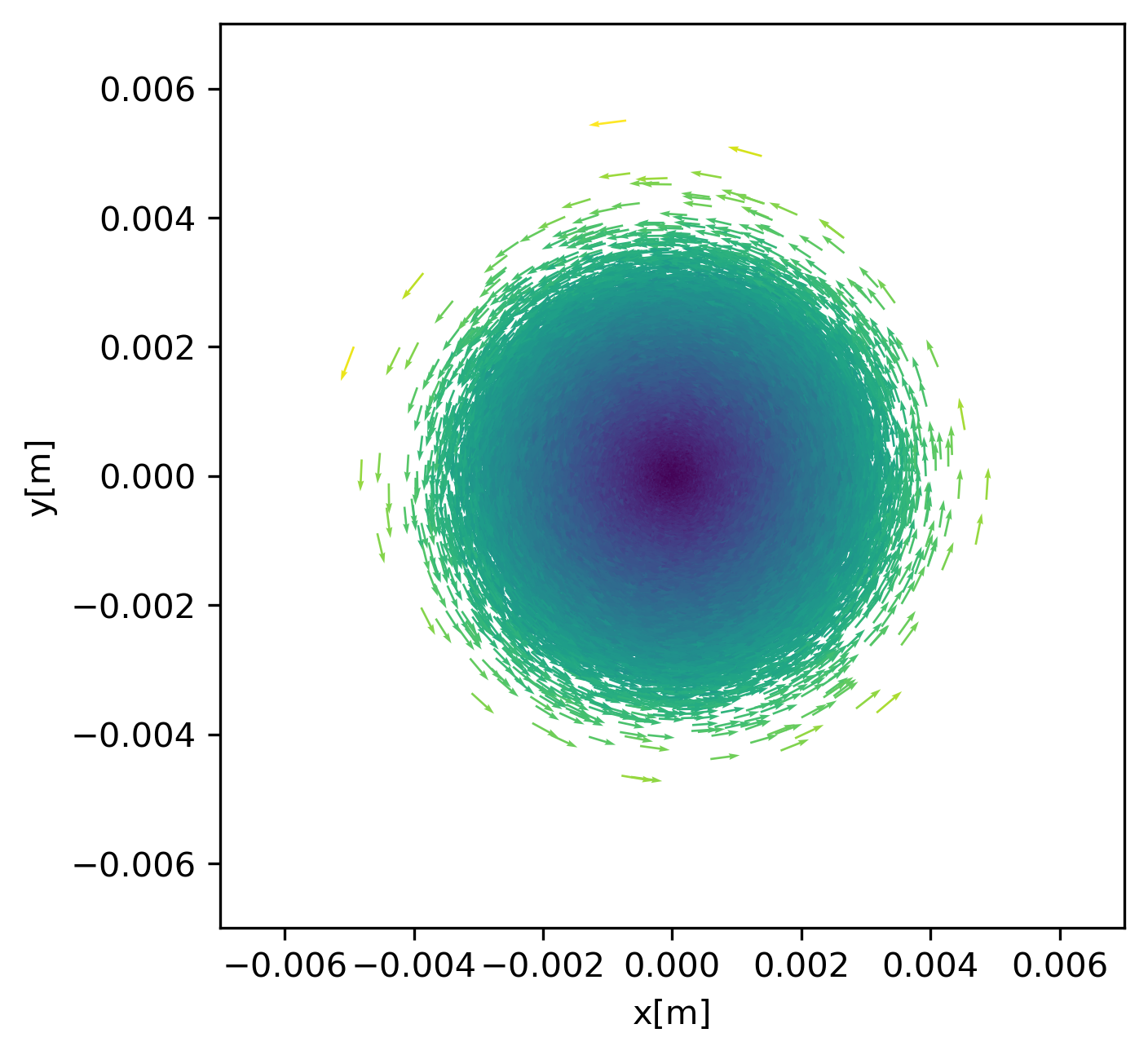}
    \includegraphics[width=0.49\linewidth]{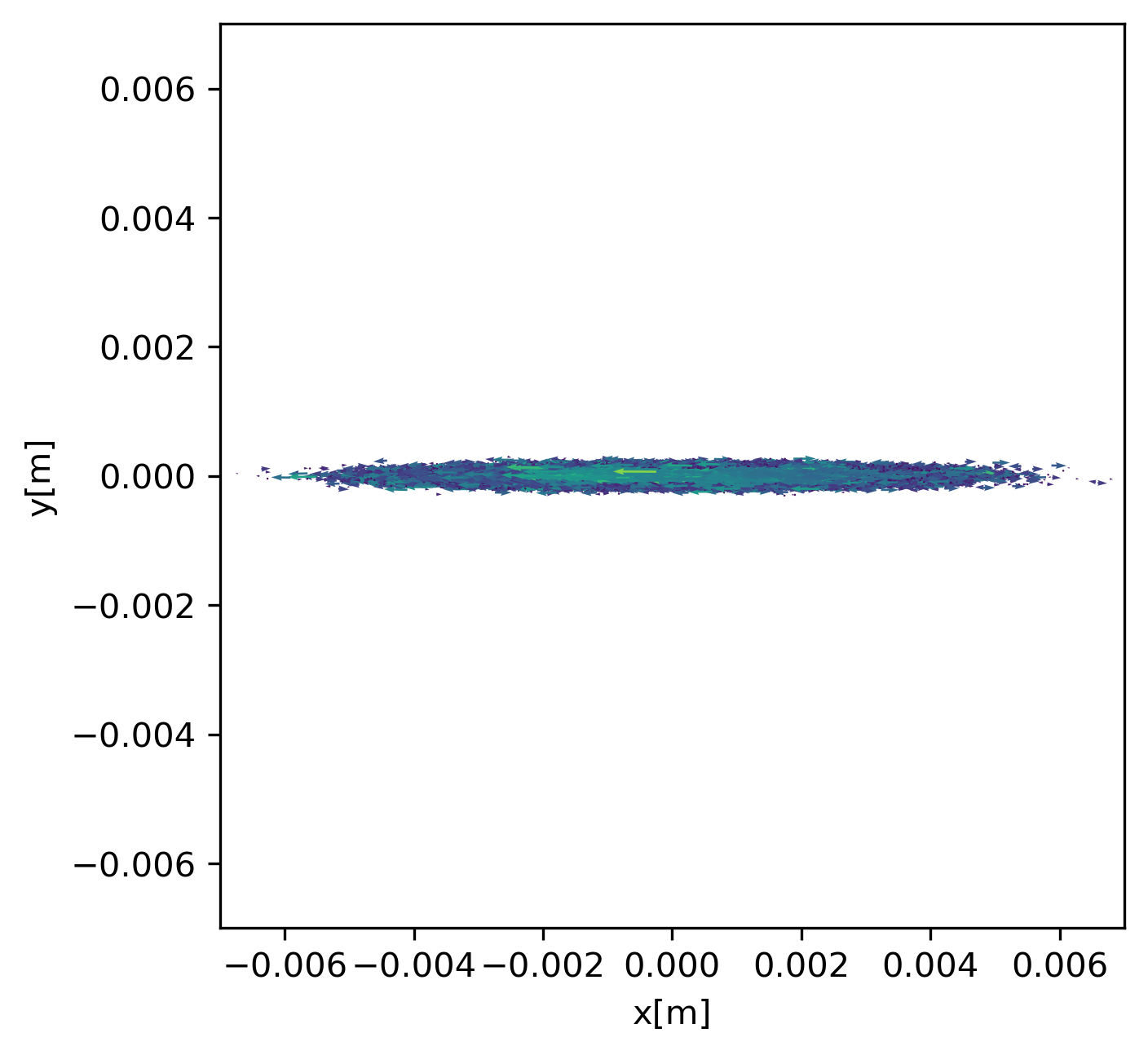}
    \caption{Particle tracking results with $Q_{\mathrm{bunch}}=55\mathrm{\,nC}$. Eigenmode parameters are shown in the top-left plot, and the eigenmode emittance ratio is shown in the top-right plot. The beam profile in the $(x,y)$ plane at the end of 20,000 turns is shown in the bottom-left plot, and the decoupled beam profile in the $(x,y)$ plane is shown in the bottom-right plot. ImpactX code is used for tracking with 20000 turns.}
    \label{fig:quadsynchresults}
\end{figure}

After the booster simulation is performed, it is assumed that an intrinsic flatness ratio of 500 is maintained and successfully transported to a larger synchrotron with an injection energy of $3.2$\,GeV. The beam is then accelerated to $23.2$\,GeV, where the space-charge effects are no longer dominant. At the $3.2-$\,GeV synchrotron injection energy, the rms bunch length and energy spread are $\sigma_z=2.05$\,m and $\sigma_{\Delta E}=3.42$\,MeV, respectively, corresponding to $\sigma_\delta=8.7\times 10^{-4}$. The acceleration ramp crosses the transition energy, and the resulting longitudinal response is retained in the full six-dimensional tracking and in the self-consistent three-dimensional space-charge calculation. The beam crosses transition at approximately 8000–9000 turns. In the RF-phase convention used here, the synchronous phases of the five RF cavities are reversed above transition to maintain stable longitudinal motion. The crossing produces the small transient visible in the rms eigenmode emittances in Fig.~\ref{fig:quadsynchresults}, but it does not result in subsequent secular degradation of the intrinsic flatness ratio, which remains approximately \(\mathcal R=490\) after 20000 turns.

\begin{figure}[tbp]
    \centering
    \includegraphics[width=0.49\linewidth]{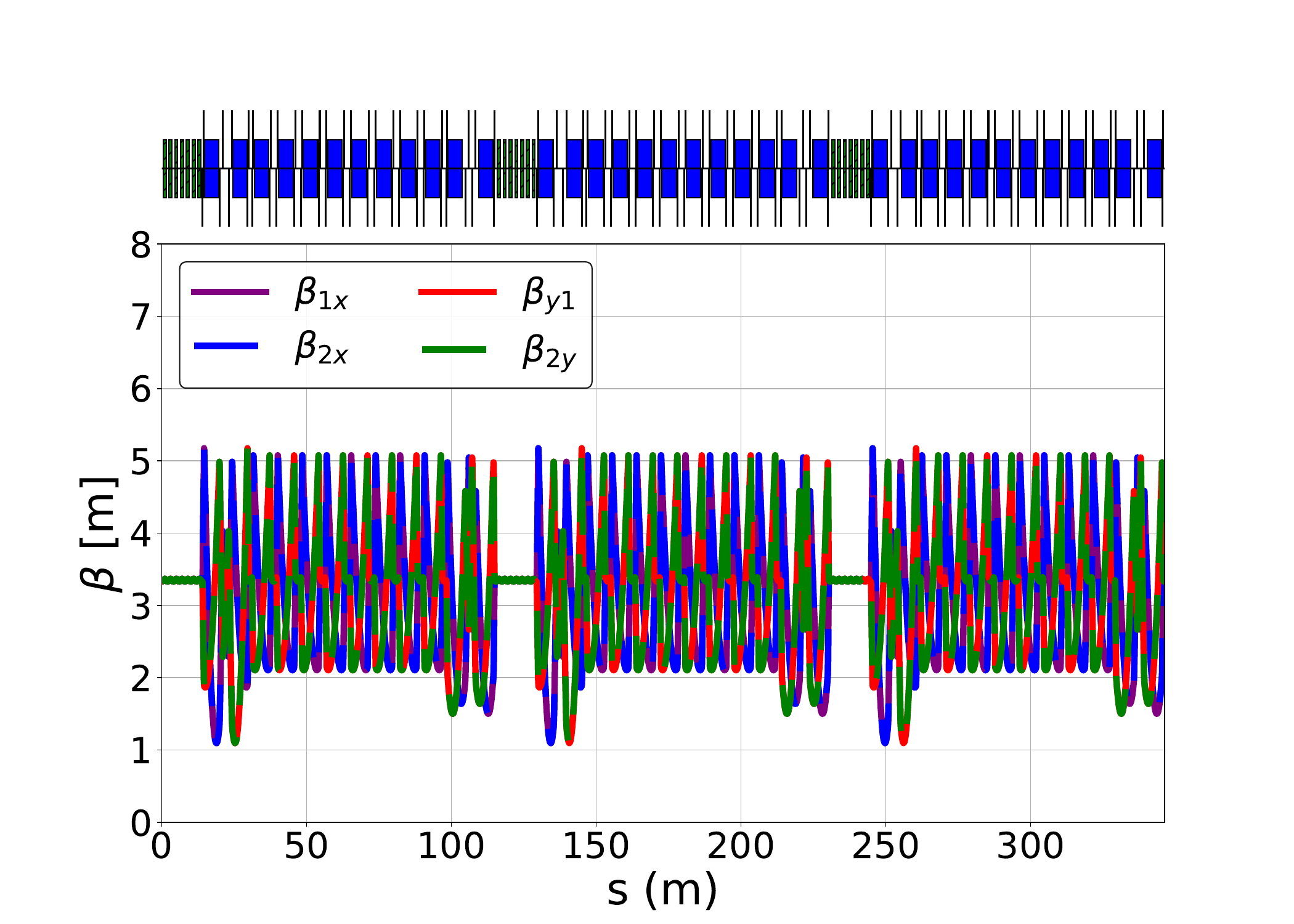}
    \includegraphics[width=0.49\linewidth]{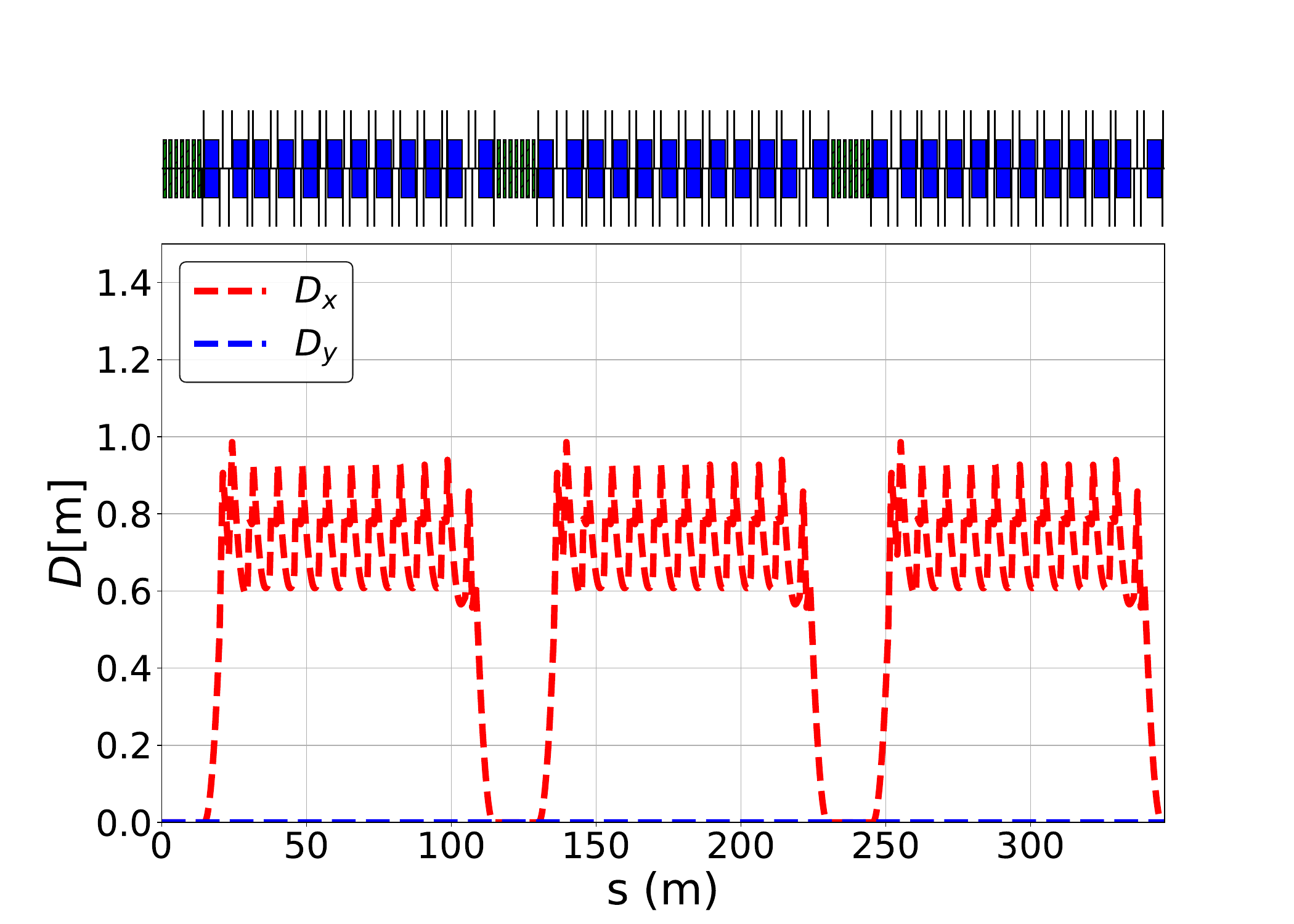}
    \includegraphics[width=0.49\linewidth]{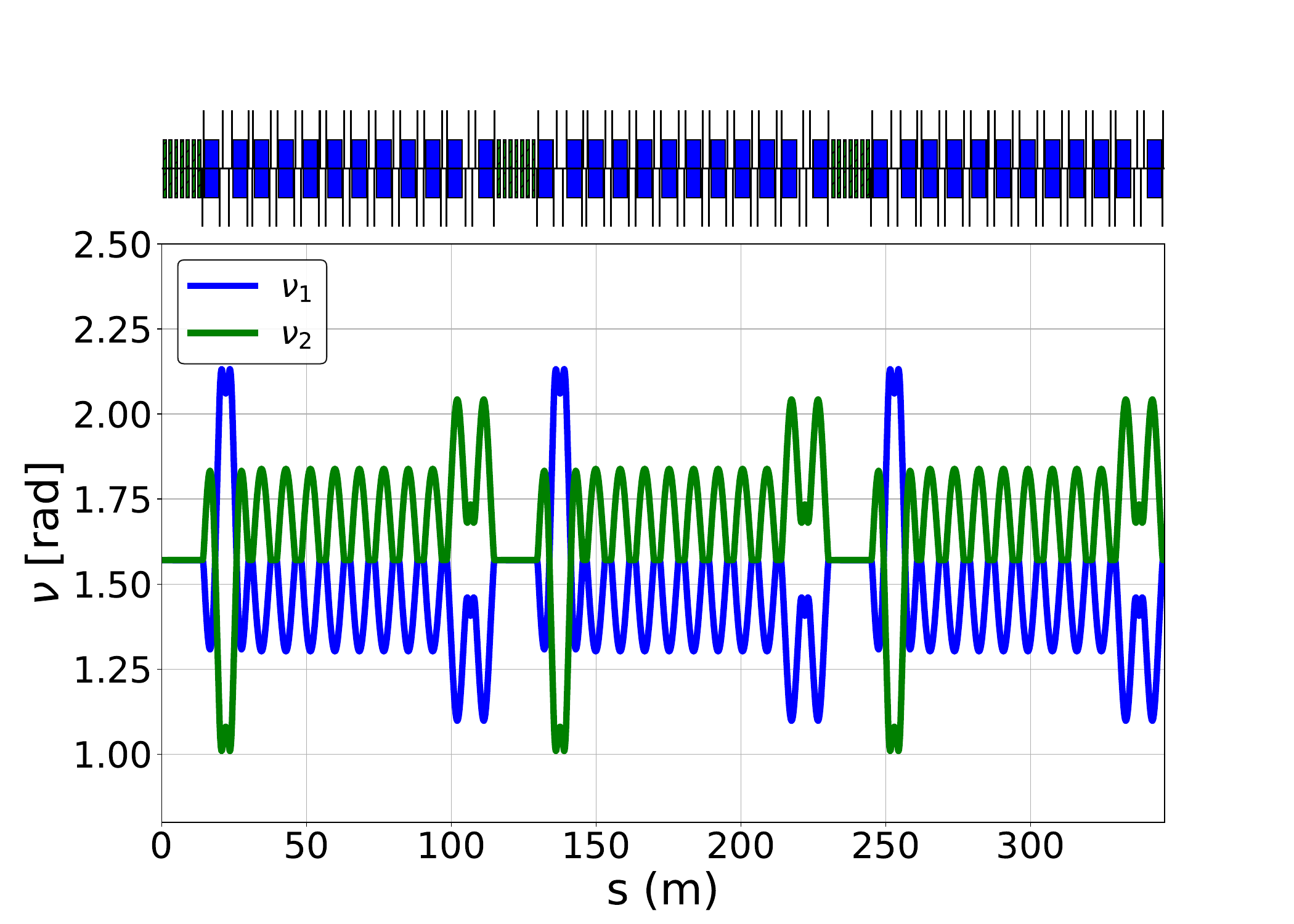}
    \includegraphics[width=0.49\linewidth]{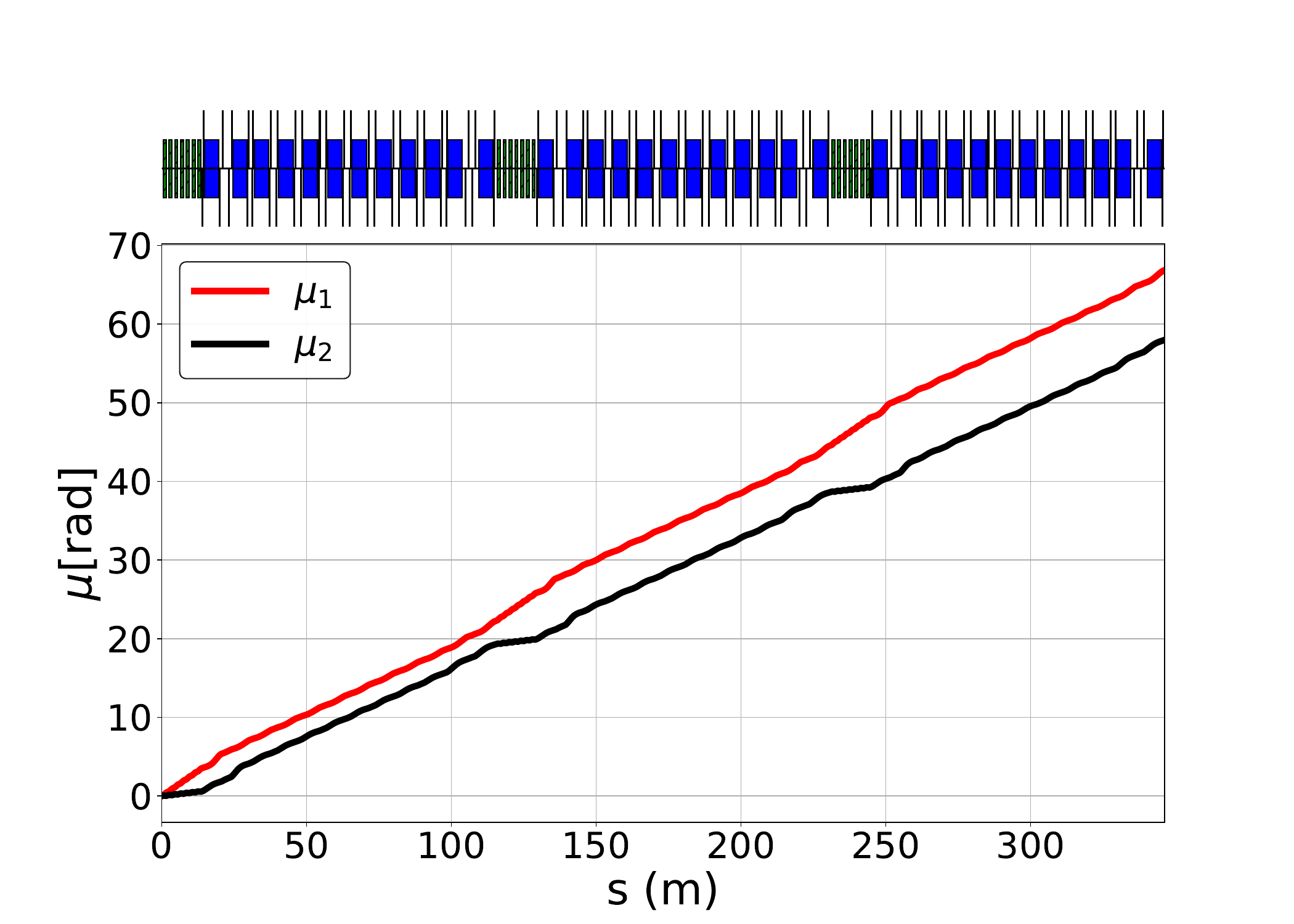}
    \caption{Quadrupole-based synchrotron ring: coupled $\beta$ functions are shown in the top-left plot, dispersions are shown in the top-right plot, the coupling phases are shown in the bottom-left plot, and phase advances are shown in the bottom-right plot.}
    \label{fig:quadsynchoptics}
\end{figure}

\begin{figure}[tbp]
    \centering
    \includegraphics[width=0.6\linewidth]{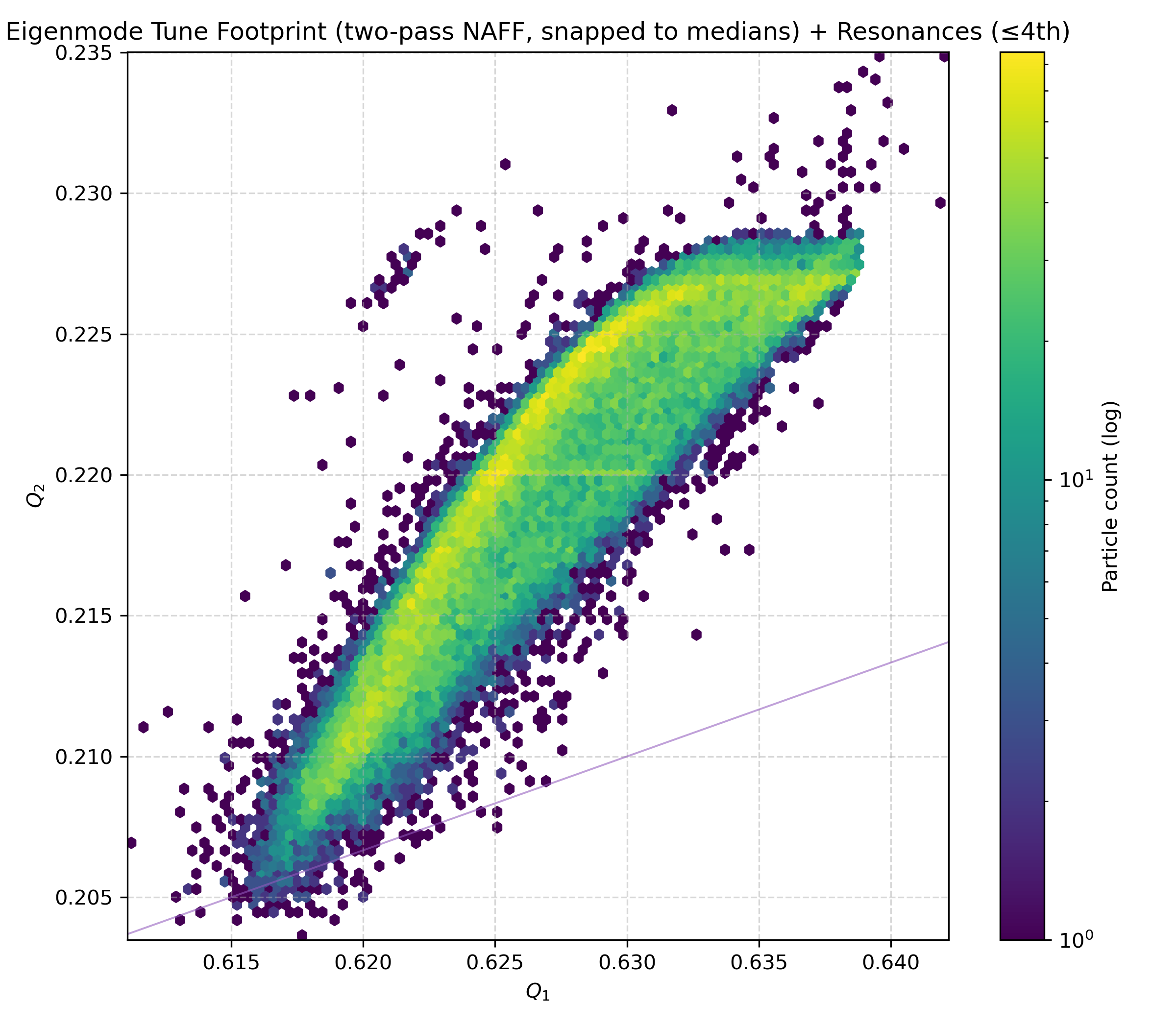}
    \caption{Computed space-charge-induced tune spread for the synchrotron. $Q_1$ is tune of eigenmode~1 and $Q_2$ is the tune for eigenmode~2. The tune spread is computed by a NAFF algorithm using turn-by-turn monitor data. Only fractional tunes are plotted.}
    \label{fig:tunespreadsynch}
\end{figure}

The synchrotron design is shown in Fig.~\ref{fig:quadsynchoptics} with tunes $Q_1=10.64$ and $Q_2=9.23$. Similar to the quadrupole booster design, solenoids are used in the straight sections for tune splitting and minor focusing. The coupling phases are maintained close to the ideal value of $\pi/2$ throughout the transport for maintaining a round beam. The particle tracking results with a 55-nC beam are shown in Fig.~\ref{fig:quadsynchresults}. As seen in the tracking results, the space-charge effect is no longer dominant, leading to only minor demagnetization. At the end, an intrinsic flatness ratio of 490.0 is preserved, where the beam is also decoupled with a Derbenev adapter, revealing the flat beam. The tune footprint obtained by applying the NAFF algorithm to the simulated turn-by-turn data is presented in Fig.~\ref{fig:tunespreadsynch}. The simulated footprint exhibits the same one-sided structure predicted for a mode-1-dominated circular-mode beam in the right panel of Fig.~\ref{fig:coupledspreadcircmode}, rather than the complete necktie structure of the equal-emittance round beam. It therefore provides a qualitative numerical consistency check of the predicted eigenmode-footprint topology. No point-by-point comparison of the analytical and simulated footprint boundaries is intended.

In summary, we have designed lattices capable of maintaining circular-mode beams. Among the configurations considered, the solenoid-based ring performs best in preserving the circular mode and maintaining intrinsic flatness at high intensity; for $Q_{\mathrm{bunch}}=60$\,nC, the final flatness ratio is $\mathcal{R}\approx 470.0$. However, solenoidal focusing becomes impractical at high energies, making quadrupole-based lattices necessary. Guided by the mid-plane symmetry conditions described in Sec.~\ref{subsec:preservingangmom}, we designed a quadrupole-based lattice that preserves the circular mode. Its performance is nevertheless inferior to that of the solenoid-based lattice, because the angular momentum is conserved only discretely from cell to cell rather than continuously, which leads to demagnetization and a loss of intrinsic flatness. Finally, the quadrupole-based synchrotron design performs best among the quadrupole-based options, since space-charge effects are reduced by relativistic dilution during acceleration. At extraction, the circular-mode beam is decoupled and the intrinsic flatness is restored.

\section{Conclusion}

In this manuscript, we studied circular-mode optics as a framework for transporting and accelerating intrinsically flat beams while maintaining a round transverse projection in physical space. The central motivation is that conventional flat beams, although attractive for increasing beam brightness and collider luminosity, are strongly degraded by direct space-charge forces at low energy. Circular modes provide a way to preserve intrinsic flatness in the eigenmode structure while maintaining a round beam in the plane where the direct space-charge force acts.

Using the coupled-optics formalism, we analyzed the formation and preservation of circular modes and derived lattice-design conditions for conserving canonical angular momentum and coupling phases. These considerations lead naturally to rotationally invariant transport and ring cells based on solenoids, symmetric alternating-gradient structures, and indexed dipoles. In quadrupole-based lattices, the symmetric alternating-gradient design enables discrete conservation of canonical angular momentum from cell to cell. We also developed the corresponding space-charge tune-shift and tune-spread formalisms for circular-mode beams in both uncoupled and coupled lattices. The analysis shows that, to leading order, circular-mode beams retain the space-charge behavior of round beams while preserving intrinsic eigenmode flatness, and that their tune-spread footprint is reduced compared with conventional round-beam cases. 

The particle-in-cell simulations performed with TRACK and ImpactX confirm that circular modes can be generated, transported, accelerated, and decoupled in representative lattice designs. Among the cases considered, the solenoid-based ring provides the best preservation of circularity and intrinsic flatness under strong space-charge conditions. For the test case considered here, a beam with $Q_{\mathrm{bunch}}=60$\,nC and initial intrinsic flatness ratio $\mathcal{R}_i=1000$ retains a final ratio of approximately $\mathcal{R}_f\approx470$ at the end of tracking. By contrast, the quadrupole-based lattice with solenoids in the straight sections for tune splitting yields a smaller final ratio, $\mathcal{R}_f\approx200$, because space charge breaks the ideal angular-momentum conservation and demagnetizes the beam. In the synchrotron design, however, the higher beam energy weakens the space-charge effect through relativistic dilution, thereby reducing demagnetization and improving the preservation of intrinsic flatness. Finally, we showed that the intrinsic flatness can be restored in configuration space by decoupling the beam after the space-charge-dominated region is passed. Overall, these results show that circular-mode optics is a promising approach for high-intensity hadron beams and for applications in which flat beams are desired downstream.

\section{Acknowledgements}

The authors thank Alexey Burov, Valeri Lebedev, Sergei Nagaitsev, Nick Evans, and Austin Hoover for fruitful discussions. 

This work was supported by the U.S. Department of Energy, under Contract No. DE-AC02-06CH11357. Work funded by DOE's Office of Nuclear Physics and Argonne contract No. 3J-60004-0003A to Illinois Tech, formerly Illinois Institute of Technology.

\clearpage

\appendix
\section{Relations Between Circular Modes and Lebedev-Bogacz Parameters}\label{sec:appendixA}
This appendix provides the derivations supporting circular-mode properties summarized in Sec.~\ref{sec:theory}. We first derive the vortex relation geometrically, then obtain the circular-mode optical conditions from the Lebedev--Bogacz parametrization, and finally derive the projected covariance matrix and second moments of a circular-mode beam. 

\subsection{Geometric derivation of the vortex relation}
\label{subsec:appendixAvortexrel}

Circular modes describe rotating beams with nonzero canonical angular momentum and intrinsically strong coupling between the transverse degrees of freedom. For intuition, consider a particle executing planar motion in polar coordinates $(r,\theta)$, where overdots denote derivatives with respect to time:
\begin{equation}
\begin{split}
x &= r\cos\theta, \qquad \dot{x}=\dot{r}\cos\theta - r\dot{\theta}\sin\theta, \\
y &= r\sin\theta, \qquad \dot{y}=\dot{r}\sin\theta + r\dot{\theta}\cos\theta.
\end{split}
\end{equation}
The particle's transverse motion is confined on the circle with non-zero angular speed, $\dot{r}=0$ and $\dot{\theta}\ne 0$, which reduces the polar parametrization to
\begin{equation}
    \begin{split}
        x&=r\cos\theta, \qquad \dot{x}= -r\dot{\theta}\sin\theta, \\
        y&=r\sin\theta, \qquad \dot{y}=r\dot{\theta}\cos\theta.
    \end{split}
    \label{eq:polarformcoordinaterep}
\end{equation}
Equation~\eqref{eq:polarformcoordinaterep} can be written in a different form which yields the vortex relation
\begin{equation}
\begin{pmatrix}
y\\ \dot{y}
\end{pmatrix}
=
\begin{pmatrix}
0 & -1/\dot{\theta} \\
\dot{\theta} & 0
\end{pmatrix}
\begin{pmatrix}
x\\ \dot{x}
\end{pmatrix}.
\label{eq:polarformvortex}
\end{equation}
Using Courant-Snyder parameters, Eq.~\eqref{eq:polarformcoordinaterep} becomes:
\begin{equation}
    \begin{split}
        x&=\sqrt{\epsilon_{r}\beta_{r}}\cos\psi_{r}, \qquad x'=-\sqrt{\frac{\epsilon_{r}}{\beta_{r}}}\sin\psi_{r} \\
        y&=\sqrt{\epsilon_{r}\beta_{r}} \sin\psi_{r}, \qquad y'=\sqrt{\frac{\epsilon_{r}}{\beta_{r}}}\cos\psi_{r}.
    \end{split}
    \label{eq:polarformCSformat}
\end{equation}
Here, $\beta_{x}=\beta_{y}$, $\alpha_{x}=\alpha_{y}=0$ and $\psi_{y}=\psi_{x}-\dfrac{\pi}{2}$ conditions are used. Equation~\eqref{eq:polarformCSformat} also yields vortex relation similar to Eq.~\eqref{eq:polarformvortex} as:
\begin{equation}
\begin{pmatrix}
y\\ y'
\end{pmatrix}
=
\begin{pmatrix}
0 & -\beta_{r} \\
1/\beta_{r} & 0
\end{pmatrix}
\begin{pmatrix}
x\\ x'
\end{pmatrix}.
\label{eq:phasespacecoordvortex}
\end{equation}
Vortex relation in Eq.~\eqref{eq:phasespacecoordvortex} is also given in~\cite{burov2002circular,derbenev1998adapting,groening2021particle}. 

\subsection{Circular-mode conditions in the Lebedev--Bogacz parametrization}
\label{subsec:appendixAcircularmodecondLB}

The eigenvectors in Eq.~\eqref{eq:eigenvectorsgen} are normalized with respect to symplectic unit matrix as:
\begin{equation}
    \begin{split}
        \vec{v}_{1}^{\dagger}\mathbf{S}\vec{v}_{1}&= -2i, \quad \vec{v}_{2}^{\dagger}\mathbf{S}\vec{v}_{2}=-2i, \\
        \vec{v}_{1}^{T}\mathbf{S}\vec{v}_{1}&=0, \quad \vec{v}_{2}^{T}\mathbf{S}\vec{v}_{2}=0, \\
        \vec{v}_{1}^{T}\mathbf{S}\vec{v}_{2}&=0, \quad \vec{v}_{1}^{\dagger}\mathbf{S}\vec{v}_{2}=0,
    \end{split}
    \label{eq:eigenvectornormalization}
\end{equation}
where, $S$ is $4\times 4$ symplectic unit matrix and given as:
\begin{equation}
    \mathbf{S}= \begin{pmatrix}
            0 & 1 & 0 & 0 \\
            -1 & 0 & 0 & 0 \\
            0 & 0 & 0 &  1 \\
            0 & 0 & -1 & 0
        \end{pmatrix}.
        \label{eq:symplecticunitmatrix4b4}
\end{equation}
Using the normalization conditions in Eq.~\eqref{eq:eigenvectornormalization}, the coupling strength parameter is 
\begin{equation}
	\begin{split}
		\kappa_{x} &= \sqrt{\frac{\beta_{2x}}{\beta_{1x}}}, \qquad \kappa_{y} = \sqrt{\frac{\beta_{1y}}{\beta_{2y}}}, \\
		A_{x}&= \kappa_{x}\alpha_{1x} - \kappa_{x}^{-1}\alpha_{2x}, \\
		A_{y}&= \kappa_{y}\alpha_{2y} - \kappa_{y}^{-1}\alpha_{1y}, \\
		u&= \frac{-\kappa_{x}^{2}\kappa_{y}^{2}\pm \sqrt{\kappa_{x}^{2}\kappa_{y}^{2}(1+\frac{A_{x}^{2}-A_{y}^{2}}{\kappa_{x}^{2}-\kappa_{y}^{2}}(1-\kappa_{x}^{2}\kappa_{y}^{2}))}}{1-\kappa_{x}^{2}\kappa_{y}^{2}}. 
	\end{split}
    \label{eq:ucalculationLB}
\end{equation}
For circular-mode beams with $\alpha_{1x}=\alpha_{2x}=\alpha_{1y}=\alpha_{2y}=0$, the expression in Eq.~\eqref{eq:ucalculationLB} simplifies to:
\begin{equation}
    u= \frac{\kappa_{x}\kappa_{y}}{1+\kappa_{x}\kappa_{y}}.
\end{equation}
Equal beta functions in decoupled plane $\beta_{x}=\beta_{y}$ also implies the coupled functions are equal $\beta_{1x}=\beta_{2x}=\beta_{1y}=\beta_{2y}=\beta_{0}$. With equal beta functions, the coupling strength parameter is $u=\dfrac{1}{2}$.

\subsection{Projected covariance matrix and second moments}
\label{subsec:appendixAsecondmoments}

The projection of the transverse ellipsoid onto the $(x,y)$ plane can be computed from the Gaussian distribution by taking the level set of the inverse covariance matrix, $\vec{r}^T\Sigma_{xy}^{-1}\vec{r} = 1$ for $\vec{r}=(x,y)$. The covariance matrix for $(x,y)$ plane is given as:
\begin{equation}
    \Sigma_{xy} = \begin{pmatrix}
        \sigma_{x}^2 & \sigma_{xy} \\
        \sigma_{xy} & \sigma_{y}^2
    \end{pmatrix}, \qquad \text{with}\,, \qquad \mathrm{det}(\Sigma_{xy}) = \sigma_x^2\sigma_y^2(1-\tilde{\alpha}^2).
    \label{eq:covariancematrixsigmaxy}
\end{equation}
Computing $\vec{r}^T\Sigma_{xy}^{-1}\vec{r}=1$ yields:
\begin{equation}
    \frac{x^2}{\sigma_x^2} - \frac{2\tilde{\alpha}}{\sigma_{x}\sigma_y}xy + \frac{y^2}{\sigma_y^2} = 1-\tilde{\alpha}^2, \qquad \tilde{\alpha} = \frac{\sigma_{xy}}{\sigma_x\sigma_y}.
    \label{eq:xy_projection}
\end{equation}
Here, $\tilde{\alpha}$ is the $(x,y)$ correlation (tilt) parameter that sets the strength of the $xy$ cross term. In the circular-mode limit the beam is round, $\sigma_x=\sigma_y\equiv\sigma_b$, and the $(x,y)$ projection is untilted, i.e., $\tilde{\alpha}=0$. A sufficient condition for $\tilde{\alpha}=0$ is $\cos\nu_{1}=\cos\nu_{2}=0$, e.g., $\nu_{1}=\nu_{2}=\pi/2$ (mod $\pi$).

Furthermore, second-order moments from~\cite{lebedev2010betatron} are
\begin{equation}
	\begin{split}
		\sigma_{xx} &= \epsilon_{1}\beta_{1x} + \epsilon_{2}\beta_{2x}, \\
		\sigma_{xx'} &= -\alpha_{1x}\epsilon_{1} - \alpha_{2x}\epsilon_{2},\\
		\sigma_{x'x'} &= \epsilon_{1}\frac{(1-u)^{2} + \alpha_{1x}^{2}}{\beta_{1x}} + \epsilon_{2}\frac{u^{2} + \alpha_{2x}^{2}}{\beta_{2x}}, \\
		\sigma_{yy}&= \epsilon_{1}\beta_{1y} + \epsilon_{2}\beta_{2y}, \\
		\sigma_{yy'} &= -\alpha_{1y}\epsilon_{1} - \alpha_{2y}\epsilon_{2}, \\
		\sigma_{y'y'} &= \epsilon_{1}\frac{u^{2}+\alpha_{1y}^{2}}{\beta_{1y}} + \epsilon_{2}\frac{(1-u)^{2} + \alpha_{2y}^{2}}{\beta_{2y}}, \\
		\sigma_{xy}&= \epsilon_{1}\sqrt{\beta_{1x}\beta_{1y}}\cos\nu_{1} + \epsilon_{2}\sqrt{\beta_{2x}\beta_{2y}}\cos\nu_{2}, \\
		\sigma_{xy'} &= \epsilon_{1}\sqrt{\frac{\beta_{1x}}{\beta_{1y}}}(u\sin\nu_{1} - \alpha_{1y}\cos\nu_{1}) - \epsilon_{2}\sqrt{\frac{\beta_{2x}}{\beta_{2y}}}((1-u)\sin\nu_{2}+\alpha_{2y}\cos\nu_{2}), \\
		\sigma_{yx'} &= -\epsilon_{1}\sqrt{\frac{\beta_{1y}}{\beta_{1x}}}((1-u)\sin\nu_{1}+\alpha_{1x}\cos\nu_1) + \epsilon_{2}\sqrt{\frac{\beta_{2y}}{\beta_{2x}}}(u\sin\nu_{2} - \alpha_{2x}\cos\nu_{2}), \\
		\sigma_{x'y'} &= \epsilon_{1}\frac{1}{\sqrt{\beta_{1x}\beta_{1y}}}((\alpha_{1y}(1-u)-\alpha_{1x}u)\sin\nu_{1} + (u(1-u) + \alpha_{1x}\alpha_{2y})\cos\nu_{1}) \\
		&+ \epsilon_{2}\frac{1}{\sqrt{\beta_{2x}\beta_{2y}}}\left( 
				(\alpha_{2x}(1-u) - \alpha_{2y}u)\sin\nu_{2} + (u(1-u) + \alpha_{2x}\alpha_{2y})\cos\nu_{2}
		\right).
	\end{split}
    \label{eq:RMSquantitiesLB}
\end{equation}
Here, $\epsilon_{1,2}$ are the RMS eigenmode emittances. For circular-mode beam optics, the rms quantities are 
\begin{equation}
    \begin{split}
    \sigma_{xx}&=\sigma_{yy}=(\epsilon_{1}+\epsilon_{2})\beta_{0}, \\
    \sigma_{xx'}&=\sigma_{yy'}=\sigma_{x'y'}=0, \\
    \sigma_{x'x'}&=\sigma_{y'y'}= (\epsilon_{1}+\epsilon_{2})\frac{1}{4\beta_{0}}, \\
    \sigma_{xy}&=0, \\
    \sigma_{xy'}&= \frac{1}{2}(\epsilon_{1}-\epsilon_{2}), \quad \sigma_{yx'}=-\frac{1}{2}(\epsilon_{1}-\epsilon_{2}),
    \end{split}
    \label{eq:circmodeRMSquantitiesLB}
\end{equation}

\section{Importance of Coupling Phases}\label{sec:importanceofcouplingphases}

\begin{figure}[tbp]
    \centering
	\includegraphics[width=0.49\linewidth]{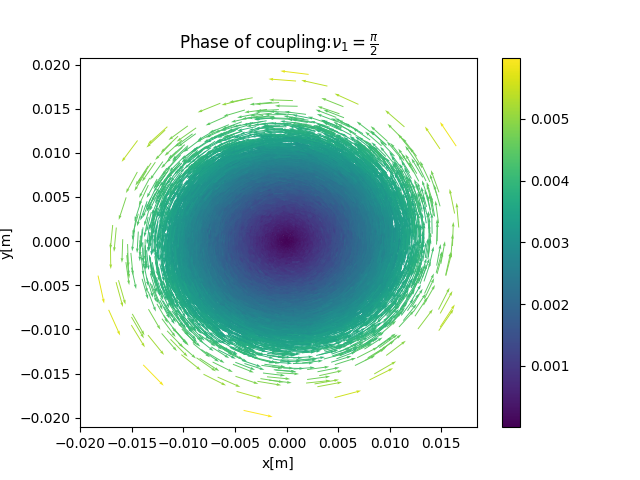}
	\includegraphics[width=0.49\linewidth]{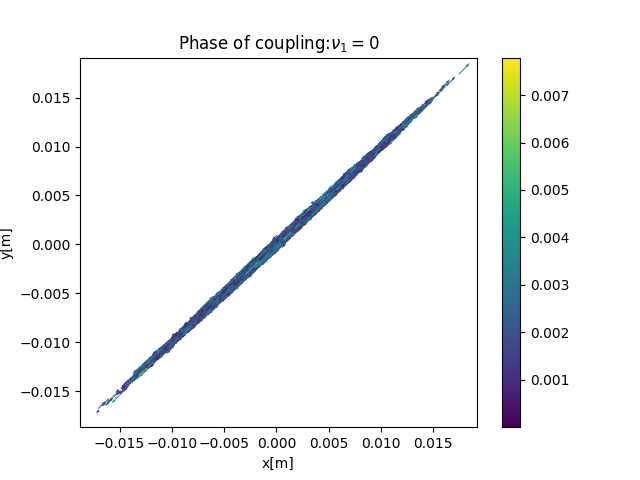}
    \caption{$(x,y)$ projection of circular-mode beam with phases of coupling values: $\nu_{1}=\pi/2$ (left), $\nu_{1}=0$ (right).}
    \label{fig:phaseofcouplingapp}
\end{figure}

The phases of couplings dictate the orientation of the projected ellipses in the phase planes. For circular-mode beams, the coupling phase of mode~1 determines which plane the intrinsic flatness is projected. The projection equation in Eq.~\eqref{eq:xy_projection}, for mode~1 dominance is
\begin{equation}
\frac{x^{2}}{\sigma_{x}^{2}} - \frac{2xy}{\sigma_{x}\sigma_{y}}\cos\nu_{1} + \frac{y^{2}}{\sigma_{y}^{2}} = \sin^{2}\nu_{1}.
\label{eq:xyprojmode1dom}
\end{equation}
For circular-mode beam parameters, the coupling phases are $\nu_{1,2}=\tfrac{\pi}{2}$, which yields the projection for $\sigma_{x}=\sigma_{y}\equiv\sigma_{b}$
\begin{equation}
    x^{2}+y^{2} = \sigma_{b}^{2}.
    \label{eq:roundbeamproj}
\end{equation}
Interestingly, if the phase of coupling is zero or $\pi$, i.e, $\nu_{1}=0,\pi$ then the projection becomes a straight line equation
\begin{equation}
	\begin{split}
		\frac{x^{2}}{\sigma_{x}^{2}} - \frac{2xy}{\sigma_{x}\sigma_{y}} + \frac{y^{2}}{\sigma_{y}^{2}}&=0, \\
		(\frac{x}{\sigma_{x}}-\frac{y}{\sigma_{y}})^{2}&=0, \\
		y&=\pm\frac{\sigma_{y}}{\sigma_{x}}x.
	\end{split}
\end{equation}
For other values of coupling phases, the projection is a tilted ellipse. The projection of circular-mode beam for $\nu_{1}=\tfrac{\pi}{2}$ and $\nu_{1}=0$ is shown in Fig.~\ref{fig:phaseofcouplingapp}. Figure~\ref{fig:phaseofcouplingapp} shows the importance of coupling phase in circular-mode beams. If the coupling phase of the dominant mode is not controlled, the beam's intrinsic flatness projects onto $(x,y)$ plane.

\section{Effect of Acceleration on Angular Momentum}\label{sec:appendixC}
Hill's equation in the case of acceleration includes damping in the transverse plane, such as:
\begin{equation}
	x'' + \frac{(\gamma\beta)'}{\gamma\beta}x' + \kappa_{x}x=0 , \quad
	y'' + \frac{(\gamma\beta)'}{\gamma\beta}y' + \kappa_{y}y=0.
	\label{Eq:Hillsequationwithacc}
\end{equation}
Here, $\beta$ and $\gamma$ are the relative velocity and Lorentz's relativistic factor, respectively, while $\kappa$ is the periodic focusing function. The propagation of angular momentum through acceleration is calculated as:
\begin{equation}
\begin{split}
    \frac{d}{ds}L_{z}&=\frac{d}{ds}(mc\gamma\beta(xy'-yx')), \\ 
                    &= mc(\gamma\beta)'(xy'-yx'), \\
                    & + mc\gamma\beta(x'y' + xy'' -y'x'-yx''), \\
                    &= mc(\gamma\beta)'(xy'-yx') + mc\gamma\beta(xy''-yx''). \\
\end{split}                    
\label{Eq:angmomderivative}
\end{equation}
It is straightforward to show that using Eq.~\eqref{Eq:Hillsequationwithacc}, Eq.~\eqref{Eq:angmomderivative} is zero where angular momentum is conserved. Re-normalizing the coordinates as, $\tilde{x}=\frac{x}{\sqrt{\beta\gamma}}$ and $\tilde{y}=\frac{y}{\sqrt{\beta\gamma}}$, leads to:
\begin{equation}
\begin{split}
    L_{z}&=(xy'-yx'), \\
        &=(\frac{\Tilde{x}}{\sqrt{\beta\gamma}}(-\frac{(\gamma\beta)'}{2(\beta\gamma)^{3/2}}\Tilde{y}+\frac{1}{\sqrt{\beta\gamma}}\Tilde{y}'), \\
        & -\frac{\Tilde{y}}{\sqrt{\beta\gamma}}(-\frac{(\gamma\beta)'}{2(\beta\gamma)^{3/2}}\Tilde{x}+\frac{1}{\sqrt{\beta\gamma}}\Tilde{x}')), \\
        &=-\frac{(\gamma\beta)'}{2(\beta\gamma)^{2}}\Tilde{x}\Tilde{y} + \frac{\Tilde{x}\Tilde{y}'}{\gamma\beta} +\frac{(\beta\gamma)'}{2(\beta\gamma)^{2}}\Tilde{x}\Tilde{y} - \frac{\Tilde{y}\Tilde{x}'}{\beta\gamma}, \\
        &= \frac{1}{\gamma\beta}(\Tilde{x}\Tilde{y}'-\Tilde{y}\Tilde{x}').
\end{split}
\label{Eq:normalizedangmom}
\end{equation}
Eq~\eqref{Eq:normalizedangmom} shows that angular momentum is normalized the same way as the emittance, $L_{zN}=\gamma\beta L_{z}$. As a result of this, the normalized angular momentum is invariant under acceleration, the same as the normalized emittance assuming linear forces.

\section{Space-Charge Induced Tune Spread Computation}\label{sec:appendixD}
Space charge potential for Gaussian distribution in rotated principal axes $(\eta,\tau)$ is
\begin{equation}
    \begin{split}
        V_{sc}(\eta,\tau,\bar{z})&= \int_{0}^{1}dt \frac{1}{t\sqrt{(\xi^{2}-1)t+1}}\left( \frac{t}{(1-\chi^{2})t+\chi^{2}} \right)^{1/2} \exp \left ( -\frac{\eta^{2}t}{2\sigma_{\eta}^{2}} - \frac{\tau^{2}tF_{\xi}}{2\sigma_{\tau}^{2}} - \frac{\bar{z}^{2}tF_{\chi}}{2\sigma_{\bar{z}}^{2}} \right ).
    \end{split}
    \label{eq:spacechargepotentialgauss}
\end{equation}
Here, $\xi=\sigma_{\tau}/\sigma_{\eta}$, $\chi=\sigma_{\eta}/\sigma_{\bar{z}}$, $F_{\xi}=\tfrac{\xi^{2}}{(\xi^{2}-1)t+1}$, $F_{\chi}=\tfrac{1}{(1-\chi^{2})t+\chi^{2}}$, and $\bar{z}=\gamma z$, with $\sigma_{\bar{z}}=\gamma \sigma_{z}$. The principal rms beams sizes are computed by diagonalizing the projected $(x,y)$ plane beam matrix:
\begin{equation}
	\begin{split}
		\Sigma_{xy}&= \begin{pmatrix}
			\sigma_{x}^{2} & \sigma_{xy} \\
			\sigma_{xy} & \sigma_{y}^{2}
		\end{pmatrix}, \\
		\sigma_{\eta}^{2} &= \frac{1}{2}\sigma_{x}^{2} + \frac{1}{2}\sigma_{y}^{2} + \frac{1}{2}\sqrt{(\sigma_{x}^{2}-\sigma_{y}^{2})^{2}-4\sigma_{xy}^{2}}, \\
		\sigma_{\tau}^{2} &= \frac{1}{2}\sigma_{x}^{2} + \frac{1}{2}\sigma_{y}^{2} - \frac{1}{2}\sqrt{(\sigma_{x}^{2}-\sigma_{y}^{2})^{2}-4\sigma_{xy}^{2}}.
	\end{split}
    \label{eq:principalrmsbeamsizes}
\end{equation}

Space charge induced tune spread computation for an uncoupled beam in uncoupled lattice is given in~\cite{sen2023density}. Following similar methods in~\cite{sen2023density}, the perturbed Hamiltonian for coupled lattices is
\begin{equation}
    \mathcal{H} = Q_{1}J_{1} + Q_{2}J_{2} + V_{sc}(x,y,\bar{z}).
\end{equation}
In the case of circular-mode beams, the beam is aligned in the projected transverse $(x,y)$ plane, $\theta=0$. The coordinates are parametrized using Lebedev-Bogacz parametrization
\begin{equation}
    \begin{split}
        x&=\sqrt{2J_{1}\beta_{1x}}\cos\psi_{1} + \sqrt{2J_{2}\beta_{2x}}\cos(\psi_{2}-\nu_{2}), \\
        y&=\sqrt{2J_{1}\beta_{1y}}\cos(\psi_{1}-\nu_{1}) + \sqrt{2J_{2}\beta_{2y}}\cos\psi_{2}.
    \end{split}
\end{equation}
In the circular-mode limit, the parametrization reduces
\begin{equation}
    \begin{split}
        x&=\sqrt{2J_{1}\beta_{0}}\cos\psi_{1} + \sqrt{2J_{2}\beta_{0}}\sin\psi_{2}, \\
        y&=\sqrt{2J_{1}\beta_{0}}\sin\psi_{1} + \sqrt{2J_{2}\beta_{0}}\cos\psi_{2}.
    \end{split}
    \label{eq:circmodecoordparam}
\end{equation}
To compute the tune spread the space charge potential is differentiated with respect to the mode actions, $J_{1}$ and $J_{2}$, which yields
\begin{equation}
    \begin{split}
        \Delta Q_{1} &= -C_{sc}\int_{0}^{1}dt\frac{1}{\sigma_{b}^{2}} \left( x\frac{\partial x}{\partial J_{1}} + y\frac{\partial y}{\partial J_{1}} \right)\exp \left( -\frac{x^{2}t}{2\sigma_{b}^{2}} - \frac{y^{2}t}{2\sigma_{b}^{2}} - \frac{\tilde{z}^{2}t}{2\sigma_{z}^{2}}  \right), \\
        \Delta Q_{2} &= -C_{sc}\int_{0}^{1}dt\frac{1}{\sigma_{b}^{2}} \left( x\frac{\partial x}{\partial J_{2}} + y\frac{\partial y}{\partial J_{2}} \right)\exp \left( -\frac{x^{2}t}{2\sigma_{b}^{2}} - \frac{y^{2}t}{2\sigma_{b}^{2}} - \frac{\tilde{z}^{2}t}{2\sigma_{z}^{2}}  \right).
    \end{split}
    \label{eq:tunespreadintegralbefore}
\end{equation}
The partial derivatives for circular-mode limit are given as:
\begin{equation}
    \begin{split}
        x\frac{\partial x}{\partial J_{1}}&= \beta_{0}\cos^{2}\psi_{1} + \beta_{0}\sqrt{\frac{J_{2}}{J_{1}}}\sin\psi_{2}\cos\psi_{1}, \\
        y\frac{\partial y}{\partial J_{1}}&= \beta_{0}\sin^{2}\psi_{1} + \beta_{0}\sqrt{\frac{J_{2}}{J_{1}}}\sin\psi_{1}\cos\psi_{2}, \\
        x\frac{\partial x}{\partial J_{2}}&= \beta_{0}\sin^{2}\psi_{2} + \beta_{0}\sqrt{\frac{J_{1}}{J_{2}}}\sin\psi_{2}\cos\psi_{1}, \\
        y\frac{\partial y}{\partial J_{2}}&= \beta_{0}\cos^{2}\psi_{2} + \beta_{0}\sqrt{\frac{J_{1}}{J_{2}}}\cos\psi_{2}\sin\psi_{1}.
    \end{split}
    \label{eq:differentials}
\end{equation}
Introducing particle amplitudes $a_{1}=\sqrt{2J_{1}/\epsilon_{1}}$ and $a_{2}=\sqrt{2J_{2}/\epsilon_{2}}$, lets us to write Eq.~\eqref{eq:circmodecoordparam} as:
\begin{equation}
    \begin{split}
        x&= a_{1}\sigma_{1}\cos\psi_{1} + a_{2}\sigma_{2}\sin\psi_{2}, \\
        y&=a_{1}\sigma_{1}\sin\psi_{1} + a_{2}\sigma_{2}\cos\psi_{2}.
    \end{split}
    \label{eq:amplitudebasedeq}
\end{equation}
Here, all beta functions are equal; therefore, the rms beam sizes within each eigenmode are equal: $\sigma_{1x}=\sigma_{1y}\equiv\sigma_{1}$ and $\sigma_{2x}=\sigma_{2y}\equiv\sigma_{2}$. The tune-shift expression in Eq.~\eqref{eq:tunespreadintegralbefore} must be betatron-phase averaged. To phase average effectively, we write the Gaussian exponent in Eq.~\eqref{eq:tunespreadintegralbefore} coordinates using the formulas in Eq.~\eqref{eq:amplitudebasedeq},
\begin{equation}
    \begin{split}
        \exp\left( -\frac{x^{2}t}{2\sigma_{b}^{2}} - \frac{y^{2}t}{2\sigma_{b}^{2}} \right)\exp\left( -\frac{\bar{z}^{2}t}{2\sigma_{\bar{z}}^{2}} \right) &= \exp \left( -\frac{t}{2\sigma_{b}^{2}} (a_{1}^{2}\sigma_{1}^{2} + a_{2}^{2}\sigma_{2}^{2} + 2a_{1}a_{2}\sigma_{1}\sigma_{2}\sin(\psi_{1}+\psi_{2})) \right)\\
        &\times \exp\left(-\frac{t}{2}a_{z}^{2}\cos^{2}\psi_{z}\right), \\
        &= \exp\left(-\frac{t}{2\sigma_{b}^{2}}(a_{1}^{2}\sigma_{1}^{2} + a_{2}^{2}\sigma_{2}^{2}) \right)\\
        &\times \exp\left( -\frac{t}{\sigma_{b}^{2}}(a_{1}a_{2}\sigma_{1}\sigma_{2}\sin(\psi_{1}+\psi_{2}))\right) 
        \exp\left(-\frac{t}{2}a_{z}^{2}\cos^{2}\psi_{z} \right). 
    \end{split}
    \label{eq:gaussianexponent}
\end{equation}
To perform betatron phase average of Eq.~\eqref{eq:tunespreadintegralbefore}, we employ the following identities:
\begin{equation}
    \begin{split}
        &\frac{1}{2\pi}\int_{0}^{2\pi}d\psi \exp\left( a\cos^{2}\psi \right)= e^{a/2}I_{0}[\frac{a}{2}], \\
        & \frac{1}{2\pi}\int_{0}^{2\pi}d\psi \sin(\psi)e^{a\sin\psi} = I_{1}[a],
    \end{split}
\end{equation}
Here, $I_{0}$ and $I_{1}$ are modified Bessel functions. Combining Eqs.~\eqref{eq:tunespreadintegralbefore}, \eqref{eq:differentials}, and \eqref{eq:gaussianexponent} yields the tune-spread expressions:
\begin{equation}
	\begin{split}
		\Delta Q_{1}&= -C_{sc}\int_{0}^{1}dt\frac{\beta_{1}}{\sigma_{b}^{2}}\left(I_{0}(2\tilde{a}_{1}\tilde{a}_{2}t) - \frac{\tilde{a}_{2}}{\tilde{a}_{1}}I_{1}(2t\tilde{a}_{1}\tilde{a}_{2}) \right)I_{0}(\frac{a_{z}^{2}t}{2})\exp\left(-\frac{t}{2}(\tilde{a}_{1}^{2} + \tilde{a}_{2}^{2}+2a_{z}^{2})\right), \\
		\Delta Q_{2}&= -C_{sc}\int_{0}^{1}dt\frac{\beta_{2}}{\sigma_{b}^{2}}\left(I_{0}(2t\tilde{a}_{1}\tilde{a}_{2}) - \frac{\tilde{a}_{1}}{\tilde{a}_{2}}I_{1}(2t\tilde{a}_{1}\tilde{a}_{2}) \right)I_{0}(\frac{a_{z}^{2}t}{2})\exp\left(-\frac{t}{2}(\tilde{a}_{1}^{2} + \tilde{a}_{2}^{2}+2a_{z}^{2})\right),
	\end{split}
\end{equation}
where $\tilde{a}_{1} = a_{1}\sigma_{1}/\sigma_{b}$ and $\tilde{a}_{2}=a_{2}\sigma_{2}/\sigma_{b}$.

\clearpage

\bibliographystyle{unsrt}      
\bibliography{main}

\begin{thebibliography}{10}

\bibitem{nagaitsev2021accelerator}
S~Nagaitsev, Z~Huang, J~Power, J-L Vay, P~Piot, L~Spentzouris, J~Rosenzweig, Y~Cai, S~Cousineau, M~Conde, et~al.
\newblock Accelerator and beam physics research goals and opportunities.
\newblock {\em arXiv preprint arXiv:2101.04107}, 2021.

\bibitem{burov2013circular}
Alexey Burov.
\newblock Circular modes for flat beams in the lhc.
\newblock {\em Physical Review Special Topics-Accelerators and Beams}, 16(6):061002, 2013.

\bibitem{derbenev1998adapting}
Ya~Derbenev.
\newblock Adapting optics for high energy electron cooling.
\newblock {\em University of Michigan, UM-HE-98-04}, 1998.

\bibitem{kim2025experimental}
Seongyeol Kim, Philippe Piot, Gongxiaohui Chen, Scott Doran, Wanming Liu, Charles Whiteford, Eric Wisniewski, and John Power.
\newblock Experimental demonstration of cascaded round-to-flat and flat-to-round beam transformations.
\newblock {\em Nuclear Instruments and Methods in Physics Research Section A: Accelerators, Spectrometers, Detectors and Associated Equipment}, 1072:170206, 2025.

\bibitem{groening2021particle}
L~Groening, C~Xiao, and M~Chung.
\newblock Particle beam eigenemittances, phase integral, vorticity, and rotations.
\newblock {\em Physical Review Accelerators and Beams}, 24(5):054201, 2021.

\bibitem{danilov2003self}
V~Danilov, S~Cousineau, S~Henderson, and J~Homles.
\newblock Self-consistent time dependent two dimensional and three dimensional space charge distributions with linear force.
\newblock {\em Physical Review Special Topics- Accelerators and Beams}, 6(9):094202, 2003.

\bibitem{holmes2018injection}
JA~Holmes, T~Gorlov, NJ~Evans, M~Plum, and S~Cousineau.
\newblock Injection of a self-consistent beam with linear space charge force into a ring.
\newblock {\em Physical Review Accelerators and Beams}, 21(12):124403, 2018.

\bibitem{courant1958theory}
Ernest~D Courant and Hartland~S Snyder.
\newblock Theory of the alternating-gradient synchrotron.
\newblock {\em Annals of physics}, 3(1):1--48, 1958.

\bibitem{burov2002circular}
Alexey Burov, Sergei Nagaitsev, and Yaroslav Derbenev.
\newblock Circular modes, beam adapters, and their applications in beam optics.
\newblock {\em Physical Review E}, 66(1):016503, 2002.

\bibitem{kim2003round}
Kwang-Je Kim.
\newblock Round-to-flat transformation of angular-momentum-dominated beams.
\newblock {\em Physical Review Special Topics-Accelerators and Beams}, 6(10):104002, 2003.

\bibitem{lebedev2010betatron}
Valeri~A Lebedev and SA~Bogacz.
\newblock Betatron motion with coupling of horizontal and vertical degrees of freedom.
\newblock {\em Journal of Instrumentation}, 5(10):P10010, 2010.

\bibitem{edwards1973parametrization}
D.~A. Edwards and L.~C. Teng.
\newblock Parametrization of linear coupled motion in periodic systems.
\newblock {\em IEEE Transactions on Nuclear Science}, 20(3):885--888, 1973.

\bibitem{willeke1989methods}
Ferdinand Willeke and Gerhard Ripken.
\newblock Methods of beam optics.
\newblock In {\em AIP Conference Proceedings}, pages 758--819. American Institute of Physics, 1989.

\bibitem{gilanliogullari2025formalism}
Onur Gilanliogullari, Brahim Mustapha, and Pavel Snopok.
\newblock A formalism for the transport and matching of coupled beams in accelerators.
\newblock {\em arXiv preprint arXiv:2505.08987}, 2025.

\bibitem{groening2018extension}
Lars Groening, C~Xiao, and M~Chung.
\newblock Extension of busch’s theorem to particle beams.
\newblock {\em Physical review accelerators and beams}, 21(1):014201, 2018.

\bibitem{hannon2019transverse}
Fay Hannon and Mark Stefani.
\newblock Transverse uncorrelated emittance diagnostic for magnetized electron beams.
\newblock {\em Physical Review Accelerators and Beams}, 22(10):102801, 2019.

\bibitem{TRACKweb}
Argonne~National Lab.
\newblock {\em TRACK Simulation}, 2025.
\newblock Available at \url{https://www.phy.anl.gov/atlas/TRACK/}.

\bibitem{brinkmann2001low}
R~Brinkmann, Ya~Derbenev, and K~Fl{\"o}ttmann.
\newblock A low emittance, flat-beam electron source for linear colliders.
\newblock {\em Physical Review Special Topics-Accelerators and Beams}, 4(5):053501, 2001.

\bibitem{halavanau2017magnetized}
Aliaksei Halavanau, Jibong Hyun, Daniel Mihalcea, Philippe Piot, Tanaji Sen, Jayakar Thangaraj, et~al.
\newblock Magnetized and flat beam experiment at fast.
\newblock In {\em 8th Int. Particle Accelerator Conf.(IPAC'17), Copenhagen, Denmark, 14{\^a} 19 May, 2017}, pages 3876--3879. JACOW, Geneva, Switzerland, 2017.

\bibitem{groening2011concept}
Lars Groening.
\newblock Concept for controlled transverse emittance transfer within a linac ion beam.
\newblock {\em Physical Review Special Topics-Accelerators and Beams}, 14(6):064201, 2011.

\bibitem{groening2014emtex}
L.~Groening, M.~Maier, C.~Xiao, L.~Dahl, P.~Gerhard, O.~K. Kester, S.~Mickat, H.~Vormann, M.~Vossberg, and M.~Chung.
\newblock Experimental proof of adjustable single-knob ion beam emittance partitioning.
\newblock {\em Physical Review Letters}, 113(26):264802, 2014.

\bibitem{cheon2021mitigation}
Yoo-Lim Cheon, Seok-Ho Moon, Moses Chung, and Dong-O Jeon.
\newblock Mitigation of space-charge-driven resonance and instability in high-intensity linear accelerators via beam spinning.
\newblock {\em Nuclear Instruments and Methods in Physics Research Section A: Accelerators, Spectrometers, Detectors and Associated Equipment}, 1013:165647, 2021.

\bibitem{hoover2021computation}
Austin Hoover, Nicholas~J Evans, and Jeffrey~A Holmes.
\newblock Computation of the matched envelope of the danilov distribution.
\newblock {\em Physical Review Accelerators and Beams}, 24(4):044201, 2021.

\bibitem{hoovereigenpainting}
Nicholas~J. Evans, Austin Hoover, Timofey Gorlov, and Vasiliy Morozov.
\newblock Demonstration of a novel phase-space painting method in a coupled lattice to mitigate space charge in high-intensity hadron beams.
\newblock {\em Phys. Rev. Lett.}, 136:075002, Feb 2026.

\bibitem{franchi2007emittance}
A~Franchi, E~M{\'e}tral, and R~Tom{\'a}s.
\newblock Emittance sharing and exchange driven by linear betatron coupling in circular accelerators.
\newblock {\em Physical Review Special Topics—Accelerators and Beams}, 10(6):064003, 2007.

\bibitem{montague1968fourth}
Brian William St~Leger Montague.
\newblock Fourth-order coupling resonance excited by space-charge forces in a synchrotron.
\newblock Technical report, CERN, 1968.

\bibitem{huebl2022next}
Axel Huebl, Remi Lehe, Chad~E Mitchell, Ji~Qiang, Robert~D Ryne, Ryan~T Sandberg, and Jean-Luc Vay.
\newblock Next generation computational tools for the modeling and design of particle accelerators at exascale.
\newblock {\em arXiv preprint arXiv:2208.02382}, 2022.

\bibitem{sen2023density}
Tanaji Sen.
\newblock Density distributions of tune shifts from space charge or beam-beam interactions in gaussian bunches.
\newblock {\em Phys. Rev. Accel. Beams}, 26:080101, Aug 2023.

\bibitem{xiao2024periodic}
C~Xiao and L~Groening.
\newblock Periodic four-dimensional solution for transport of intense and coupled coasting beams through quadrupole channels.
\newblock {\em Physical Review Accelerators and Beams}, 27(3):031602, 2024.

\end{thebibliography}

\end{document}